\documentclass[5p]{elsarticle}

\usepackage{lineno,hyperref}

\usepackage{subcaption}
\usepackage{longtable}
\usepackage{placeins}
\usepackage{caption}
\usepackage{xcolor}
\usepackage{ifthen}

\newcommand{\balanceRationale}{`balance'}
\newcommand{\shapesRationale}{`shapes'}
\newcommand{\algorithmRationale}{`algorithm'}

\newcommand{\pageNameIntro}{Introductory text}
\newcommand{\pageNameInstructions}{Instructions}
\newcommand{\pageNameStory}{Philosopher story}
\newcommand{\pageNameTask}{Programming task}
\newcommand{\pageNameExplanation}{Self explanation of choice}
\newcommand{\pageNameAlternative}{Alternative explanations}
\newcommand{\pageNameRanking}{Ranking life-aspects}
\newcommand{\pageNameWords}{Words suggestions}
\newcommand{\pageNameDemographics}{Demographics}
\newcommand{\onlyForUs}[1]{}
\newcommand{\onlyCHIsubmission}[1]{}
\newcommand{\onlyARXIV}[1]{#1}

\newboolean{withcomments} 
\setboolean{withcomments}{false}
\newcommand{\jj}[1]{\ifthenelse{\boolean{withcomments}}{{\color{blue}JJ: #1}}{}}

\modulolinenumbers[5]

\journal{a journal, under review.}

\biboptions{authoryear}

\begin{document}

\begin{frontmatter}

\onlyARXIV{
\title{Studying the Transfer of Biases from Programmers to Programs}
}
\onlyCHIsubmission{
\title{Programming Like Your Mama Taught You: \\Revealing Human-to-Computer Bias Transfers \\(long version)}
}


%
%

\onlyARXIV{
\author[johannamainaddress]{Johanna Johansen\corref{mycorrespondingauthor}}
\cortext[mycorrespondingauthor]{Corresponding author}
\ead{johanna@johansenresearch.info}
\address[johannamainaddress]{Department of Informatics, University of Oslo}

\author[toremainaddress,toresecondaryaddress]{Tore Pedersen}
\ead{tore.pedersen@feh.mil.no}
\address[toremainaddress]{Center for Intelligence Studies, Norwegian Defence Intelligence School}
\address[toresecondaryaddress]{Department of Psychology, Bj\o{}rknes University College}

\author[christianmainaddress]{Christian Johansen}
\ead{cristi@ifi.uio.no}
\address[christianmainaddress]{Department of Technology Systems, University of Oslo}
}
\onlyCHIsubmission{
\author{Anonymous Authors}
}

\begin{abstract}
It is generally agreed that one origin of machine bias is resulting from characteristics within the dataset on which the algorithms are trained, i.e., the data does not warrant a generalized inference. We, however, hypothesize that a different `mechanism', hitherto not articulated in the literature, may also be responsible for machine's bias, namely that biases may originate from (i) the programmers' cultural background, such as education or line of work, or (ii) the contextual programming environment, such as software requirements or developer tools. Combining an experimental and comparative design, we studied the effects of cultural metaphors and contextual metaphors, and tested whether each of these would `transfer' from the programmer to program, thus constituting a machine bias. The results show (i) that cultural metaphors influence the programmer's choices and (ii) that `induced' contextual metaphors can be used to moderate or exacerbate the effects of the cultural metaphors. This supports our hypothesis that biases in automated systems do not always originate from within the machine's training data. Instead, machines may also `replicate' and `reproduce' biases from the programmers' cultural background by the transfer of cultural metaphors into the programming process. Implications for academia and professional practice range from the micro programming-level to the macro national-regulations or educational level, and span across all societal domains where software-based systems are operating such as the popular AI-based automated decision support systems. 
\end{abstract}


\end{frontmatter}

\section{Introduction}

Biases are manifestations of incorrect judgments resulting from cognitive tendencies that humans exhibit in situations of uncertainty or when there is insufficient information but a judgment must be made nevertheless. Biases are often difficult to study because of the complex thinking `machinery' that makes up the human brain and because of the human's interaction with its complex social environment. Moreover, people are usually unaware of their own biases and they may even be prone to rationalize their own biased tendencies. Nevertheless, by employing carefully designed experiments, both the specific psychological mechanisms that lead to biases and the occurrences of specific human biases have successfully been identified and well established (e.g. \cite{gilovich2002heuristics,tversky1974judgment,oliver2014satisfaction,wilson2003affective}).

Due to the increase of power and societal penetration of artificial intelligence (AI) algorithms, the occurrence of artificial biases have been observed and described, something which is rather counter to our intuition that  machines are unbiased and objective. Biases in AI have a strong negative impact in society, prompting organizations like ACM and the European Parliament to issue strong statements \cite{council2018statement,STOA2019} and prominent researchers to publish lengthy reports \cite{brundage2018malicious} warning against biased AI.

The occurrence of biases in AI have been observed to appear due to biased training data on which these algorithms are built \cite{feldman2015certifying,Mittelstadt2016TheEthics,caliskan2017semantics,silva2019algorithms}. For example, if the limited sample-data that the algorithms are trained on are not sufficiently representative of the larger population-data that the algorithms are subsequently unleashed upon, then the algorithmic judgments would not be valid but instead biased. Or, for example, if the training data shows a strong relationship between two variables, say, ethnicity and crime \cite{zou2018AI,dressel2018accuracy}, a bias may occur because it is assumed that the relationship is causal when the relationship between the two variables may in fact be non-causal and instead be caused by a third variable, say, poverty.

However, we have reasons to believe that there is a second source of biases in algorithms/programs, which has been little investigated \cite{Kirkpatrick2016Battling}. We believe (and argue in this paper) that biases may also be transferred from the human programmer into the final artifact, i.e., the program. Transfer (or contagion) of biases between humans is well-known, such as, for example, the conformity bias (e.g., \cite{moscovici1972social}). Additionally, a transfer of biases due to influence on humans from social and cultural institutions like media or education is equally known (e.g., \cite{bourdieu1977Reproduction,lakoff2008metaphors}). Given the human cognitive tendencies (explained in Section~\ref{sec_CognitiveBiases}) to employ inappropriate mental judgment-modes in situations that are ``uncertain'', combined with influences from institutional agendas, human biases are ubiquitous. There is, however, yet no evidence to support the related possibility of programmer biases being encoded, in some way, in their programming artefacts. In fact, some argue against this view by claiming that programmers are ``immune'' to biases due to their technical training and tools/methods used.

\paragraph{Target Audience}

The work reported in this paper is relevant for researchers from several fields. First of all, people working in AI and machine learning can be interested in our proposal that biases in machine learning can come not only from the data but also from the people programming the algorithms. We study this to some considerable detail, as we explain in the rest of this introduction. Second, people working in psychology and cognitive sciences can be interested in this new application that we propose, where they can apply their skills and methods to study this new form of human bias and its transfer to machines. Third, practitioners working with software engineering or managing software development teams can be interested in studying more various programming environments and tools to see how much human bias is transferred to the programs in each situation. Finally, at a macro level, both governments, private business enterprises, and NGOs would become aware of machine bias originating from human programmers who unknowingly transfer the influences from their own cultural backgrounds to the machine programs. Thus, the target audiences are diverse and would benefit both on a micro level, e.g., in research and development, and in (computer science) education, as well as on a macro level, e.g., in issuing improved knowledge-informed national regulations on the domains where automated decision-support systems operate.

\subsection{Two Main Questions}

The present study investigates methodically, experimentally, and empirically the hypothesis of bias transfer in programming, providing a first convincing argument for inspiring more empirical studies to be taken in the same direction. As such, in this study our main focus is to find support for (or against) the hypothesis that people may unknowingly and inadvertently transfer biases to computer programs that they build. In particular, this work supports the hypothesized existence of an alternative mechanism than the one that is already known, that may render AI biased, i.e., in addition to the well-studied cases of biases originating from data. However, we do not study specific biases (s.a., gender or racial), nor do we test or suggest specific programming methods and tools that could counteract bias transfers, as this would be a task for future research.

The two main hypotheses that we set to investigate are:
\begin{enumerate}[I.]
\item\label{mainQuestionTransfer} \emph{Are biases being transferred from the human programmer to the program artefact?}\\
This is studied in a basic form with a bias revealing test that we detail in Section~\ref{sec_BiasRevealing}, which we impose on the subjects of our study, as described in Section~\ref{sec_DesigningTheStudy}. The biases that we study are of both cultural and contextual nature.
\item\label{mainQuestionPriming} \emph{Can programmers be manipulated, i.e., primed  by inducing a new bias, and is this new bias then transferred to the program?}\\
In Section~\ref{sec_MetaphorsAsPrimingMethod} we describe the method that we use to prime our study subjects towards the same biases studied for the first question. Subsequent sections then describe how we used the priming in our studies and their outcomes.
\end{enumerate}

One can see several immediate benefits of the present study alone. For example, in education one could measure how well programming courses train the students, by measuring the bias transfer-rate at the start and end of the courses. Another example is to measure how effective some technology quality assurance method is at removing or identifying programmer's biases, like testing frameworks, peer programming, abstract/detailed specifications, code generators, etc. Moreover, regarding the growing population of `lay' programmers in the smart-living and IoT-ubiquitous programming environments of today (i.e., almost everyone in technologically `modern' societies) both business companies and consumers would benefit from more insight into the non-conscious influence of culture and context on the programming choices that are made by the `novice' programmer that has no formal training. In terms of education and learning, we argue that this insight could be used to help consumers become more aware of the cultural and contextual influences that shape their cognitive tendencies when they are programming.

\subsection{Key Methodological Aspects}\label{subsec_Intro_Methods}

We employ several different methods to help us divide and detail these two general hypotheses. In short, here are the main aspects that are specific to our studies; details about our methods can be found throughout the paper and especially in Section~\ref{sec_PerformingTheStudy}.

\begin{enumerate}[A.]
\item\label{methodBiasRevealingTask} We develop a cognitive task intended to reveal biases originating from the cultural background of a programmer, such as education, line of work, and free-time activities. 
\item\label{methodSkillLevels} We investigate users with different programming skill levels, i.e., from professional to amateur. 
\item\label{methodGeneralBias} We investigate a very general form of bias, but well hidden inside a programming task. This is because the programming task needs to be appropriate also for people with little or no programming skills. 
\item\label{methodEffectInducingBias} We investigate whether \emph{inducing} a bias is effective. 
\item\label{methodBiasProgrammers} We investigate whether people educated in programming exhibit less biases and are less prone to manipulation.
\end{enumerate}

\paragraph{Aspect \ref{methodBiasRevealingTask}} To reveal the existence or nonexistence of cultural biases, we decided that the respondents in our studies should have different educational/professional backgrounds, i.e., from  social and natural sciences, as well as from arts and cultural studies; see Section~\ref{subsec_TheParticipants}.

\paragraph{Aspect \ref{methodSkillLevels}} is motivated by the observation that increasingly more lay people (wrt. programming) are interacting and designing rather complex systems. Nowadays it is not only expert developers that program, but people with all levels of expertise carry out various programming-like tasks, from simple configurations of IoT systems in their smart home, to more complex configurations of technology systems in their work, to more unconventional forms of programming using visual languages and/or domain specific languages, and even assembling ready-programmed components into a final software system as done, e.g., in the IBM's IoT development environment\footnote{IBM Watson IoT Platform  \url{https://www.ibm.com/internet-of-things/solutions/iot-platform/watson-iot-platform}}. This is happening mainly because of the proliferation of simple (abstract, graphical, etc.) programming languages and interfaces aimed at non-programmers to design domain specific information systems, e.g.: a biologist programming a DNA search or an oil-engineer programming a complex database search. Therefore, in our study we use a simple programming task presented as fictitious, i.e., imagining to be programming. This allows us to perform our study both on programmers and non-programmers. 

\paragraph{Aspect \ref{methodGeneralBias}} is important in order to avoid any intrinsic de-biasing, and thus we must ``hide'' from the subjects the real goal of the study behind a seemingly unrelated goal, i.e., the ``study of natural language in programming''. We must avoid giving any indications or hints to the study subjects about our intention to study their cognitive tendencies during the programming task. 

\paragraph{Aspect \ref{methodEffectInducingBias}} is a standard study approach in research on biases because many types of biases can also be experimentally induced, sometimes termed priming \cite{tulving1990priming,yonelinas2002nature}. We thus also study the influence of contextual metaphors, in addition to the cultural bias, from \emph{\ref{methodBiasRevealingTask}}. Once we have established in this paper whether or not priming also works in the setting of programming biases, other works can carry out more detailed studies about whether such priming already exists ``out there'' (intentionally or not) and what types of priming would work and to what degree.

\paragraph{Aspect \ref{methodBiasProgrammers}} is meant to investigate a considerable opposition that our idea has encountered, i.e., that programmers cannot be biased when writing code. Moreover, we also aim to study whether it is possible to experimentally induce a bias on this category of users, or if this particular category is more resistant to priming and bias-transfer to programs than other user categories. 

\vspace{2ex}
Our present work is motivated by the need to prove or disprove the idea that human biases could be transferable to the programming artifacts. However, which types of biases and how dangerous these might be are not the subject of  this study. Other specific studies would have to be devised, maybe similar to the research on human biases developed in the psychology field. 


\subsection{Overview of the Paper}

\begin{figure*}[t]
\centering
\includegraphics[width=\textwidth]{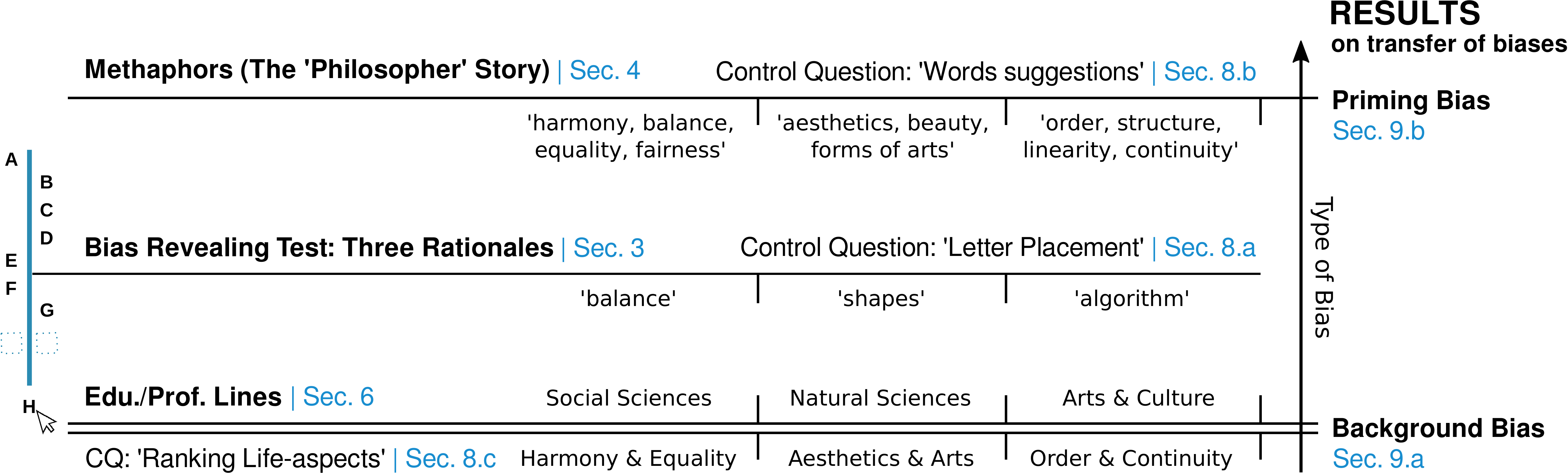}
\caption{Overview of the contributions of the paper with references to relevant sections.} 
\label{fig_paperOverview}
\end{figure*}

The contributions of this paper are presented schematically in the diagram from Figure~\ref{fig_paperOverview}. The vertical axis of the diagram represents types of biases (or degrees of influence) with the arrow indicating a direction from more general, long-term, and strong biases to more contextual, short-lived, and weak forms of biases. In this paper we study two extremes, namely biases coming from the background of a person (including socio-cultural, educational, and professional influences), and biases coming from manipulations such as priming. However, in between these two extremes one may study other forms of biases coming, e.g., from working cultures (think of programmers working for Google compared to a startup), or from media or propaganda. These biases form over a considerable period, e.g., several years, which is shorter than a life-time or childhood period, but longer than minutes as it is the case with priming.

Our results will be presented in the end of the paper, in Section~\ref{sec_Results}, showing how these two types of biases are being transferred from humans to programs, providing evidence that transfer exists in both cases. In order to support such a new claim, the rest of the paper is devoted to the development of our studies and analysis of the data. Importantly, Section~\ref{sec_ControlQuestions} contains the analysis from our control questions, which were meant to help us understand better whether the main elements (and assumptions) of our study were correct. The first important element of our study is presented in Section~\ref{sec_BiasRevealing} where we develop a bias revealing test, which is used for both types of biases.
This cognitive task allows respondents to answer only with one of the three rationales that are listed on the middle line of the diagram. 
In Section~\ref{subsec_LeftRightPlacement} we analyse how well our test worked, using one of our control questions. 
Section~\ref{sec_MetaphorsAsPrimingMethod} details the manipulation method that we used, involving priming participants with metaphors hidden in a fictitious `Philosopher' story. How well these manipulations worked is studied again with a control question involving listing of `similar words' in Section~\ref{subsec_WordsSuggestions}. We created three metaphors to match the three rationales, which in turn match with three kinds of educational lines (or views on life). This correspondence is reflected in the vertical alignment of these elements in the diagram.
We have thus chosen our participants to represent these three different backgrounds, as detailed in Section~\ref{sec_PerformingTheStudy}.
To test the assumptions about our participants' backgrounds we used one control question (analysed in Section~\ref{subsec_LifeAspects}) asking them to rank the three `life-aspects' listed on the bottom line of the diagram. 
The rest of the paper is devoted to presenting in Section~\ref{sec_DesigningTheStudy} the major phases of designing our surveys and our studies using usability testing, and analysing the data and demographics, in Section~\ref{sec_AnalysingTheData}.

\section{Cognitive Biases in Human Judgment}\label{sec_CognitiveBiases}

Numerous robust experiments in psychological science studying human judgment and decision making over nearly five decades have unanimously arrived at the now well-established fact that humans have a tendency to exhibit cognitive biases. Contrary to making an error, which represents a single incident in which one makes an incorrect judgment, \emph{a bias} arises from a systematic tendency to commit the same type of error over time or in different situations. Thus, a cognitive bias can be understood as a sustained tendency to make an incorrect judgment. The tendency to exhibit a cognitive bias is particularly prominent in situations or contexts that are characterized by uncertainty, e.g., when processing information that are either too voluminous or too complex for the human brain to handle, or when forced to make a rapid judgment in a time-frame that is too short to review the information at hand, or when there is insufficient information for making the required decision, like in underspecified software requirements in programming. This last aspect, i.e., underspecification, is what we study empirically in this paper.

Uncertain situations have a tendency to prompt the use of automatic and non-conscious cognitive processing. Whereas controlled and conscious cognitive processing, often termed Analytic thinking, System 2-thinking, or ``slow thinking'', is the preferred cognitive mode for arriving at a correct judgment when the situation is in fact uncertain, this mode is not always easily attained. This is because the brain's preferred cognitive mode is the automatic and non-conscious System 1, also termed Intuitive thinking or ``fast thinking''. In situations characterized by certainty, ``fast thinking'' usually works well because we are on a ``familiar terrain''. In uncertain situations, however, System 1 has a tendency to kick in and spark our automatic thinking mode even if it should not. But how does System 1 actually arrive at incorrect judgments? The reason is simply because our brains employ mental shortcuts, which are very useful in  situations characterized by \emph{certainty}. The problem is that our brain employs System 1 also in situations that are characterized by \emph{uncertainty}, when it should in fact consciously activate System 2 \cite{kahneman2011thinking}.

Two key concepts in mental shortcuts, also termed heuristic processing, are mental \emph{accessibility} (closely related to the availability heuristic) and mental \emph{representativeness} \cite{tversky1974judgment,gilovich2002heuristics,thaler2009nudge}. When something is easily retrievable from short-term or long-term memory, we have a tendency to wrongfully regard it as something that is also occurring frequently, although it may not be occurring frequently at all. We may also make an incorrect judgment about an unfamiliar phenomenon by identifying superficial resemblances to a familiar phenomenon. Because it is cognitively effortful to identify substantial similarities between two phenomena, particularly in situations characterized by incomplete information and uncertainty, superficial similarities are more easily identified. In many instances this results in an incorrect judgment. To summarize, situations or contexts characterized by uncertainty in one way or another, prompts an inappropriate cognitive processing in System 1-mode, where the employment of mental shortcuts leads us to arrive at an incorrect judgment, thus exhibiting a cognitive bias.

\subsection{Algorithmic Bias: the Data or the Programmer}

Media and the general public seem to have the assumption that machines and algorithms are in themselves neutral and objective. However, it has been known for quite some time (and recently also came to the attention of several public actors) that complex algorithms, such as those from artificial intelligence among others, may exhibit biases s.a.: racial bias \cite{SchlesingerLetsTalk2018}, gender discrimination \cite{zou2018AI} and other socially relevant types of biases \cite{friedman1996bias,boyd2012critical,jobin2019global}, when processing information in the support of human and institutional decision making \cite{corbett2017algorithmic,dressel2018accuracy}. This phenomenon is commonly labeled as machine biases or algorithmic bias, and has been confirmed in different areas (published in respective top venues), e.g., in big data \cite{hajian2016algorithmic}, web \cite{baeza2016data,baeza2018bias}, autonomous systems \cite{danks2017algorithmic}. Among institutions that have raised concerns about the existence of ``biased algorithms'' are: 
the ACM US Public Policy Council%
\footnote{ACM U.S. Public Policy Council and ACM Europe Policy Committee (2017). Statement on Algorithmic Transparency and Accountability. \url{https://www.acm.org/binaries/content/assets/public-policy/2017_joint_statement_algorithms.pdf}}; 
the EU Parliament\footnote{EU Parliament (2016). EU Framework on algorithmic accountability and transparency. \url{https://www.europarl.europa.eu/doceo/document/E-8-2016-007674_EN.pdf}}; 
the New York City Council which passed a bill on ``Accountability and transparency in algorithms for public agency support''\footnote{New York City Council (2018). A Local Law in relation to automated decision systems used by agencies. \url{http://legistar.council.nyc.gov/LegislationDetail.aspx?ID=3137815&GUID=437A6A6D-62E1-47E2-9C42-461253F9C6D0}}; 
ERCIM \cite{rauber2019transparency}; 
World Wide Web Foundation\footnote{World Wide Web Foundation (2017). ``Algorithmic Account\-ability: Applying the concept to different country contexts''. \url{https://webfoundation.org/docs/2017/07/Algorithms_Report_WF.pdf}.}
and many more \cite{cath2018artificial}. An influential research report \cite{brundage2018malicious} has raised even more concerns about harmful algorithms, and has recently been joined by articles in major publication venues such as Science and Nature \cite{obermeyer2019dissecting,zou2018AI,gianfrancesco2018potential} and by scholarly books \cite{margaret2008mind,o2016weapons}.

All the works above focus on the data that AI algorithms train on, and show how the data contains biases. We are not aware of works that study empirically the transfer of biases from the human programmer to the algorithm, although we have found related ideas mentioned in recent articles: 
\cite{silva2019algorithms} describe nine types of biases (present at five different algorithmic stages: input, algorithmic operations, output, users, and feedback), some of which can be studied in conjunction with the general bias transfer that we demonstrate in this paper (we consider it useful to demonstrate a type of bias through empirical studies as we carry out here).
\cite{baeza2018bias} brings up the users and producers of the web content as sources of bias related to the data, but also points out different forms of bias coming from the user interface made by interaction designers whom could be regarded as `programmers'. We consider it particularly useful to study empirically the different forms of biases described by \cite{baeza2018bias} in the light of our hypothesis of `bias transfer' and using methods similar to what we present in this paper, especially so since the author recognizes in the conclusion the same general sources of biases as we study here, i.e.:
\emph{``each program probably encodes the cultural and cognitive biases of  their creators''}, and points in the introduction ``measuring bias'' as a major challenge, which is what we do here.

To state that algorithms are biased, or to assert that algorithms systematically produce an output that is biased, must as a consequence lead scholars to pose the question of whether a biased output could be proven and whether potential causal mechanisms leading to the bias could also be tested and studied. 

Although an awareness of biases in algorithms has arisen, including the awareness of biases originating from data, no research programs seem to have undertaken the aim to study empirically the (cognitive) mechanisms that may lead to biases in algorithms, i.e., biases that do not originate from data itself, but from the programmers' cultural backgrounds or from contextual influences in the immediate programming environment.

Thus, rather than pointing once more to the problem itself, the present study will instead provide explanations and insights, derived from our scientific study containing empirical evidence of actual human programming behaviour, into why this phenomenon may occur. We operate within the same paradigm and with the same agenda as those who study human behaviour, that is, we follow in the path of other multidisciplinary research themes such as Behavioral Economics \cite{tversky1974judgment,kahneman1991anomalies}, Behavioral Transportation Research \cite{pedersen2011affective,garling2014handbook} and \onlyARXIV{our own}\onlyCHIsubmission{the} recent contributions termed Behavioral Artificial Intelligence \cite{pedersen2019behavioural} and Behavioural Computer Science \cite{pedersen2018behavioural}.

As mentioned in the introduction, some may argue that robust quality assurance procedures eliminate any instances of biases in algorithms, at least in professional programming environments. We leave out for now testing whether the quality assurance procedures themselves have inherent biases or miss some forms of biases, and focus only on proving that \emph{programmers may be a source of biases}, and not only the data given to the program. 

Because heuristic thinking is seen as the main psychological ``engine'' for generating cognitive biases, our experiments will also employ a heuristic approach, that is, relying on mental shortcuts such as ``accessibility/availability'' when inducing a bias on the participants in the study.

\section{A Bias Revealing Cognitive Task}\label{sec_BiasRevealing}

\begin{figure}[h]
\centering
\includegraphics[width=0.06\textwidth]{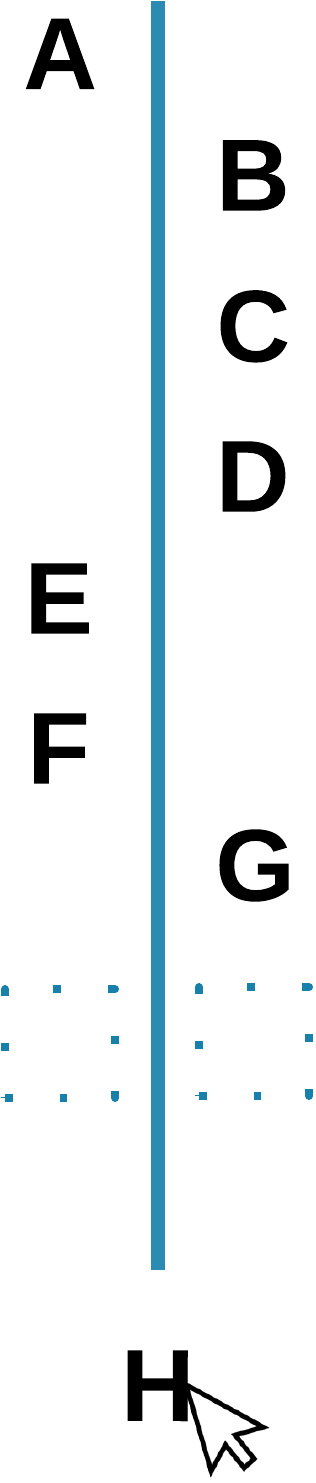}
\caption{The puzzle game.} 
\label{fig_puzzle}
\end{figure}

Over the years, the first author has used a simple cognitive task \cite{townsend2003curious} (originally called Alice's Alphabet Puzzle) in lectures on judgment and decision making. In Townsend's book, this particular puzzle lists all the letters in the English alphabet (i.e., latin, roman) on a horizontal line, where straight lined letters are placed above and curved lined letters are placed below the line. We changed the puzzle in a vertical position to trigger more the infinity of the line, and the balance of the sides. The idea is to make students and professionals in various disciplines reflect on the possibility that facts that we ``see'' in the real world are generally a result of a ``theory'' of how we view the world. In the task, the audience is first shown, as in Figure~\ref{fig_puzzle}, a correct sequence of the letters A to G -- divided by a vertical line where A, E, and F is on the left side, and B, C, D, and G is on the right side -- and then asked to decide on which side of the vertical line (left or right) the next letter H should be placed and \emph{why} it should be there. 

Although some want to place H on the left side of the line, whereas others want to place it on the right side, when asked \emph{why}, they provide distinctly different rationales for their decision. Their rationales always seem to fall into three categories, which we categorized as: (\ref{rationaleBalance}) \balanceRationale, (\ref{rationaleShapes}) \shapesRationale, or (\ref{rationaleAlgorithm}) \algorithmRationale. The quintessence of their arguments are as follows: 

\begin{enumerate}[I.]
\item\label{rationaleBalance} Some argue that there should be an equal number of letters on each side of the line: since there are already four letters on the right side and only three on the left side, the next letter, H, should go on the left side, thus indicating a sense of \emph{`balance'}. 
\item\label{rationaleShapes} Others argue that the straight-lined letters A, E, and F are on the left side of the line, whereas the curved-shaped letters B, C, D, and G are on the right side, something which makes it perfectly reasonable that H should be together with its ``kin'' on the left side, given the different \emph{`shapes'} of the letters.
\item\label{rationaleAlgorithm} Yet others argue that there is an inherent order (or pattern) in the sequence: e.g., some indicate the sequence Left-1, Right-3, Left-2, Right-4, thus suggest placing H on the right side due to perceiving the image/puzzle as having the characteristics of an \emph{`algorithm/pattern'}.
\end{enumerate}

This exercise is simple enough to reveal cognitive tendencies
of System 1, instead of triggering the System 2 analytical thinking, as usually done by more complicated tasks.
More importantly, the bias cannot be avoided because there is nothing else in the picture to help them when making the decision, and any placement is correct; therefore only something from the background of the subject, alternatively, an experimentally induced prime, could help with making the decision. Moreover, since the subjects are asked ``Why?'', they most often find an explanation in the memory related to the decision, or make up an explanation after the fact; though sometimes they answer that they choose at random.

\section{Metaphors as Priming Method}\label{sec_MetaphorsAsPrimingMethod}

We use the above task to force revealing a bias, albeit an innocent one (compared to racial or gender), which would have its origin in the cultural background of the person (e.g., education, line of work, hobbies).
This will be used to test our first main hypothesis that we explained in the introduction, namely that cultural metaphors would influence the programmers' choices. Besides that, in order to test our second hypothesis, we want to test whether we can prime the subjects 
to non-consciously make a decision in one specific ``direction'' -- being one of the three rationales that we identified in Section~\ref{sec_BiasRevealing}.

Our motivation for hypothesizing that programmers would non-consciously be affected by the prime, comes from the well-known effect of the availability heuristic and the representativeness heuristic, together with the anchoring heu\-ristic \cite{gilovich2002heuristics}.
Under conditions of uncertainty, where one does not \textit{know}, but nevertheless has to make a judgment or a choice, one will non-consciously base one's judgments either on instances that spring easily to mind (i.e., the cultural background or the contextual prime triggers the availability heuristic), or on instances that resemble the current problem (the cultural background or the contextual prime triggers the representativeness heuristic).
The judgment can also be anchored as an approximation to the most recent, the most related or the most relevant information; the context of the prime triggers the use of the anchoring heuristic. Thus, as regards cognitive processing, heuristic thinking in System 1 mode is very much an associative reasoning mode influenced by cognitive availability and perceived representativeness. However, one needs to also consider the \textit{content} of the heuristic processing mode, e.g., what is actually easily accessible in memory. In terms of associative reasoning, the metaphors and metaphorical thinking are a strong source of influence on how we as humans view the world.

The essence of a metaphor is, according to \cite{lakoff2008metaphors}, simply that ``we understand and experience one kind of thing in terms of another''. For example, an argument may be understood and experienced in terms of the metaphor \textsc{war}, where we may ``attack weak points in others' arguments'', we may ``shoot down'' others' arguments, and we may ``win or lose'' arguments. Similarly, we may perceive the concept of time in terms of the metaphor \textsc{money}, where we may ``waste each other's time'', or `` save time'', or even ``budget our time'', and ``borrow'' time. In fact, metaphors are so pervasive and ubiquitous in our lives that we simply cannot do without them. Think for example of how the term ``happy'' is understood, experienced, and communicated in terms of \textsc{up}, whereas ``sad'' is understood in terms of \textsc{down}, showing that affective states, and even human health, consciousness, or control, are understood and experienced in terms of directionality. The good things are \textsc{up} and the bad things are \textsc{down}. 

The concept of metaphor, ``understanding one kind of thing in terms of another'' as Lakoff and Johnson put it, particularly the way we perceive the world in terms of metaphors, is something that can even be manipulated. Consider for example \cite{thibodeau2011metaphors} study of the effect of metaphor on the general public's preferences for crime-prevention measures. When reporting crime-rates in a fictitious city, crime was either described in terms of ``a beast'' or in terms of ``a virus''. When exposed to the metaphor \textit{crime is a beast}, the general public argued for harsher and more severe crime-preventing measures than what was the case when they were exposed to the metaphor \textit{crime is a virus}. Thus, not only are these ``metaphors that we think with'' \cite{lakoff2008metaphors} something that we employ constantly in order to make sense of the world, but the way we use metaphors to experience the world is also something that can be manipulated. This can make us alter our view of the world, and most of the time we are not aware, neither of the fact that we think metaphorically, nor that our metaphorical thinking can be manipulated by governments, media, our employers, or others, either for commercial or political purposes. 

The employment of metaphors, as well as the manipulation of our employment of metaphors, are processes that mostly go on outside our conscious awareness. Thus, since metaphorical thinking is so pervasively and ubiquitously present in our understanding, our experiencing, and our sense-making of the world, metaphors are exceptionally well suited for studying how the ultimate purpose of a computer program may be affected both by the programmer's cultural metaphors, as well as whether one can influence the programmer's initial understanding of the program's purpose by a manipulation, i.e., inducing new metaphors that would alter the initial metaphors that the programmer employs. 

In the case that the programming specifications are too sparse to completely fulfill the formal purpose of the computer program, how does the programmer `fill in the blanks'? We suggest that this is done by employing meta\-phorical thinking, either in terms of ingrained metaphors, i.e., cultural background, or in terms of cues in the proximate contextual environment that triggers metaphorical thinking, i.e., our  manipulation of contextual metaphors.

\subsection{Experimental Manipulation Using Metaphors}\label{subsec_MetaphorsAsPrimingMethod_Experimental}

The experimental manipulation that we use consists of three experimental conditions in the form of `a story about a philosopher who invented a puzzle' which we manipulated by varying the embedment of a different `life-aspect', i.e., one specific metaphor, forming three different versions of the story. We also had one control condition, i.e., a story that did not contain any life-aspects/metaphor, intended for a comparison to the experimental groups. The three different life-aspects were:

\begin{enumerate}[A.]
\item\label{lifeAspectHarmony} \textsc{harmony and equality}
\item\label{lifeAspectArts} \textsc{aesthetics and arts}
\item\label{lifeAspectOrder} \textsc{order and continuity}
\end{enumerate}

The metaphor, which is given in detail below, included four words, placed in two groups of two words, one in the beginning of the story, and the other in the end, following indications from relevant literature \cite{lakoff2008metaphors,thibodeau2011metaphors}. The words were:

\begin{enumerate}[A.]
\item\label{metaphorHarmony} \textit{harmony and balance;} then \textit{equality and fairness}
\item\label{metaphorArts} \textit{aesthetics and beauty;} then \textit{forms of arts}
\item\label{metaphorOrder} \textit{order and structure;} then \textit{linearity and continuity}
\end{enumerate}

\textit{We hypothesized} that each of the above life-aspects would metaphorically influence the participants in the respective group A/B/C to provide an explanation that could be interpreted as one of the rationales from Section~\ref{sec_BiasRevealing}, respectively \textit{rationale \ref{rationaleBalance}/\ref{rationaleShapes}/\ref{rationaleAlgorithm}} (i.e., \balanceRationale, \shapesRationale, \algorithmRationale). 

The metaphorical primes were embedded in the following fictional brief story about the philosopher who was presented as the one who originally created the puzzle from Section~\ref{sec_BiasRevealing}. Each subject will read a story that differs only in the words shown inside square brackets below, i.e., the first word pair alternative is provided to Group 1, the second pair to Group 2, and the third pair to Group 3.

\begin{quote}\begin{itshape}
\textit{``A philosopher who lived a life filled with \emph{[}~harmony and balance $|$ aesthetics and beauty $|$ order and continuity~\emph{]} created the riddle used in the game that we ask you to imagine that you program on the next page. Although the philosopher is nearly forgotten today, we know that the philosopher influenced many contemporary philosophers' view of the world. The most prominent influence seems to have been the importance of maintaining \emph{[}~equality and fairness $|$ forms of arts $|$ linearity and continuity~\emph{]} in life and in society.''}
\end{itshape}\end{quote} 

We also had a control group in order to compare the effect of the different primes in each of the three experimental groups to a neutral condition without any metaphor-priming. In this control group, `ethics' was embedded in the story, as this concept is unrelated to the three metaphors. The following story was given to the control group.

\begin{quote}\begin{itshape}
\textit{``A philosopher created the riddle used in the game that we ask you to imagine that you program on the next page. Although the philosopher is nearly forgotten today, we know that the philosopher influenced many contemporary philosophers' view of the world. The most prominent influences seem to have been the importance of ethics in life and society.'' }                                                                                                                                                                                                                                                                                                                                                                       \end{itshape}\end{quote}

\section{Designing the Study}\label{sec_DesigningTheStudy}

We have spent considerable effort on designing our studies. Since the hypothesis that we want to test (i.e., that programmers can transfer their biases to the programming artefacts) seems to be quite hard to accept (at least we have seen this as the common reaction), we needed to build a case backed by strong, data-driven evidence. 

First, we incorporate the two instruments that we described before, i.e., the bias revealing cognitive task and the priming metaphor, into a programming task. For now, we focus on a ``paper-task'', where the subject imagines to be programming, so that we can easily involve non-programmers, since part of our hypothesis is that in the near future (if not already so) people with various backgrounds (outside computer science) would be involved in various ``types of programming''. Examples that we are aware of include, e.g., ``configuration'' tasks as when managing a cloud environment with AWS or when using the IBM's Watson IoT Platform for building IoT systems from software components plugged together on a graphical interface; or DSLs (domain specific languages) used in many disciplines that interact with software; or visual programming languages such as Fraunhofer's IoT Programming Language NEPO\footnote{``Fraunhofer IAIS IoT Programming Language NEPO Roberta$^{\textregistered}$ Lab'' by Thorsten Leimbach and Daria Tomala. In ERCIM News 120, January 2020, Special theme: Educational Technology. \url{https://ercim-news.ercim.eu/en120/special/fraunhofer-iais-iot-programming-language-nepo-in-the-open-roberta-lab}} or Google's Blockly. 

We then continue to describe in Section~\ref{subsec_TheSurvey} the survey in which we have incorporated the programming task described in Section~\ref{subsec_TheProgrammingTask}. We explain each of the survey's questions and their purposes, as well as our rationale for ``hiding well'' the priming metaphor that is hypothesised to result in a bias.
Besides the programming task, the survey contains additional questions meant to collect information intended for different purposes, from identifying `unserious subjects', i.e., subjects that did not pay sufficient attention to the task, but instead responded more or less randomly, to information intended to help with the interpretation of the subjects' explanations of their rationales or their background (see also Section~\ref{sec_AnalysingTheData}).

We carried out our work in two stages. First we performed pilot studies, which we used to learn of flaws and to make improvements to the design. In this section we describe the steps and techniques that we used to arrive at our final studies. In particular, we first carried out specific usability testing in one pilot survey (described in Section~\ref{subsec_PilotTesting}). Then we improved the survey by using eye tracking technology to make sure that the priming is being read and to see more of how the users/subjects would interact with the survey (see Section~\ref{subsec_EyeTracking}). 

In the second stage, three studies were done in full-scale with all the planned groups/cohorts from different cultural backgrounds: people with background in -- or/and working in a field related to -- social sciences (psychology), natural sciences (informatics), and visual arts \& design, and performing arts (theatre and music). The results that we report in Section~\ref{sec_PerformingTheStudy} are from these three full-scale studies. We adapted the questionnaire to the respective target group of subjects; particularly, after the first full-scale study, we made a few improvements that helped reduce the number of uninterpretable and nonsensical answers that needed to be discarded.

\subsection{The Programming Task}\label{subsec_TheProgrammingTask}

We designed a fictitious programming task in which actual programming was not undertaken during the session, but where the focus was on the subject's \emph{reasoning} about the programming task. Thus, we informed the subject that she should place herself in the \emph{role} of a programmer, i.e., \emph{imagine} being the programmer to whom this task is given to.

The task was to \emph{program a game for children} where the image from Figure~\ref{fig_puzzle} would be the game board. The game would consist of the player (which would be different from the subject/programmer) having to place the next letter H into one of the two designated empty boxes, i.e., on either left/right side of the vertical line. Upon correct placement, the game (i.e., the programmer) would reward the player. The boxes designated for placing the letter H are drawn with a dotted line with a large space between the dots. This design choice was made in order for the game to be perceived as continuing downwards and thus not blocked by a solid and continuous line/box. This was done to reduce the risk of being confounded by unintended biases (i.e., in this case, to avoid the subject to perceive the game board as being finite).

What we gave to the subjects as task description can be seen as the ``requirements'' that programmers receive from their clients (or elicited during a requirements engineering process); sometimes these include, so called, ``user stories'', which are realistic descriptions of the functionality of the software in terms of how a user (in our case, a player) would interact/work (in our case, play) with the software (in our case, the game). Our requirement contains one major intended omission (i.e., it is incomplete) in that it does not say what would constitute a ``correct'' placement of the letter H. In consequence, the subjects (as the programmer) need to decide for themselves to which side of the line they should give the reward (i.e., in other words, we wanted to elicit on which side they would prefer to place the letter H, themselves). As one can read further below, the subjects would be asked to explain their choice in a subsequent question of the survey.

We hypothesized that the uncertainty inherent in the task would elicit heuristic thinking prompted by either (i) cultural or (ii) contextual metaphors, in the same way people are influenced by metaphors in real-life situations. 
Specifically, the subjects' cultural metaphors (from their cultural background) would influence the choices they would make in the programming task, and also that contextual metaphors (coming from our experimental manipulations described in Section~\ref{sec_MetaphorsAsPrimingMethod}) would have the potential to moderate their choices. We expect the choices to be one of the three different rationales from Section~\ref{sec_BiasRevealing}, i.e., the `balance', `shapes', or `algorithm' rationale.

In order to introduce the priming metaphor, the game board image was linked to the story of the philosopher by saying that this ``puzzle'' was created by the philosopher. This link was made after the pilot testing (see details in the respective subsection below). We hypothesized that, if the participants were offered a simple explanation of the origins of the puzzle, then the philosopher story, containing one of the primes, would prompt the subject to non-consciously choose an explanation similar to the inherent rationale in the respective prime. 

The task description that we used is the following.

\begin{quote}\begin{itshape}
\textbf{``First, spend one minute imagining how you would be programming the simple task below. Then proceed to answer the following questions.}

Imagine that you are a non-expert programmer who is developing a simple puzzle game. The game is based on a riddle made by the philosopher that you read about previously.
Imagine that you have already drawn the game board that you can see below:

[The image in Figure~\ref{fig_puzzle}]

Now you are going to program the player's interaction with the game.

The player (\textsc{not you}, you are the programmer of the game) has to solve the puzzle by drag-and-dropping the letter H on to one of the two dotted boxes. The player is rewarded if the program accepts the placement of the letter H as the correct placement.'' 
\end{itshape}\end{quote} 


We want to conceal the bias revealing task behind the programming task in order to avoid debiasing. We want that the subjects focus on the programming aspect, and do not realize that we ask them to make a choice that has no wrong answer. This is relevant for what we discussed in \ref{subsec_Intro_Methods}.\ref{methodGeneralBias}.

\subsection{The Survey}\label{subsec_TheSurvey}

The survey is created in SurveyMonkey\footnote{Platform for creating online surveys:  \url{https://www.surveymonkey.com}}.  The respondents are directed by a link to this online platform to complete the questionnaire. SurveyMonkey was also used for the respondents from mTurk. The survey was created bilingual, in both English and Norwegian; the Norwegian respondents had the possibility to choose between English and Norwegian (we considered the possibility of international students being enrolled in the studies). 
Screenshots of all the pages of the survey are given in \ref{app_TheQuestionnaire}.

The survey starts with \emph{`Page 1: \pageNameIntro'}, presenting the goal of the survey and how the data is going to be dealt with. The goal of the experiment is only partially disclosed, and the true hypothesis remains completely undisclosed: 

\begin{quote}\begin{itshape}
``There is now a growing number of programmers using programming languages that do not require programming skills or education. We are therefore conducting research regarding programming done by non-experts and we invite you to participate in a brief exercise about the use of natural language to explain the function of computer programs.''\end{itshape}\end{quote} 

We did not disclose the fact that we are studying biases. This nondisclosure was done in order to not influence the effect of priming and in order not to make subjects aware that we are interested in the potential relation between their cultural background and the choices they make in the programming task. We also inform the subjects about how long, on average, the completion of the survey should take and that the study is completely anonymous. Since the respondents have various backgrounds, from other disciplines 
than computer science, it was also important to mention that no prior knowledge of computer programming is required for taking part in the survey/task/exercise. 

\paragraph{`Page 2: \pageNameInstructions'} contains information that we consider important for the respondents to know before starting the survey. There is also a reason for choosing to have such guidelines on a separate page, which we will expand on later in the Section~\ref{subsec_PilotTesting}. The respondents are informed that: 

\begin{quote}\begin{itshape}
``The \textbf{back button} is \textbf{disabled}. You will not be able to go back to a previous question, so we ask you to \textbf{read} each question \textbf{carefully}, because some depend on the previous ones. 

Please \textbf{put effort} into reading carefully everything on each page.''\end{itshape}\end{quote} 

Note that some text is being emphasized. In the case of skim-reading the minimum information the respondent perceives is: ``back button disabled'', ``read carefully'', and ``put effort''.

We need the participants to actually read the texts in the survey for the primes to work and for understanding the requirements in the programming task. For the mTurk and SurveyMonkey respondents, who were paid, we also added information about required minimum time for completion.

\paragraph{`Page 3: \pageNameStory'} contains our story intended for priming, which we have detailed in Section~\ref{sec_MetaphorsAsPrimingMethod}. We experienced during the pilot test that the participants might not read a text if the information there cannot be used for answering questions in the survey. Therefore, we added one question meant as extra motivation for the respondents to read the text that contained the primes that were hypothesized to lead to a biased programming choice. (See \ref{app_TheQuestionnaire}.)

\paragraph{`Page 4: \pageNameTask'} contains the text from the previous Section~\ref{subsec_TheProgrammingTask} (see also \ref{app_TheQuestionnaire}). 
We also added three questions to this page, partly meant as extra motivation for the respondents to read the text with the programming task. One of these questions is important because it is directly related to the bias, i.e., it asks about the placement of the letter H. In order to conceal the importance of this question we added two more questions completely irrelevant for our experiment. However, all three questions are made to look like questions that concern the programming task, i.e., it makes the task realistic. If we would have left only the question about the choice of placement then the subject could have observed the missing information in the requirements that we gave and thus perceived the task as less realistic.

\paragraph{`Page 5: \pageNameExplanation' and `Page 6: \pageNameAlternative'} is where the respondents give (or choose) an explanation for the choices they made in the programming task. We detail these two pages in the next subsections.

The rest of the questions on the following pages are meant to gather more information that could influence the results of the experiment, i.e., one's view on life, hobbies, educational background, and demographics (age, gender).

\paragraph{`Pages 7: \pageNameRanking'} where the three alternatives from Section~\ref{sec_MetaphorsAsPrimingMethod} could be ranked. We ask the subjects to 

\begin{quote}\begin{itshape}
"Please rank the following three pairs of life-aspects in the way that best reflects how \textbf{you} view life yourself (where 1 is the highest while 3 is the lowest).\\ 
$[$~Options: harmony and equality $|$ aesthetics and arts $|$ order and continuity.~$]$'' \end{itshape}\end{quote} 

This is a form of self-evaluation, where the subjects have the chance to express directly their order of preference for the three instances of priming metaphors (this is done after they have completed the main task, and they are not aware that they were themselves randomly exposed to one of the metaphors).  If they rank the prime that they were exposed to highest, this might indicate that the prime has had an influence. If so, this would strengthen the results (i.e., the prime has influenced both programming and subsequent statements). If they do not rank ``their own'' prime highest, this could mean that the prime had an effect on the programming task, even if it did not have an effect on the participant's statements. 

The UI for ranking questions is made well by SurveyMonkey so that when the question is required, then the subject must indeed provide a ranking, and not just leave the default, since some action is required before the next button is enabled (e.g., either provide a ranking number or move the choices with the mouse).

\paragraph{`Pages 8: \pageNameWords'} where the subjects could suggest one to three words characterizing each of the three life-aspects, from the ``Ranking life-aspects'' question. The open-ended format chosen for this page has several reasons. 
\begin{itemize}
 \item We wanted to have a way to identify unserious subjects or robot-generated answers (as detailed later in Section~\ref{subsec_TransitioningFromVolunteersToProfessionals}). 
 \item We also wanted to have another way to check the metaphorical priming effect by looking whether, and how often, our priming words appear among the answers. 
 \item We also wanted to gather more data for future studies; i.e., others could use some of these words in future metaphor studies.
\end{itemize}

\paragraph{`Pages 9: \pageNameDemographics'} where the subjects had 5-7 questions about age, gender, years of education, field of study, and leisure activities (i.e., hobbies).

In the following section we explain the reasoning behind the way the questions are composed, as a result of the discoveries we have made during the pilot testing.

\subsection{Pilot Testing of Usability}\label{subsec_PilotTesting}

To test the usability of the survey we performed several pilot tests.

The first pilot test used the method of usability testing \cite{dumas1999practical}, with our survey being the product under test. One goal with testing the survey for its usability is to see whether the explanatory texts, requirements and questions are written in a clear and easy-to-understand language. Moreover, since we intended to prime the subjects, we needed to make certain that the story of the philosopher (which contained our primes) was read carefully and not just skimmed through. 

The general goal for this first iteration was to find obvious problems with the survey. The usability study involved five participants. Four of these participants have a background in computer science and one in arts \& design. Subjects with these two types of backgrounds were going to be used in our full-scale experiments as well.  

The participants were asked to take the survey while being observed by us, sitting next to them (one of us took the role of a moderator, while another researcher was only an observer). The test was run 
with one participant at the time. Before taking the survey, the subjects were explained verbally the purpose with the test session, which was to help us improve the face-value quality of the survey, although the hypothesis was not revealed. They were also presented the order of the tasks: first they were to take the survey, without any interruption, and then were supposed to answer questions meant to elicit suggestions for improvements of the survey.

After a participant completed the survey, s/he was asked to verbally answer a set of open-ended questions (post-test questionnaire/interview). The questions listed below represented the starting point for a discussion where the participants were free to give comments on the usability problems s/he encountered, while the moderator was recording their comments and asking supporting and clarifying questions.

\begin{quote}\begin{itshape}
\begin{itemize}
\item Were the questions/tasks easy or difficult to understand? Which of these?
\item Did you find the instructions in the programming task clear or confusing?
\item Was the text of the questions or explanatory text too long or too short? Which of these?
\item Would you have needed/wanted to go back and read the previous question/texts? In which case? (we had the back button disabled so that they could not navigate back to the previous questions).
\item How much time did you need to spend on reading the task about programming the puzzle? (they were asked to  spend 30 seconds before answering the questions related to the programming task).
\item Did you spend more time on certain questions than on others? Which of these? Why?
\end{itemize}
\end{itshape}\end{quote} 

The participants were asked to recall what was experienced as difficult or unclear by retrospectively going through each page in the survey and reading again the content of the page.

The findings were marked with a severity level of high or medium. An example of a ``high severity'' is related to the \emph{`\pageNameRanking'} question where the participants were asked to rank the three pairs of life-aspects. We wanted the participants to do the rating so that it reflects their own way of viewing life, but in the first pilot we did not emphasize this. By not specifying this, the subjects' answers could potentially reflect their perceived view of others. In the case of one participant, s/he chose the answer that reflected the view of the philosopher presented in the story on the second page, as s/he surmised that this is what we might have wanted.

In general the findings from this first pilot test helped with shortening and simplifying the text and making the requirements/questions more clear. 

The test also helped with validating our initial decision of disabling the back button in the survey. For the priming to work, the participants should not realize the connection between their choice in the programming task and the \emph{`\pageNameStory'}. If the participants would understand at a later stage in the survey that such a connection existed, they should not be allowed to navigate back and read the `Philosopher story' again. In this pilot study one of the participants had the back button purposely left enabled. This participant did just what we expected, s/he navigated back to the `Philosopher story', read it again, and adjusted her/his answers to reflect the view of the philosopher and not her/his own as the question required.

Another observation from this first pilot test was that on the \emph{`\pageNameStory'} page
the participants scrolled down quickly to the questions and then read only the part of the text that helped them to answer those questions. This was however difficult to establish with accuracy only through observation and the participants might not want to recognize that they did not put enough effort into reading the whole text. Only one of them acknowledged that. We concluded that the best way to reveal the exact behaviour that the participants have in reading the information, and also which flow they follow, was by using eye tracker technology \cite{bojko2013eye}.

\subsection{Eye Tracking for Better Insights}\label{subsec_EyeTracking}

\begin{figure*}[tp]
    \centering
    \begin{subfigure}[t]{0.5\textwidth}
        \centering
\includegraphics[width=\textwidth]{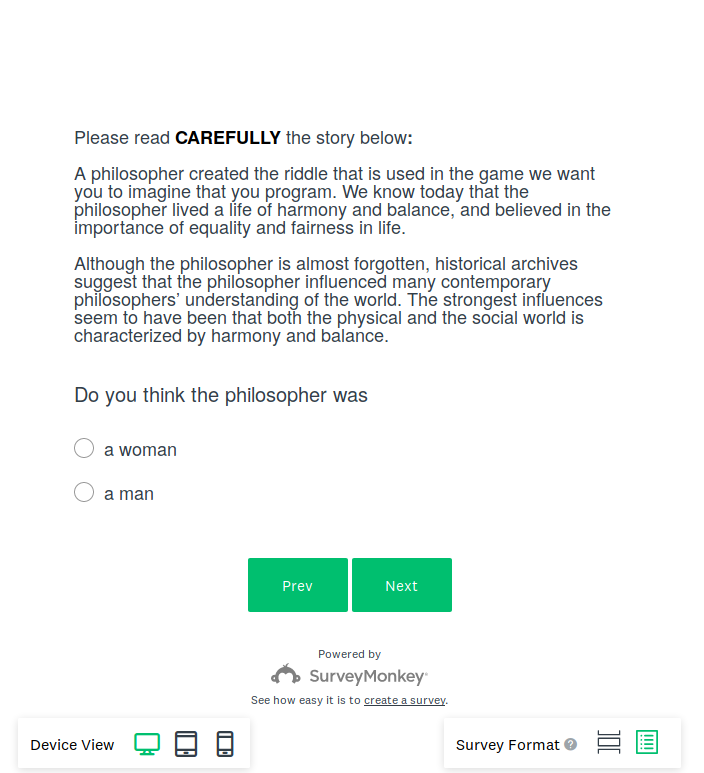}
\caption{`Short story'.} 
\label{fig_shortStory}
    \end{subfigure}%
    \begin{subfigure}[t]{0.5\textwidth}
        \centering
\includegraphics[width=\textwidth]{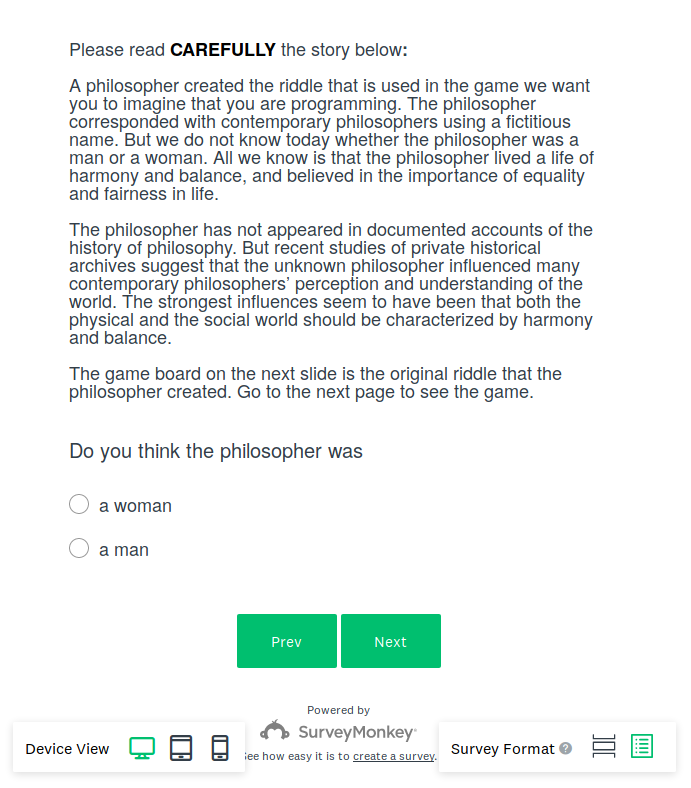}
\caption{`Long story'.} 
\label{fig_longStory}
   \end{subfigure}
    \caption{The `Philosopher story' tested with eye tracking.}
\label{fig_eyetrackerStory}
\end{figure*}

We created two versions of the \emph{`\pageNameStory'} -- a `Short story' (Figure~\ref{fig_shortStory}) and a `Long story' (Figure~\ref{fig_longStory}). We employed summative research\footnote{Summative research implies comparing an interface or product to its other versions, competitors, or benchmarks \cite{bojko2013eye}.} in combination with eye tracking methods for comparing the two versions and deciding which was more effective.
 
The test was done in a usability laboratory set up with eye tracking equipment. The survey was displayed to the participants on a desktop computer to which a remote eye-tracker was attached. Using a remote type of eye-tracker, required to calibrate the device before starting the testing with each of the participants, as well as instructing the participants to find a comfortable and stable position which to maintain during the whole testing, so to avoid head and especially body movement. To test the effectiveness of the two versions, a combination of single-subject and  between-subject design was used, where each participant (ten in total) was exposed to only one of the test stimuli -- five to the  `Short story' and five to the `Long story'. This was to avoid any carryover effects between the stories. For their help, the participants (students from Karlstad University) received a small reward in terms of a coupon usable in the student cafeteria.

Both stories contained the same priming words. The `Long story' was created with the purpose of helping the reader to immerse in the story -- by giving more background information on the philosopher -- and prepared the participant for the task on the next page of the survey, the \emph{`\pageNameTask'}. Since we intended to prime the subject, we needed to hide the priming words well in the story, so that unintended debiasing (e.g., reactance) would not occur. At the same time, a too long story could make the subject not read the whole text and thus possibly skip the priming words. A shorter version of the story would also reduce the cognitive burden on the subject. The eye tracking testing was thus meant to help us identify whether the subjects skip our priming words, and also how much cognitive effort (i.e., how much time) they puts into reading the stories.

The heatmaps and gase plots visualizations\footnote{The gaze was offset vertically by approximately one line. This was due to the mismatch between the Operating System version and the version of Eye tracking software at the time of testing. The offset has been consistent across the participants and did not affect our interpretations.} provided both spatial and temporal insight into how the participants interacted with the text on each page of our survey. 
We obtained information about which areas of the text were fixated and for how long, the number of fixations and the order in which the fixations occurred. More specifically, we obtained insight into which lines and words were read, whether some of these were reread, and also how many times.

\begin{figure*}[tp]
\centering
\includegraphics[width=\textwidth]{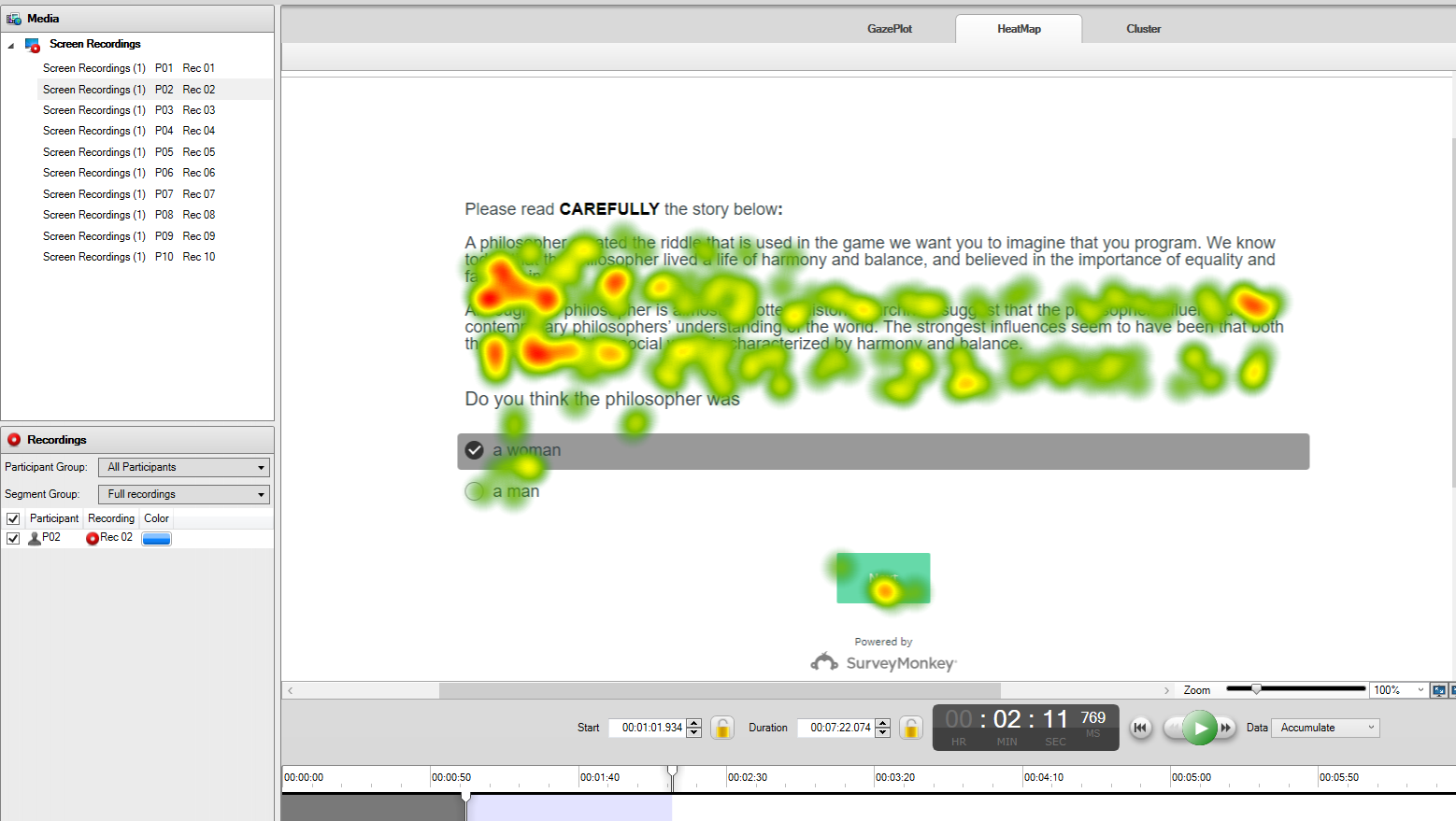}
\caption{Heat map with offset gaze.} 
\label{fig_heatMap}
\end{figure*}

Interpreting this data we concluded that there was no noticeable difference in how the text, and especially the priming words, were read between the long and short version of the story. For both cases, the participants read the text thoroughly, line by line (Figure~\ref{fig_heatMap}). 

This shows that the instruction on the \emph{`\pageNameStory'} page about reading the story ``\textsc{carefully}'' had the wished effect. Thus, this requirement was carried over to the full-scale experiments as well. The difference between the reading of the two stories was that the long one required more time and effort, 1:10 minutes compared to 40 seconds for the short one. Reducing cognitive load and time by almost 50 percent would be of help to the subjects, and thus we decided to use the `Short story' in our full-scale studies.

Another aspect that we analysed with the help of eye tracking was whether the question about the philosopher being a man or a woman works as extra motivation for the participants to read the story. We found out that in order to answer this question, the participants returned to reading the story several times. In addition to the motivational aspect, questions such as this one will help in drawing potential attention away from the true hypotheses.

\begin{figure*}[tp]
\centering
\includegraphics[width=\textwidth]{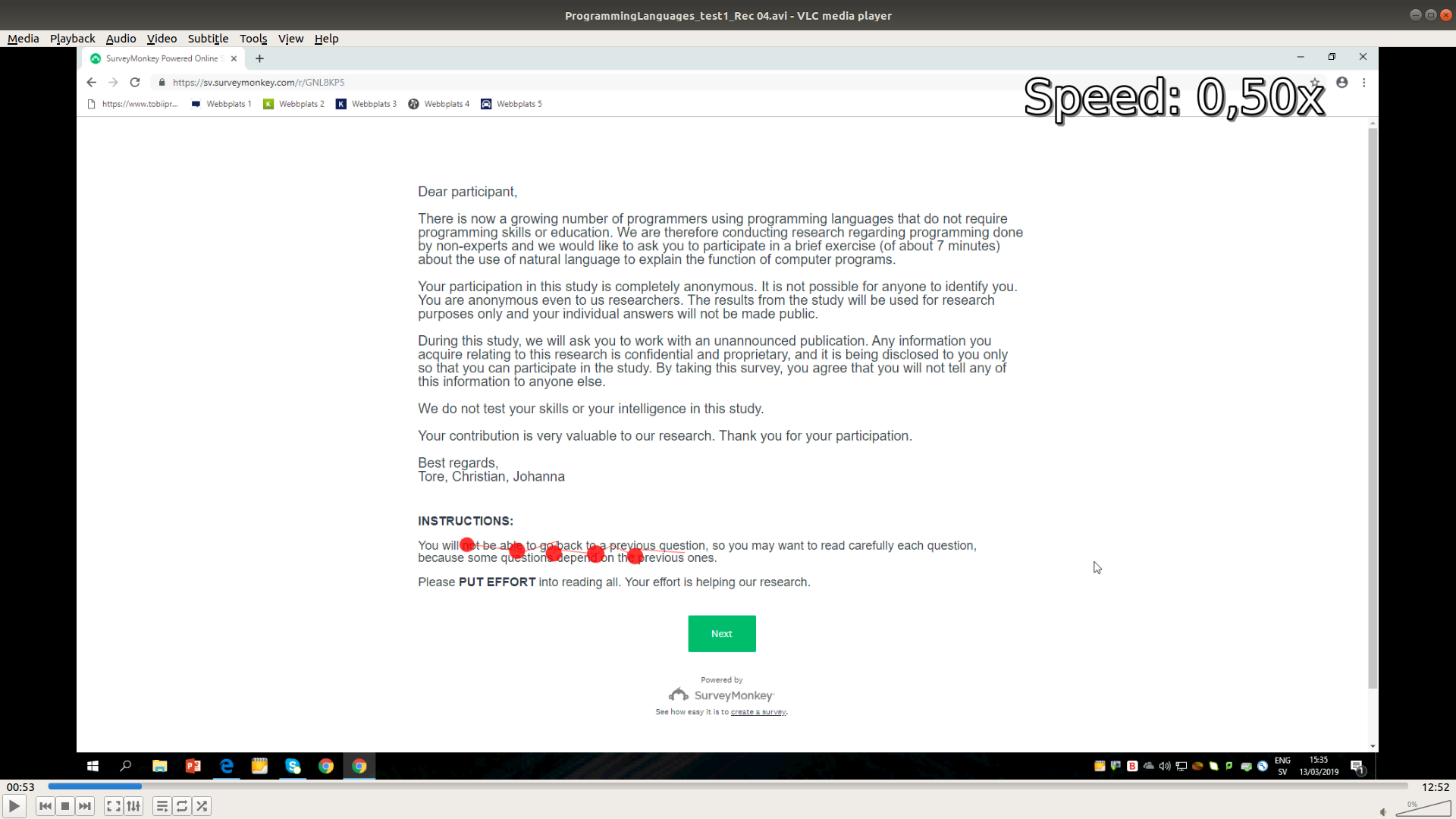}
\caption{Image of one of the recordings, watched in half of the real speed, showing that the instructions were carefully read, with an even distribution of fixations.} 
\label{fig_fixations}
\end{figure*}

We also tracked the reading of the \emph{`\pageNameInstructions'} text, as this is where we ask the participants to ``put effort into reading'' the texts in the survey. During the eye tracking phase, the instructions were part of the \emph{`\pageNameIntro'} page, as a separate paragraph at the end of the page, separated by the rest of the text by a capitalized and emphasized header ``\textsc{instructions}'' (Figure~\ref{fig_fixations}). In addition, we also included a sentence at the start of the \emph{`\pageNameStory'} asking the respondents: ``Please read \textsc{carefully} the story below:'', again with capitalized and emphasized letters (Figure~\ref{fig_eyetrackerStory}). 

The findings from the recordings revealed that one participant did not read the instructions text, but only looked at the words ``\textit{instructions}'' and ``\textit{put effort}''. These words were standing out through their typography design. This participant was found to not put effort in several other places. Three of the participants were only reading parts of the instructions, while the remaining six were reading carefully, line by line. In a laboratory context the participants are known to be putting in more effort than they might do on their own in other more natural contexts. 

By analysing the behaviour exhibited by the participants in reading the instructions on both pages, we concluded that we should create a dedicated page for the instructions, just before the \emph{`\pageNameStory'} where we should include the \textit{``put effort into reading''} text. Thus, we drew attention to the importance of the instructions by creating a separate page for them. Here we also included the \textit{``... \textbf{put effort} into reading carefully ...''} text, in order to avoid repeating it on the \emph{`\pageNameStory'} page. This was done with the intention of preventing the possibility of the participants guessing our priming intention by marking the story as something to pay special attention to. We also decided to typographically emphasize the important words or sentences used for instructions throughout the whole survey, as we have seen that they are always read by the participants.

\section{Performing the Study}\label{sec_PerformingTheStudy}
\subsection{The Participants}\label{subsec_TheParticipants}

Here, we describe the reasons behind the way we chose and grouped our subjects, and how the different cohorts of participants are meant to help study the research questions that we elaborated on in the Introduction. 

The participants were chosen based on their educational or occupational background, to span three main educational and professional domains, which is relevant for the 
\ref{subsec_Intro_Methods}.\ref{methodBiasRevealingTask} aspect
presented in the Introduction. This is also meant to cover well different computer programming skill levels as well as socio-cultural influences, properties and preferences. We reason that, when enrolled in a certain study line or field of work, people have already developed predominant skills and characteristics needed for the specific education or occupation. 

We will thus talk about three main cohorts of respondents:

\begin{enumerate}[A.]
\item \emph{`Social sciences' cohort} -- composed of students studying psychology; 
\item \emph{`Natural sciences' cohort} -- composed of students studying computer science; and 
\item \emph{`Arts and Culture' cohort} -- composed of a group of participants working in the field of arts and design, a group of students studying theatre, and another group studying music. The last two subgroups we term the \emph{`Cultural studies'} participants.\end{enumerate}

This categorization based on the educational and professional background of the respondents is confirmed by the analysis of the data obtained from the control question on the \emph{`\pageNameDemographics'} page, i.e., specifically about which field of study or/and line of work the respondents affirm their background to be mainly consistent with. 

Based on the conditions to which the respondents were exposed, we can also categorise the three cohorts into:

\begin{enumerate}[I.]
\item\label{notHelpedNotConfined}  `not helped' and `not confined', 
\item\label{helpedNotConfined} `helped' and `not confined', 
\item\label{helpedConfined} `helped' and `confined'. 
\end{enumerate}

The `confined' or `not confined' categorization describes the environment where the respondents were at the time of taking the survey. The `not confined' respondents took the survey in the environment of their choice, which was unknown to us, whereas `confined' means taking the survey in a more controlled environment (i.e., the university auditorium). The `helped' or `not helped' classification refers to whether the respondents were supported or not by being given alternative explanations to pick from, in order to explain their choice for the placement of the letter `H'.  Starting with the second full-scale study we introduced an extra page in the survey, offering such alternative explanations with possible answers to choose from, meant to reduce the number of uninterpretable answers. The respondents that were given the `Alternative explanations' to help them with explaining their choice are termed as `helped'. The \emph{`Social sciences'} cohort belongs to the `non helped' and `not confined' category (\ref{notHelpedNotConfined}), as the survey they were given did not contain the `Alternative explanations' page and they could take the survey at the time and place of their choice.  In the case of the \emph{`Arts and Culture'} cohort we had a combination of `helped/confined' and `helped/not confined' (\ref{helpedNotConfined}, \ref{helpedConfined}) with a predominance of the latter. The mTurk and SurveyMonkey respondents were helped with `Alternative explanations' and were free to choose the environment where to take the survey. The `Cultural studies' students were also `helped' but confined to a classroom, where the course leader and one of the authors were also present. Similarly, the \emph{`Natural sciences'} cohort was both `helped' and `confined' (\ref{helpedConfined}) as the survey was taken as  part of their regular course-work. The `confined' / `not confined' and `helped' / `not helped' are categories that are going to be used for analysing the sensical vs. nonsensical data in Section~\ref{subsec_SensicalVsNonsensical}.

The environment of the participants and the support they received is related to three main types of environment where (future) programming activities can take place in: 

\begin{enumerate}[A.]
\item \emph{Typical professional programming environment}, where the programmer is `confined' to an office space and has to her disposal all the professional resources necessary to fulfill her tasks. In our case, for the programming task and the required explanations, we tried to reproduce this type of environment for the group of computer science students, by both confining them to the classroom and course hours and offering them helping answers.
\item \emph{Semi-professional environment}, where an expert in some technical field (other than programming, e.g., railway engineering) has professional tool support for simple programming/configuration, e.g., by using a GUI based programming tool or a graphical programming language s.a. Blockly.  However, programming is not their main task or responsibility and thus are not supposed to put too much effort into it, which we consider as `not confined'.  The mTurk and SurveyMonkey respondents were thus `not confined' but `helped'.
\item \emph{Non-professional environment}, where people, e.g, in their homes, are configuring an IoT system without any professional support nor prior knowledge. The \emph{`Social sciences'} respondents were neither `helped' nor `confined', and can thus be seen to some extent as fitting this profile.\end{enumerate}

For the purpose of studying the influence of the contextual-metaphor priming, we further group the respondents from each cohort by the priming they have been exposed to (or not), according to Section~\ref{subsec_MetaphorsAsPrimingMethod_Experimental}:

\begin{enumerate}[A.]
\item a control group which is not primed in any way,  
\item a group primed as in Section~\ref{subsec_MetaphorsAsPrimingMethod_Experimental}.\ref{lifeAspectHarmony} (which we call, `primed with harmony and equality'), 
\item a group primed as in Section Section~\ref{subsec_MetaphorsAsPrimingMethod_Experimental}.\ref{lifeAspectArts} (i.e., with `aesthetics and arts'), and 
\item a group primed as in Section~\ref{subsec_MetaphorsAsPrimingMethod_Experimental}.\ref{lifeAspectOrder} (i.e., with `order and continuity').
\end{enumerate}

The control group is meant to serve as a baseline to observe what the programmers' preferences for task-solutions are in the absence of primes, i.e., when presented with a description that is neutral with regard to the task at hand and the task's inherent possible solutions. This is relevant for our first main question. 

The three primed groups are meant to help us test whether the bias can be induced upon the programmer, and subsequently transferred from the programmer to the algorithms. This is relevant for our second main research question presented in the Introduction and the aspects discussed in \ref{subsec_Intro_Methods}.\ref{methodEffectInducingBias}.

The cohort with students from the computer science study line is meant to help us test whether programmers shut away the other two biases, i.e., resulting from the cultural metaphor or the induced contextual metaphor, except the pattern/infinite way of thinking, which is sometimes assumed that the programmers do.  This is relevant for what we discussed in \ref{subsec_Intro_Methods}.\ref{methodSkillLevels} and \ref{subsec_Intro_Methods}.\ref{methodBiasProgrammers}.

We also collect information about the hobbies, years of higher-education, age and gender of the participants. The hobbies and years of higher-education are meant to help with fine-graining the analysis of the educational and occupational background of the participants. The age 
data is presented in the Section~\ref{sec_AnalysingTheData}, with the purpose of giving an overview of the population distribution. We also look for indications of whether the bias is more pronounced in different age 
categories, and if this is reflected in how much it is transferred into the program.

\subsection{Methods Employed}

The main research \emph{hypothesis} to be tested is whether biases can be transferred from the programmer to the program. This hypothesis is tested throughout all our three experimental cohort-studies, and it contributes to answering our first main question from the Introduction.

The studies employ a combination of experimental design and comparative design. In the analyses of both (i) the comparative aspect (i.e., differences \emph{between} the three cohorts) and (ii) the experimental aspect (i.e. (differences within each cohort, resulting from the experimental manipulation), we employed both (a) inferential statistics, more specifically chi-square analyses of categorical data, as well as (b) descriptive statistics to report frequencies and percentages. 
We performed an experiment on each cohort, as well as compared the three cohorts to each other, regardless of the experimental manipulation. Since the three cohorts were different in terms of cultural and educational background, we were able to study the unique effect of background per se. 

Conforming to the true experimental design method \cite{lazar2017research,cook2002experimental}, we first assigned the participants of each cohort randomly to one of three \emph{experimental conditions} where we induced one specific type of contextual metaphorical thinking in each, or to a \emph{control condition} containing neither of the three primes. The control condition contained the neutral non-prime story from the end of Section~\ref{subsec_MetaphorsAsPrimingMethod_Experimental} and was meant to serve as a ``baseline'' to establish whether the participants, without being primed, were inclined to favor one of the three ``rationales'' over the other. 

The subjects are given the programming exercise described in Section~\ref{subsec_TheProgrammingTask}. The programming task, the educational/professional background of the subjects, and the story containing the primes, are the \emph{independent variables} in our experiment. The choice of what will be the right solution for the puzzle is the \emph{dependent variable}. We are interested in finding out if the primes and the background of the participants (the independent variables) induce changes in how the puzzle is programmed (the dependent variable), following the rationale that it is the programmer who decides to give the player a prize based on what she (i.e., the programmer) thinks qualifies as the right answer. 

The \emph{conditions} (also known as treatments) that we intend to compare are reflected in the explanations that the subjects provide, being under the influence of three contextual metaphor primes and three types of cultural background. 

A true experiment requires randomization and controlled trials (Randomized Controlled Trial -- RCT), as well as one or more distinct experimental manipulations. First, our study conforms to these requirements due to random assignment of each participant to the four conditions (i.e., three experimental conditions and one control condition). Second, the experimental conditions are controlled and kept constant to the extent that we recorded the time spent on the tasks and thus ensured that the tasks were completed within a reasonable time-frame. Thus, we excluded the effect of any seriously potentially confounding variables, such as diffusion of experimental manipulations (i.e., we reduced the possibility of participants sharing the contents of the tasks with other participants). Participants completed the task individually and received identical instructions. In addition, the hypotheses were not revealed to the participants. Such non-disclosure of hypotheses is the most robust experimental procedure, and it is employed in around 87 percent of all experimental-psychology research \cite{hertwig2008deception} because it allows for the elicitation of valid measures of behaviour instead of relying on less valid measures by means of other methods, s.a. self-reports \cite{broder1998deception,christensen1988deception,kimmel1998defense,kron1998reality,trice1986ethical,weiss2001deception}.

For analysing the second main hypothesis that we proposed in the Introduction, pertaining to the potential influence of the context, i.e., metaphors in the programming environment, the \emph{research hypothesis} is that the manipulation (``prime'') will increase the number of the corresponding explanations the participants give. In the event that the number of explanations does not increase as a result of the prime, we discard the research hypothesis and retain the null hypothesis.

Supporting evidence of whether the prime has induced a specific metaphorical thinking and thus has produced a biased judgment in the prime's direction should be shown in the participants' explanations, given after they have finished the programming task. The participants' explanations for their respective choices were qualitatively coded according to the three predefined categories. Explanations conforming to one of the three predefined categories were categorized both according to their discrete category (i.e., `balance', `shapes' or `algorithm') as well as whether they were `sensical' (i.e., eligible for inclusion in the predefined categories) or `nonsensical'. Non-interpretable explanations were thus labeled `nonsensical' and discarded (see Section~\ref{subsec_SensicalVsNonsensical} for a thorough analysis of this). We coded the explanations qualitatively and categorized them into one of the three possible rationales. If the rationale of prime-manipulation in the respective condition is chosen significantly more than the other rationales, this would imply that the participants were influenced by external features that are not relevant to the programming task itself.

We implemented one additional variable to control for the bias, and this stemmed from our assumption, resulting from an observation in our practical use of the cognitive task from Section~\ref{sec_BiasRevealing}, that the choice of placing the letter H is also an indication of the rational. Particularly, participants choosing \emph{Left} would be those using the rationales \ref{rationaleBalance},\ref{rationaleShapes} from Section~\ref{sec_BiasRevealing} for `balance' or `shapes' (similarly those primed with the life-aspects \ref{lifeAspectHarmony},\ref{lifeAspectArts} from Section~\ref{sec_MetaphorsAsPrimingMethod}); whereas participants choosing \emph{Right} would be those using the rationale \ref{rationaleAlgorithm} for `algorithm' (similarly those primed with \ref{lifeAspectOrder}). This is analyzed in Subsection~\ref{subsec_LeftRightPlacement}.

We choose the subjects based on their educational and professional background. However, in the questionnaire we ask the participants to provide information about their educational background themselves (because some have multiple) as well as information about their preferred free-time activities. This is done in order to disclose a possible relation between this particular aspect of the background of the participants and their choices in the programming task. Moreover, this information from free-text questions can also help detect respondents that did not relate seriously to the task, as well as to control our qualitative coding of their explanations and background.

We hypothesize that the results from the analyses will show a statistically significant relation between (i) cultural metaphors (i.e., the subjects' cultural background) or (ii) contextual metaphors (i.e., the experimental prime) and the choices made (and the explanations provided) regarding the programming task. The majority of the answers are expected to fall in one of these three categories: biased by the prime only, biased by the background only, or biased by both. If neither of this is the case, our hypotheses are discarded, and we conclude that biases are not being transferred from the programmer to the program.

\subsubsection{Alternative Explanatory Variables}

The age and gender of the participants are analysed as \emph{alternative explanatory variables.} 

Other alternative explanatory variables that might occur could result from the subjects not understanding the task well, the task being too difficult, or the prime not being strong enough as a result of superficial reading. However, these factors were something that we detected and removed through our \emph{pilot tests}. 

The nature of our study requires an (i) experimental and (ii) comparative \emph{between-group design}, where each participant in the experimental part is only subjected to one experimental condition. In the comparative part of the design, the three cohorts represent three distinctly different cultures. Thus we were able to study the potential effect of both the cultural metaphors and the contextual metaphors on programming choices as two different sources of bias.

In order to effectively exclude the potentially confusing \emph{noise} caused by the individual differences that may occur in small samples, as well as potentially nonsensical explanations that would need to be discarded, we decided to use a sufficient number of  participants in each condition. Thus, we assigned at least ten participants from each cohort to each condition in order to arrive at least at five sensical output-explanations and thus conform to the conventional requirements of chi-square comparisons, i.e., including at least five participants in each group.

We believe, however, that the individual differences will not affect the study results to a large extent. By implementing the questionnaire questions related to individual preferences and extracurricular activities (\emph{`\pageNameRanking'} and \emph{`\pageNameWords'} pages, and the question about hobbies on the \emph{`\pageNameDemographics'} page), we expect to be able to clearly identify if the choice was dictated by the bias. Moreover, the programming task is thought to be very simple, thus requiring very little cognitive effort. For such cases, it is empirically proven that the individual differences have a smaller impact \cite{lazar2017research}. For the same reason, i.e., reducing the factors influencing the main conditions to be compared in the experiment, the groups of respondents were composed of respondents studying/working in the same field, something which reduces potentially confounding individual differences at least to some extent.

\subsubsection{Ethical Aspects}
All three full-scale studies were done with the responses being anonymized. All answers are registered on the random ID that the system generates. It is thus not possible to identify any of the respondents.

\subsection{Learning from the First Full-scale Study}

\begin{figure}[h]
\centering
\includegraphics[width=\columnwidth]{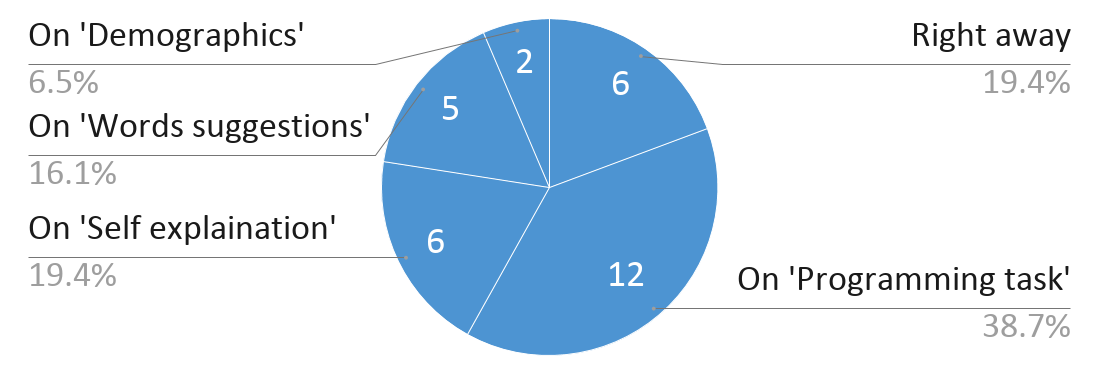}
\caption{Overview of the dropout number -- frequencies and percentages -- of respondents from the `Social sciences' cohort; including the names of the pages where the dropout happened.} 
\label{chart_dropout}
\end{figure}

The first full-scale study, also referred to as the `Social sciences' cohort, consists of undergraduate students enrolled in a psychology study program. The link to the survey on SurveyMonkey was sent through email by the study program administrators and resulted in 77 responses. Observations made after the first study helped with improving the following studies. 
An analysis of the incomplete responses (31 out of 77) from the first study shows that: 6 dropped out right away; 12 dropped out on the \emph{`\pageNameTask'} page; 6 on the \emph{`\pageNameExplanation'} page; 5 on the page with \emph{`\pageNameWords'} page; and 2 on the \emph{`\pageNameDemographics'} page. The dropout rates are respectively 19.4\%, 38.7\%, 19.4\%, 16.1\%, and 6.5\% (i.e., Figure~\ref{chart_dropout}). 

The high number of dropouts on the \emph{`\pageNameTask'} page could be explained by the fact that the students in this cohort are studying psychology and they may have never been exposed to computer programming, thus they may have deemed this task as not relevant, not interesting, or maybe too difficult. Based on this reasoning, in the second study, also referred to as the `Natural sciences' cohort, we introduced on the first page more text where we explain that ``It is not required to have any prior knowledge of computer programming.''; moreover, on the  \emph{`\pageNameTask'} page we wrote that the puzzle is simple, i.e., \textit{''... who is developing a simple puzzle game''}.

Another solution for further motivating the respondents to finish the survey was to add a progress bar indicating how much of the survey was left until completion.  For the last three pages we also added a page-footer informing, consecutively, that \textit{`there are three, two, and one pages left'}, i.e., aiming to reduce the two latter types of dropouts.

Recall that the first study was conducted with volunteer \emph{social science} students that were neither paid nor participating during their normal class hours. In contrast, the second study was run in a lecture hall, before the break, as part of a first year \emph{computer science} course. In the case of mTurk and SurveyMonkey respondents from the third study, also referred to as the `Arts and Culture' cohort (see Section~\ref{subsec_TransitioningFromVolunteersToProfessionals}), we consider the payment as an important motivating factor.

To reduce the cognitive effort required for explaining the choice for the `H' letter placement on the \emph{`\pageNameExplanation'} page we added, immediately after this page, an additional page called \emph{`\pageNameAlternative'}, containing a list of predefined explanations to choose from. This was meant to reduce the high dropout rate that we saw on the \emph{`\pageNameExplanation'} page. Moreover, adding these alternatives in the second study reduced the number of uninterpretable answers significantly. A detailed analysis of the uninterpretable answers (i.e., explanations) is done in the Section~\ref{sec_AnalysingTheData}.

For the second study, the total number of responses received was 53. Of this total number, one respondent dropped out on the \emph{`\pageNameTask'}, one on the \emph{`\pageNameExplanation'},  one on the \emph{`\pageNameRanking'}, and two on the \emph{`\pageNameWords'} page. In total there were five dropouts. The small number of dropouts in this second study indicates that the adjustments made after the first study were successful.

Regarding the language in the first study, the students could choose between English and Norwegian. The majority of participants in the first cohort were Norwegian speakers. Nevertheless, some of the students could be, for example, exchange students and thus more comfortable with English than Norwegian, therefore an English version was made available as well. Though we did not have any students choosing the English version in this case, we encountered three such cases in the second study, testifying to the relevance of providing both a Norwegian and an English version. Moreover, the English version of the survey was necessary for the mTurk and SurveyMonkey respondents in the third study.

\subsection{Transitioning from Volunteering Students to Professional Respondents}\label{subsec_TransitioningFromVolunteersToProfessionals}

For the third cohort, we planned to recruit people with a background in arts and culture in general. We intended to continue with the same style of set-up as in the second study, where the respondents take the survey during their course, confined to a classroom. It proved difficult to find a large number of respondents to comply with these requirements. We started the third study with two groups of students, studying music and theatre. Though we had no dropouts from these groups\footnote{Only one incomplete answer was removed, which was the response of the course tutor that wanted to scrutinize our survey.}, the numbers of students in the classes were too small (which is specific to these kinds of studies), i.e., 10 respondents from music and 10 from theatre. To increase the number of responses we decided to recruit respondents through specialized platforms, specifically through the Amazon Mechanical Turk and SurveyMonkey. These would no longer be volunteers but professional respondents, i.e., who are paid for their participation and do such tasks often.

The data from Amazon Mechanical Turk (mTurk)\footnote{Amazon Mechanical Turk is a crowdsourcing marketplace. https://\url{www.mturk.com/}} comes from three batches. The first batch of respondents were chosen based on the following qualification requirements: working in the Arts \& Design field  and having a Masters qualification\footnote{A Master Worker is a top Worker of the mTurk marketplace that has been granted the Mechanical Turk Masters Qualification. These Workers have consistently demonstrated a high degree of success in performing a wide range of HITs across a large number of Requesters.}. The respondents in this first batch were rewarded \$2 plus fees \$1.35 (Mechanical Turk Fee: \$0.80, Masters Fee: \$0.10, US High School Graduate Fee: \$0.05, Job Function -- Arts \& Design: \$0.40). The reward sum was decided based on what other requesters on mTurk were paying, which was generally under \$1. We received only one response. Therefore, we increased the reward to \$3 (plus appropriate fees) and reopened the mTurk HIT. However, this second batch brought only three additional responses. The number of respondents with the wished qualifications proved to be small, irrespective of the remuneration. In a third batch we kept only the job function -- Arts \& Design -- requirement, and the same payment of \$3. This generated 115 responses, which was close to the number of respondents we were aiming for (i.e., 100 respondents). 

The respondents in this batch come from the mTurk. However, given that the questionnaire was already made in SurveyMonkey, we directed the mTurk respondents to SurveyMonkey through a link.  There is a mismatch between the number of respondents registered in mTurk (115 plus four from the first two batches) and SurveyMonkey (128). This is due to the fact that we rejected (and not paid) some of the respondents, and thus their places were reopened in mTurk for new respondents, but SurveyMonkey kept generating new IDs, recording also those that we have rejected in mTurk. 

We asked the respondents (i.e., in the ``Instructions'' section of the survey) to spend at least eight minutes on their response and read all text with attention. However, quick readers could have managed to complete the survey in no less than four minutes. The respondents that spent less than four minutes could not have spent the necessary time on reading the texts in the survey and their answers cannot be considered reliable. This was established also by checking the answers in the open fields, where we could see how much effort the respondent put into writing an explanation on the \emph{`\pageNameExplanation'} page and giving examples of words on the \emph{`\pageNameWords'} page. 

Thus, from the total of 128 responses registered in SurveyMonkey, collected from all three batches, we removed 13 respondents that spent less than four minutes on completing the survey. Additionally, seven more respondents were rejected as we deemed them unserious (e.g., computer generated answers). Out of the remaining 108 responses, one participant dropped out on the \emph{`\pageNameExplanation'} page. 

Moreover, six respondents that  spent more than four minutes were still deemed unreliable and thus removed from the analysis. This was decided based on the quality of the responses given in the open-ended questions. Some examples of such `unserious' answers are: 

\begin{itemize}
\item One respondent (M:11270629093, from the third cohort) spent 18:44 minutes, but his/her answers on the \emph{`\pageNameExplanation'} page was ``none'', whereas for the \emph{`\pageNameWords'} question the answers seemed to be generated by a computer: answer \textit{``the intimately transmitted from west to east''} for life-aspect `harmony and equality' (from Sec.~\ref{subsec_MetaphorsAsPrimingMethod_Experimental}.\ref{lifeAspectHarmony}); answer \textit{``Aesthetics is branch of philosophy that examines the nature''} for life-aspect `aesthetics and arts' (from Sec.~\ref{subsec_MetaphorsAsPrimingMethod_Experimental}.\ref{lifeAspectArts}); answer \textit{``the Alternatively, this is called first-order parametric continuity''} for life-aspect `order and continuity' (from Sec.~\ref{subsec_MetaphorsAsPrimingMethod_Experimental}.\ref{lifeAspectOrder}). 
\item Another respondent (M:11270932685, third cohort) spent 11:00 minutes, but did not give any explanation on the \emph{`\pageNameExplanation'} page, and for the \emph{`\pageNameWords'} question wrote \textit{``just my opinion''}, and for the `Field of study/line of work' question on the \emph{`\pageNameDemographics'} page answered with \textit{``Letters of the words and Line''}.
\end{itemize}

To the mTurk batch we also added the responses from the study run with SurveyMonkey respondents and with the respondents from the Culture studies.

In the case of the SurveyMonkey, we requested 50 respondents  to be working in the field of Arts and Design. The country of origin of these participants was chosen to be Sweden. With the purpose of reaching the number of respondents having the required qualifications that we asked for,  SurveyMonkey created two collectors:

\begin{enumerate}[(a)]
\item The first collector generated 39 qualified responses remunerated with SEK 8,982.8 in total, and SEK 230.32 per response.
\item The second collector generated 11 qualified responses remunerated with NOK 2,332.77 in total, and NOK 212.07 per response.
\end{enumerate}
 
Out of these 50 respondents we removed 15 that spent under 4 minutes on completing the survey. Out of the remaining 35 respondents we further removed 13 unserious answers. These answers were deemed as unserious based on the evaluation of the answers in the open-field questions, e.g.: the participant with the ID: S:11174871215 (from the `Arts and Culture' cohort) answered \textit{``a e f fits my eye''} on the \emph{`\pageNameExplanation'} page  and \textit{``now / now / nos''}, \textit{``1 / 3 / 2''}, \textit{``People love us / Color field tree / Lamp tide dog''} on the \emph{`\pageNameWords'} page. Out of the remaining 22, two of the answers were incomplete, with one respondent that dropped out on the \emph{`\pageNameRanking'} page and another on the \emph{`\pageNameExplanation'} page.

\section{Analysing the Data}\label{sec_AnalysingTheData}

All the examples of answers from participants are presented by us here in English, but many of them are translations from Norwegian (including grammar corrections; though not for the English ones, which are kept verbatim, including their grammatical errors).

\subsection{Sensical vs. Nonsensical Answers}\label{subsec_SensicalVsNonsensical}

The participants' explanations (i.e., their written texts) were analyzed qualitatively and coded into one of the three rationales from Section~\ref{sec_BiasRevealing} (i.e., Section~\ref{sec_BiasRevealing}.\ref{rationaleBalance} `balance,' Section~\ref{sec_BiasRevealing}.\ref{rationaleShapes} `shapes', or Section~\ref{sec_BiasRevealing}.\ref{rationaleAlgorithm} `algorithm') depending on the fit between the text and the category. During the first full-scale study we found one answer (pID 38, first cohort) which triggered us to introduce another category or rationales, called `sounds'; the answer explained the choice of letter placement as ``If you sing the alphabet in Norwegian then the best fit with the rhythm is to place `H' to the left, because you have a small pause before singing `H' after `G'.''.

There were still many answers that could not be categorized into the above, either because they did not make much sense, or the reason given was no reason at all. However, many of these answers were recurrent, transcending even the language differences, and this allowed us to group them in categories. Some of the more generic answers were so similar between English and Norwegian that we could regard them as `universal'.  

\begin{itemize}
\item \emph{`Logical'}: \textit{``I think it would be logical put the H in the right position''} (pID M:11272137574, third cohort); or \textit{``Left seems right because it seems logical''} (pID 53, first cohort); or \textit{``because it seemed most logical''} (pID 12, first cohort).

\item \emph{`Pattern'}: \textit{``My choice was made by what I thought was a pattern''} (pID M:11282013578, third cohort); or \textit{``because of the order of the previous ones.''} (pID 47, first cohort); or \textit{``Due to previous placements above.''} (pID 50, first cohort); or \textit{``The left seems to follow the pattern''} (pID M:11270235127, third cohort); or \textit{''Because I think the pattern follows that path.''} (pID 4, first cohort).

\item \emph{`Random'}: \textit{''Just chose something''} (pID 33, first cohort); or \textit{``It seemed like the pattern of the letters would place the H on the right, but there isn't enough information for me to decide, so it is kind of a guess.''} (pID M:11270119183, third cohort).

\item \emph{`Alphabet''}: \textit{``...going in reverse alphabetical order.''} (pID M:11272389655, third cohort); or \textit{``The letters are to be placed based on the alphabet song.''} (pID M:11271323609, third cohort).

\item \emph{`Handed'}: \textit{``I'm right handed so I favor my right side and it just seemed like the `correct' answer to me.''} (pID M:11271930008, third cohort); or \textit{``Most people are right-handed, so dragging the letter to the right felt like an automatic default action. Dragging it to the left requires a more deliberate choice.''} (pID M:11270365264, third cohort); or \textit{``I chose the right because in every day society its pretty common for the right side of thing to be accepted as good, such as right handed people, the right hand of god, etc etc. I also chose the right side because its `right'.''} (pID M:11270469031, third cohort).

\item \emph{`No reason'}: \textit{``Because it looked most natural compared to what has already been done.''} (pID 36, first cohort); or \textit{``it looked natural''} (pID 7, first cohort); or \textit{``It felt reight''} (pID S:11178992036, third cohort); or \textit{``Seems better''} (pID S:11174871629, third cohort).

\item \emph{`micro-balance'}:\textit{ ``H on the right side follows the pattern of the EF on the left side, which are a pair.''} (pID M:11272137574, third cohort); or \textit{``Because it makes sense to me that H and G are grouped together, since there is a grouping on the other side as well.''} (pID T:11058678726, third cohort); or \textit{``In my opinion it looks nicer to have `H' after `G'. It has a bit to do with how `E' and `F' are positioned.''} (pID T:11058678669, third cohort).\end{itemize}

To reduce the number of uninterpretable answers (i.e., explanations, rationales), starting with the second full-scale survey, we introduced the alternative answers that respondents could choose from; see Figure~\ref{fig_AlternativePage} in the \ref{app_TheQuestionnaire}. These alternatives were formulated based on the wordings that we encountered among the responses from the first study. Thus, the first study provided a type of `saturation' of alternatives. As a result of coding and categorization we arrived at five alternative answers, as well as a sixth and seventh alternative: ``I just chose something'' and ``I already gave an explanation''\footnote{Recall that this was a `required' question/page (marked with *), as opposed to the previous \emph{`\pageNameExplanation'} page,  and thus a choice must be made on this page.}. We also used these to help us code the answers, i.e. when they did not give any explanation (it was not required) but instead chose from our list, we used that choice as the rationale. When they gave an explanation that did not make sense, but then also chose one of our example explanations, we again used the one that they chose, for our categorization. There was also the case when their explanation somewhat seemed to contradict the choice that they picked. In this case, we still used the choice for the categorization. The following are a few explanations that made no sense, but an alternative was chosen: \textit{``The left side seems like the logical, correct side when compared with the letters that came before it.''} (pID M:11270382691, third cohort) but then chose the alternative answer that sounded ``Because of the appearance/form of the letters. On the left side they have straight lines, whereas on the right side are rounded.''; or \textit{``I choose left because i think it can be very good with random letters in the left.''} (pID M:11270101280, third cohort) but then chose the alternative ``The same number of letters on each side.''; or \textit{``There seems to be a pattern. Placing the letter on the left makes the most sense to continue that pattern.''} (pID M:11271499180, third cohort) but then chose a pattern from the alternative that sounded ``It creates a pattern of the type: 1-3, 2-4, 3-5, \dots or 1-3-2, 1-3-2, \dots or 1-3-2, 2-3-1, \dots ''.

We thus define as \emph{Sensical} those answers that were interpretable and allowed for category inclusion  in one of the three rationales from Section~\ref{sec_BiasRevealing}, and we define as \emph{Nonsensical} the remainder of the answers, both those that made no sense at all, as well as  those that could be coded as `sounds' which were very few in number (i.e., one in the `Social sciences' cohort, one in the `Natural sciences' cohort, and five in the `Arts and Culture' cohort, all of which did not involve our helper alternatives, but their explanations that described the reason as `sound').

In the following we make two observations about our sensical vs.\ nonsensical perspective on the responses.

\subsubsection{Helping with the Self-explanations}

\begin{figure}[h]
\centering
\includegraphics[width=0.9\columnwidth]{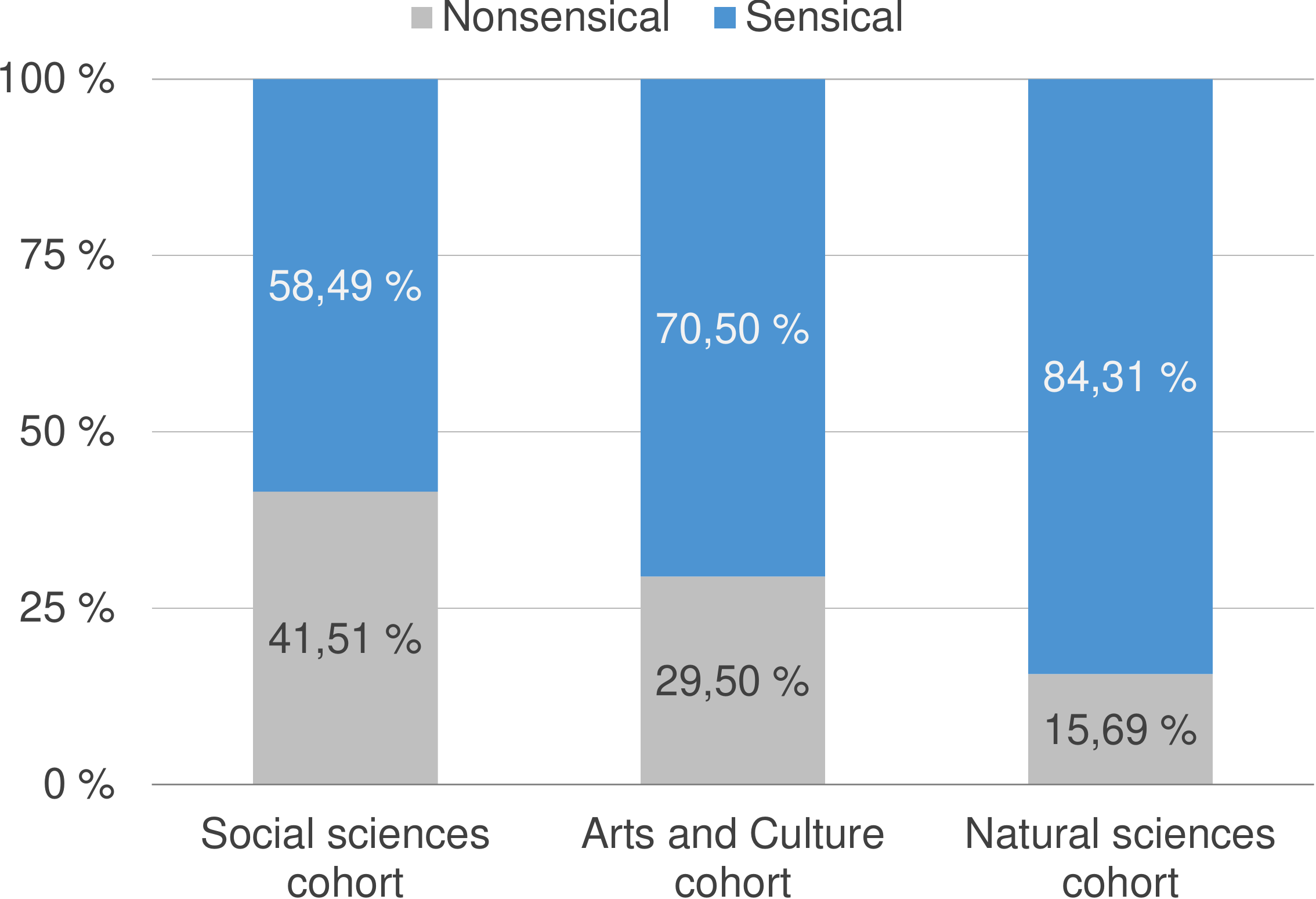}
\caption{The nonsensical answers are decreasing after improvements done to the survey after the first full-scale study.} 
\label{chart_sensical}
\end{figure}

\begin{figure*}[t]
\centering
\includegraphics[width=\textwidth]{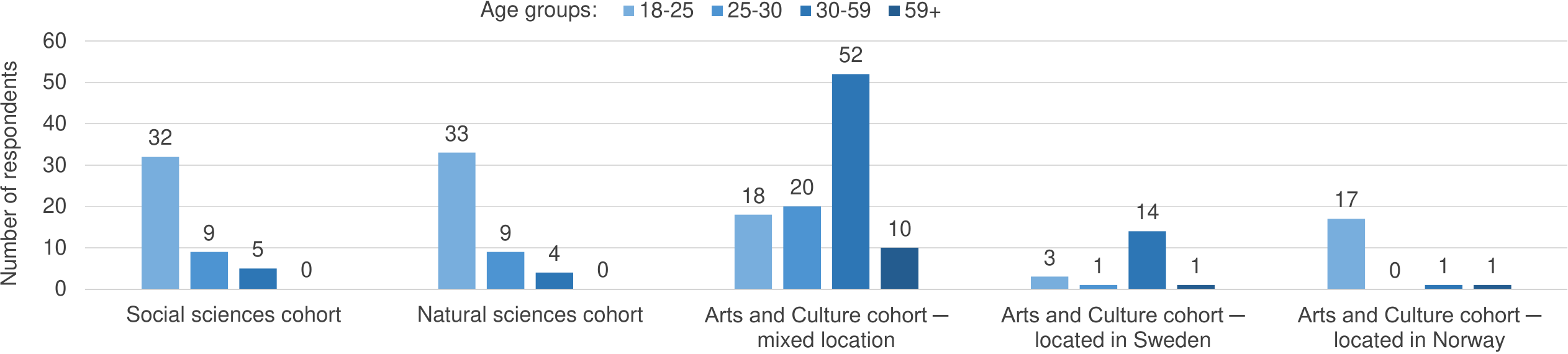}
\caption{Distribution of respondents in age groups, grouped by cohort.} 
\label{chart_age}
\end{figure*}

The first regards the level of help that the different cohorts received. We observe in the Figure \ref{chart_sensical} a significant increase of answers that allowed for interpretation when the respondents were offered the alternative explanation choices. The `Social sciences' cohort were not helped and the percentage of sensical responses is only 58.49\%. To all other respondents we allowed them to skip the \emph{`\pageNameExplanation'} question and required that they at least chose one of the alternative explanations. The sensical answers increased significantly to 70.50\% and 84.31\% for the `Arts \& Culture' and the `Natural sciences' cohorts, respectively. It is particularly noteworthy the increased level of interpretability that this choice in the design of our studies brought. We have counted 22 answers given by the participants in the `Arts and Culture' cohort that were not understandable only by themselves, but could nevertheless be coded because of their choice of alternative explanation. This would have otherwise tilted the percentage to only 54\% sensical answers. We also had 10 that chose to skip the self-explanation and only select one of the alternatives.
{\color{white} \ref{chart_age}}

\subsubsection{Programming Environment Confinement}

The second observation regards the confinement aspect that we described in Section~\ref{subsec_TheParticipants}. One can observe that as soon as the participants were confined their explanations became even more sensical. Here we look at the two cohorts that were helped (i.e., to whom alternative explanations were offered), and we notice the increase from 70.50\% to 84.31\% in the case of the respondents from the `Natural sciences' cohort who were confined to the classroom and course working-hours. This bears evidence that the transition from a non-professional towards a more professional programming environment (as explained in Section~\ref{subsec_TheParticipants}, where people are both `forced' to focus appropriately on their task as well as being helped by resources or tools) would trigger the programmers to be more careful about their choices. This could also contribute to lowering the amount of bias. Indeed, we have observed that several participants tried to think in terms of games, since the task consisted of programming a game. Examples of such explanations are: \textit{``I choose left because it's a game and i think according to the pattern gamer will choose right side psychologically. Thus he/she will lose.''} (pID M:11274822275, third cohort); or \textit{``I feel the right side would be the most common choice so if the player was thinking creatively they would choose the left side to place the h''} (pID M:11274883590, third cohort).

Another aspect of the confinement is that it triggers the System 2 thinking, which is known to result in a reduction of human biases. We have also observed instances of System 1 vs. System 2 thinking in our participants, i.e., `starting' as a System 1 response, but then `self-apprehended' and activated a System 2 response. One example is a participant who has chosen to answer the letter placement question with `Right' but then when asked to explain the choice said \textit{``I choosed right previously but actually left makes more sense.  Balancing the sides; 4 letters on the left, 4 letters on the right.''} (pID M:11272410463, third cohort). Another explanation applies this reasoning to the `players', thus also thinking in terms of the game task at hand: \textit{``The reason why I would give the reward if the player place on the left is because of both dotted boxes are the correct answer. However, I feel that most would place it to the right since it is easier to recognize that H comes after G. So figured that would place it there without knowing they could be gotten the same correct. So I concluded that the reason why I choose the left is that fewer people would pick that.''} (pID M:11272095027, third cohort).

\subsubsection{Suggestions}

Such observations should be further investigated using more controlled experiments. In any case, one piece of conclusive advice that we can offer is that it is useful for the outcome of the experiment if the respondents are given (i.e., as help) alternative choices of answers/explanations (or rationales in our case). These choices should be carefully made, preferably using answers that are observable in the target population (i.e., like we did ourselves, extracting answers from the first survey). A more controlled experiment should yield more sensical answers, e.g., by carrying out the experiment in a more strict `laboratory' setting. It seems that only paying the participants, as we did through the two platforms SurveyMonkey and Amazon's mTurk, does not increase the quality of the answers.

\subsection{Demographics}\label{subsec_Demographics}

%

\begin{figure*}[t]
    \centering
\begin{subfigure}[t]{0.45\textwidth}\vskip 0pt
        \centering
\includegraphics[width=\textwidth]{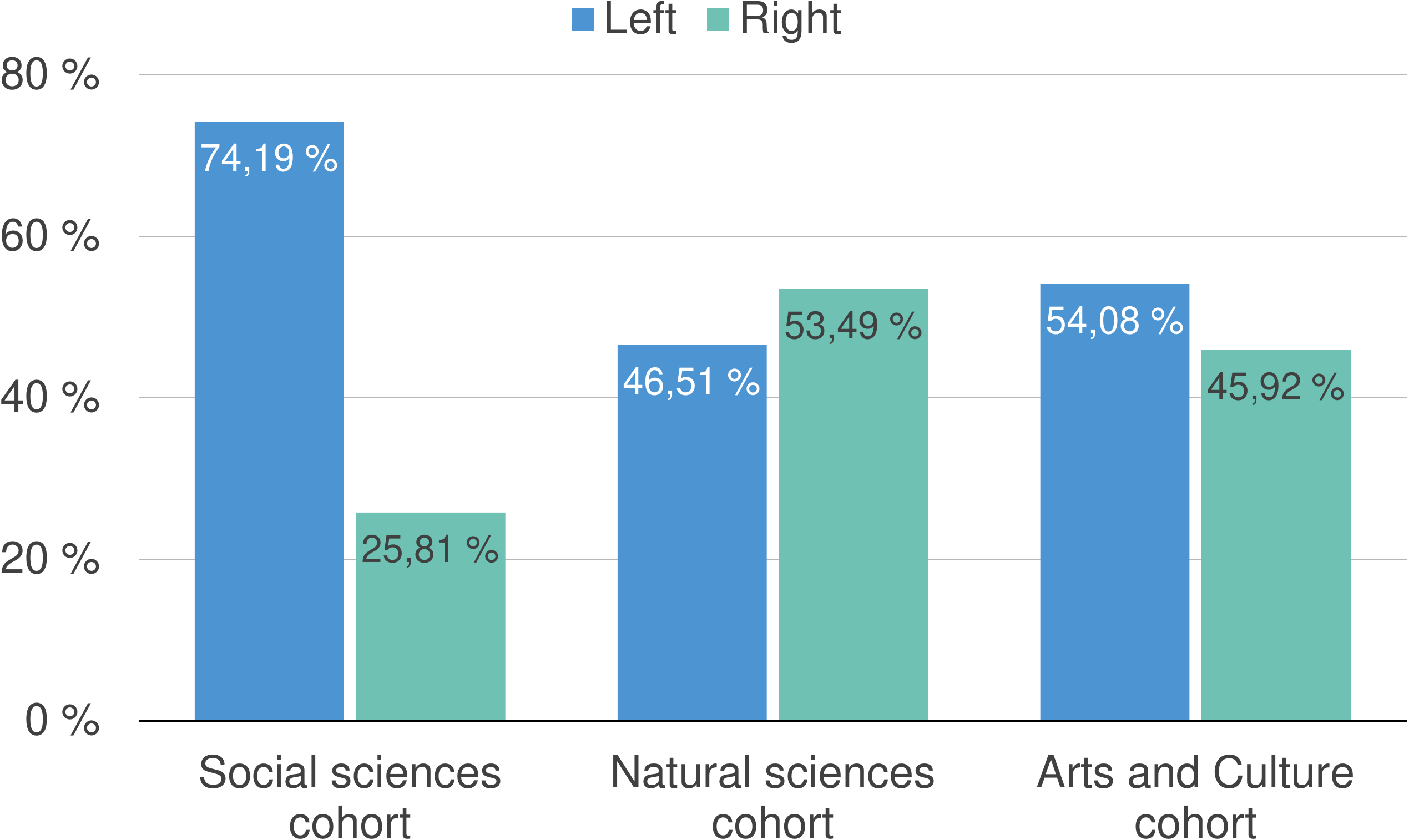}
\caption{\begin{footnotesize}Inside each cohort.         \end{footnotesize}} 
\label{chart_letterPlacement_byCohort}
    \end{subfigure}\hspace{0.06\textwidth}%
\begin{subfigure}[t]{0.49\textwidth}\vskip 0pt
\includegraphics[width=\textwidth]{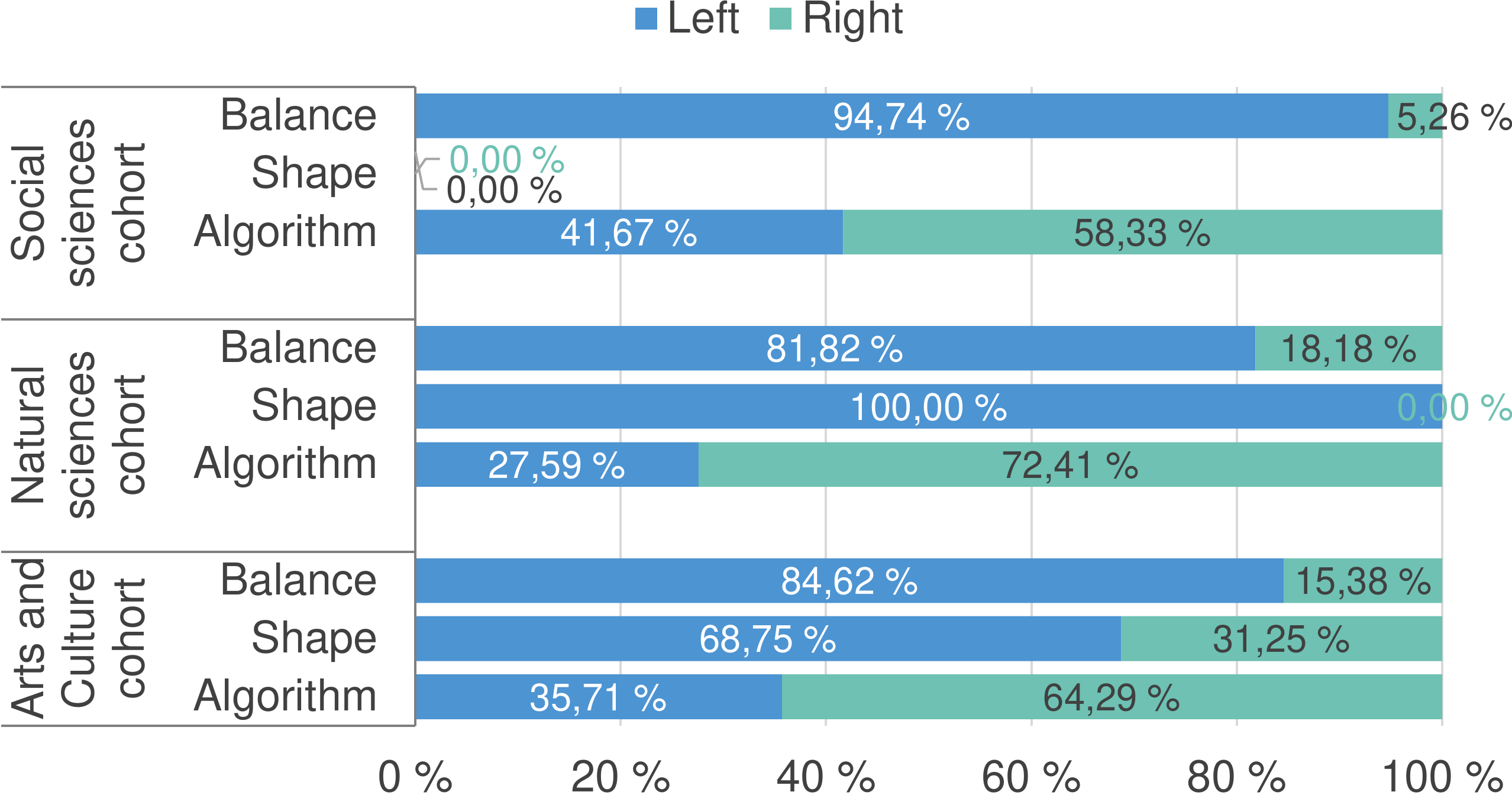}\vspace{0.2cm}
\caption{\begin{footnotesize}Distributed by the three rationales (into which the respondents' answers to the \emph{`\pageNameExplanation'} question were categorized).                                                                                                                                                \end{footnotesize}} 
\label{chart_letterPlacement_byCohortAndRationales}
   \end{subfigure}%
    \caption{Overview of the choice of placement of the letter `H' to the Left or Light.}
\label{chart_letterPlacement}
\end{figure*}

We categorized the respondents into four age-groups, which we also named: (i) 18-25 -- `younger students', (ii) 25-30 -- `older students', (iii) 30-59 -- `professionals' working in their respective fields, and (iv) older than 59 --  `approximation of the retirement age'.  Relevant for our analysis was to look at which age group is representative for each cohort. As shown in the Figure~\ref{chart_age} for `Social sciences' and `Natural sciences' cohorts, these are predominantly composed of respondents between the age of 18 and 25. 
{\color{white} \ref{chart_letterPlacement}}

In the case of the `Arts and Culture' cohort, the respondents, though they have the same educational/professional background, differ in location: (i) the group with respondents from mTurk are located in different countries, (ii) the group with respondents from SurveyMonkey are located in Sweden, while (iii) the group from `Cultural studies' are located in Norway. In addition, the mTurk and SurveyMonkey  respondents were recruited based on their professional affiliation, while the `Cultural studies' respondents were recruited based on study line affiliation. From the chart in Figure~\ref{chart_age}, we can see that for the mTurk and SurveyMonkey respondents the predominant age group is 30-59, while for the `Cultural studies' respondents is 18-25. 

Based on these age groupings we can conclude that for the `Social sciences', `Natural sciences' and `Cultural studies' cohorts the respondents are predominantly `young students', of age between 18-25, while for the mTurk and SurveyMonkey, the respondents are predominantly `professionals', of age between 30-59.


\section{Control Questions}\label{sec_ControlQuestions}

In the study we included several additional questions with the purpose to control for various aspects. As one can recall from Section~\ref{subsec_TransitioningFromVolunteersToProfessionals}, we have used the open-ended questions to identify robot/automated answers. Three questions were of particular importance, as they were meant to control, or to reinforce, three important assumptions that we have. These control questions are analysed in detail in the next subsections. Essentially, Section~\ref{subsec_LeftRightPlacement} reinforces our bias revealing test from Section~\ref{sec_BiasRevealing} as a good instrument; Section~\ref{subsec_WordsSuggestions} tests how well our priming metaphors from Section~\ref{sec_MetaphorsAsPrimingMethod} worked, since such story-based metaphors may be revealed within listings of words/synonyms; whereas Section~\ref{subsec_LifeAspects} reinforces our beliefs and categorization of the backgrounds of the three cohorts that we study, thus confirming that the categories/labels we provided in Section~\ref{subsec_TheParticipants} are appropriate, and the bias transfer studies that we conduct, as described in Section~\ref{subsec_Results_InfluencesBackground}, are well informed.

\subsection{Left/Right Placement}\label{subsec_LeftRightPlacement}

On the \emph{`\pageNameTask'} page of the survey, the respondents are asked to decide whether to reward the player for the placement of the letter `H' on the left or right side of the vertical line on the game board. 
This is one of the three questions on this page, intended as a control question for the hypothesis that we made in Section~\ref{sec_BiasRevealing}, i.e., that choosing to place the letter to the `right' should indicate a preference for the `algorithm' rationale, while choosing `left' a preference for the `balance' or `shapes' rationales. 

An analysis of the `left/right' placement wrt.\ each of the three rationales confirms this initial assumption, see Figure~\ref{chart_letterPlacement_byCohortAndRationales} for numbers. 
In particular, observe that in the case of the `algorithm' rationale the choice of placement to the `right' is overwhelming for each cohort; and similarly, `left' is the preferred choice when answering with the `balance' or `shapes' rationales in all cohorts.
%
%
%

Moreover, the analysis of the `left/right' placement overall inside each cohort, which we summarize in Figure~\ref{chart_letterPlacement_byCohort}, confirms our earlier observation that the background of the participants is reflected in their preference for one choice of placement (and thus for one type of rationale). 
The `Social sciences' cohort chose mostly `left', associated with the `balance' and `shapes' rationale, in a proportion of 74.19\%. 
The `Natural sciences' cohort chose mostly `right', associated with the `algorithm' rationale, in a proportion of 53.49\%. 
The `Arts and Culture' cohort again chose mostly `left' in a proportion of 54.08\%.
{\color{white} \ref{table_firstFiveSuggestedWords}}

\begin{table*}[t]
\centering
\includegraphics[width=\textwidth]{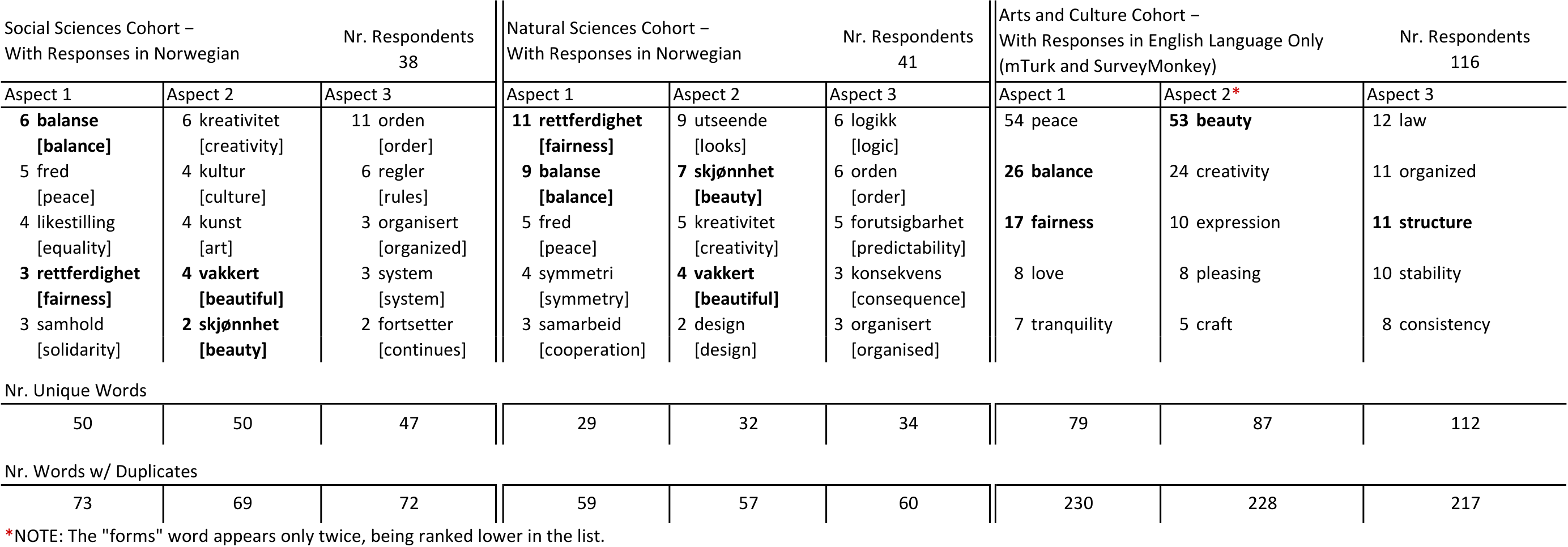}
\caption{First five most suggested words, with the number of occurrences to the left of each word.} 
\label{table_firstFiveSuggestedWords}
\end{table*}

Though there are more pattern combinations possible when the letter `H' is placed on the right, there is also one pattern combination possible with the placement of the letter `H' on the left. This is the sequence of 1-3-2 letters, reversed from right to left. For the `Natural sciences' cohort we observe that eight participants out of 20 that chose `left' (9 associated with `balance' and 3 with `shapes') chose this type of pattern combination. 
This can explain the percentage of respondents in this cohort that chose the `left' side.

There are responses identifying this type of pattern also in the `Arts and Culture' cohort. In addition, the respondents here were even more creative, by finding additional pattern combinations with the placement of the letter `H' on the left, e.g.: the initial 1-3-2-1 pattern of letters will continue with `H' as the first letter in a similar pattern of 1-3-2-1 or a pattern of the type 1-3-2-1-2-3.

\subsection{Words Suggestions vs. Priming Metaphor}\label{subsec_WordsSuggestions}

In this section we analyse the words given by the respondents to the \emph{`\pageNameWords'} question of the survey, which was meant as a control question for our priming metaphors, i.e., to reveal which of the primes worked and how well. This is relevant for our second research question: the transfer of contextual bias, which we analyse closer in Section~\ref{subsec_Results_InfluencesMetaphors}. In particular, this control question was intended to check the suitability of the priming words that we have chosen. Moreover, in the future, one can extract from these suggested words more adequate priming words, for further research on alternative priming metaphors.

The words were cleaned up for spelling and uppercases. In addition, we changed the derived words to only one syntactic form, i.e., adjective, verb, noun, or adverb. The precise syntactic form that was kept was decided based on the frequency of that form throughout all the responses. In the very few places where sentences were used instead of words, we kept only relevant individual or composed words (e.g., from \textit{``pleasant surroundings''} we kept only the word \textit{``pleasant''}; or from \textit{``being crafty''} only \textit{``crafty''}; or \textit{``appreciating beauty''} was split in two words \textit{``appreciation''} -- changed into a noun, as this was the form most used in the responses -- and \textit{``beauty''}), and removed other syntactic forms, such as conjunctions, prepositions, and pronouns that did not have a meaning by themselves or their meaning was not relevant for explaining the respective `life-aspect'. We also created compounded words where this was possible (e.g., \textit{``looking good''} was changed into \textit{``good-looking''}). We strived to be minimal in such changes, and we especially did not do semantic changes.

The participants were given three pairs of words to suggest synonyms for, each containing two of the four priming words used in the `Philosopher story', i.e.:  \textit{``Could you suggest 1 to 4 individual words that for you have similar meanings as each of the three life-aspects: 
`harmony and equality'; `balance and fairness'; `order and continuity'\hspace{0.5ex}''}. 

In analysing the responses for the \emph{`\pageNameWords'} question, we looked for the occurrence of the other two words that were used in the `Philosopher story' as primes (cf. Section~\ref{sec_MetaphorsAsPrimingMethod} also), i.e.: 
{\color{white} \ref{table_ocuurancePrimingWords}}

\begin{table*}[t]
\centering
    \begin{subfigure}[t]{\textwidth}\vskip 0pt
    \includegraphics[width=\textwidth]{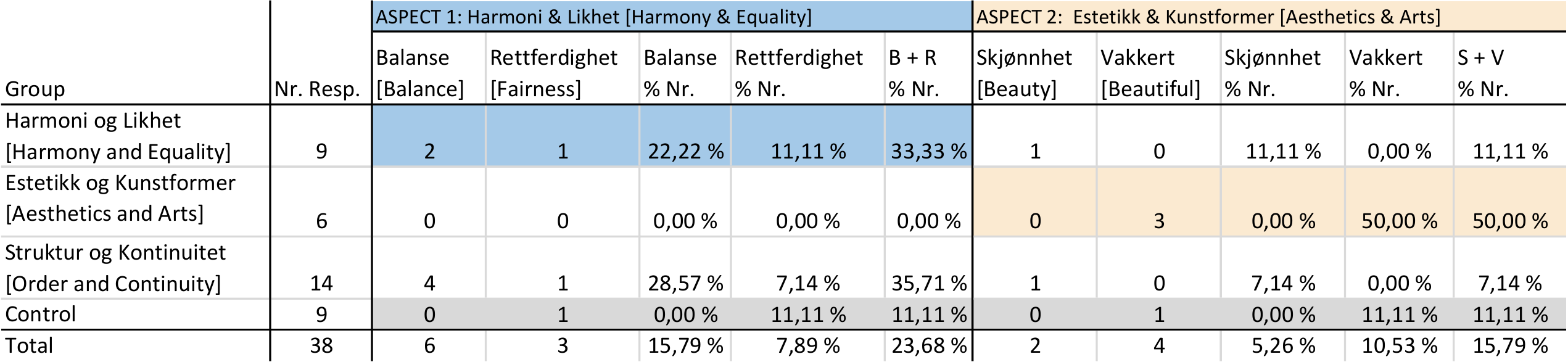}
    \caption{Occurrence of priming words in `Social sciences' cohort.} 
    \label{table_ocuurancePrimingWordsSocialSciences}
    \end{subfigure}
   \vspace{0.2cm}
    \begin{subfigure}[t]{\textwidth}\vskip 0pt
    \centering
    \includegraphics[width=\textwidth]{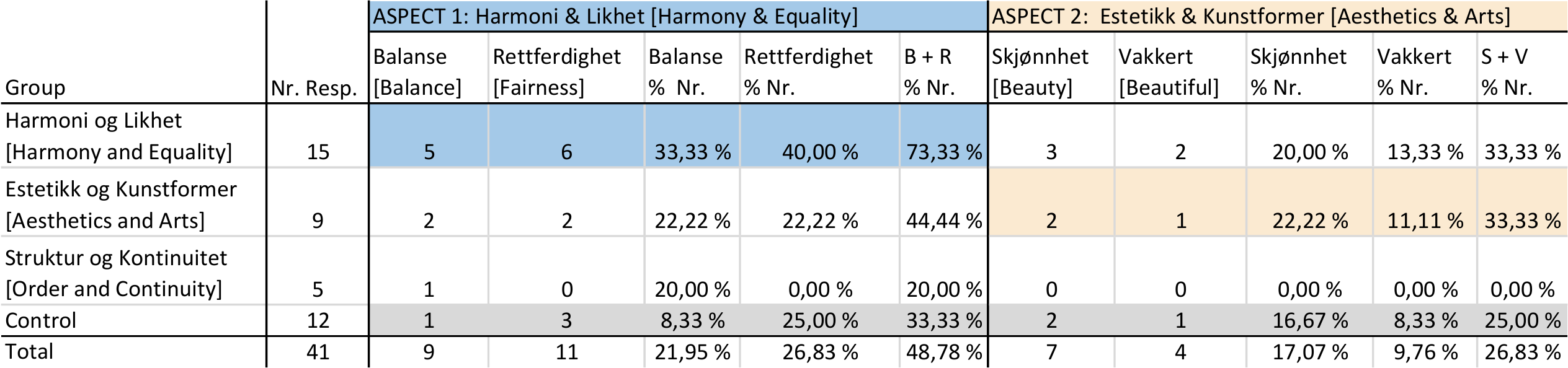}
    \caption{Occurrence of priming words in `Natural sciences' cohort.} 
    \label{table_ocuurancePrimingWordsNaturalSciences}
    \end{subfigure}
   \vspace{0.2cm}
    \begin{subfigure}[t]{\textwidth}\vskip 0pt
    \centering
    \includegraphics[width=\textwidth]{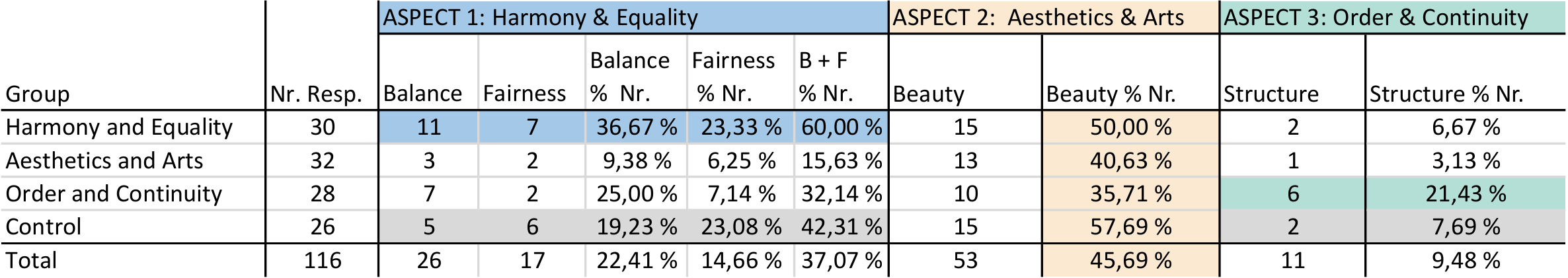}
    \caption{Occurrence of priming words in `Arts and Crafts' cohort.} 
    \label{table_ocuurancePrimingWordsArtsandCrafts}
    \end{subfigure}
\caption{Occurrence of priming words.}
\label{table_ocuurancePrimingWords}
\end{table*}
\begin{itemize}
\item For the case when the respondents were given `harmony and equality' (which we will call \emph{`Aspect 1'} in the rest of the section) as words to find synonyms for, we looked for the occurrence of `balance' and `fairness' in their answers. 
\item For `aesthetics and arts' (\emph{`Aspect 2'}) we counted occurrences of `beauty' and `forms';
\item For `order and continuity' (\emph{`Aspect 3'}) we counted `structure' and `linearity'.
\end{itemize}

Since we had Norwegian speaking respondents we prepared a Norwegian version of the \emph{`\pageNameStory'} with the corresponding words beeing respectively: `harmoni', `likhet', `balanse', `rettferdighet'; and `estetikk', `kunstformer', `skj\o{}nnhet'; and `struktur', `kontinuitet', `gjenta\-kelse', `linearitet' (see translations in Table~\ref{table_ocuurancePrimingWords}). In this case, the given words for the `Aspect 2' were actually three of the four priming words because we gave the compound word `kunstformer' which contains in translation both `forms' and `arts'. As such, only one word of the four, i.e., `beauty', remained to be counted. However, the responses contain one word that is very similar in meaning with `beauty', namely `vakkert' (beautiful). The degree of similarity was evaluated using the ``meaning relation'' of the Norwegian WordNet\footnote{National Library of Norway. 2015. Raw data: N-gram (NBdigital). Date: 2015-06-02. \url{http://www.nb.no/sprakbanken/show?serial=sbr-35&lang=nb}} which gives the extremely close numbers 0.155 for `skj\o{}nnhet' (beauty) and 0.153 for `vakkert' (beautiful).

An overview of the first five most `suggested words' for each cohort is given in Table~\ref{table_firstFiveSuggestedWords}: ``The First Five Most `Suggested Words'\hspace{0.5ex}''. The expected words for the `Aspect 1' appear among the five most occurring words, sometimes at the top, in all three cohorts. For the `Aspect 2', the word `beauty' appears highly ranked, but not the word `forms'; again this is the case for all three cohorts. However, for `Aspect 3' we only see the word `structure' appearing among the top five only in the `Arts and Culture' cohort, which had the English version. None of the two corresponding Norwegian words appear in the responses of the `Social sciences' and `Natural sciences' cohorts. The word `linearity' did not appear at all.

We have used the Norwegian WordNet (most recent year count being 2013) frequency feature to check how frequent are the two words of the `Aspect 3' in the Norwegian vocabulary. It turns out that these two words have a very low frequency: 0.0001857 for `gjentakelse' (equivalent to `structure') and 0.0000097 for `linearitet' (equivalent to `linearity'). Compare these frequencies with those for the other words: 0.001938 for `balanse', 0.001121 for `rettferdignet', which are one order of magnitude higher, or 0.002912 for `vakkert' or 0.001410 for `skj\o{}nn' (short version of `skj\o{}nnhet' which we used).

We conclude that the use of the word `structure' was successful, and can continue to be used in further research. However, another more widely known word needs to be used as Norwegian equivalent. This concerns the `linearity' word as well, for both Norwegian and English. We were not aware of the existence of WordNet at the time when we prepared our metaphors (in 2018-2019); but such tools can be valuable for choosing priming words. In our case we were not only interested in synonyms, but more in words that could prime towards the three different rationales for our programming test; e.g., the word `forms' was important, as well as `linearity'. 

Table~\ref{table_ocuurancePrimingWords} gives an overview of the numbers of occurrences (frequencies and percentages) of the priming words for each of the cohorts. The percentage numbers are calculated wrt. the number of respondents. In order to evaluate the priming effect of our metaphors we can compare the percentages from the primed group with the ones from the control group. The priming group considered is the one relevant for the respective aspect (highlighted in our tables), i.e., for the `Aspect 1' we look at the numbers of the `Harmony and Equality' group.

For the `Social Sciences' cohort we see the effect of priming on both `Aspect 1' and `Aspect 2': 33.33\% (representing the total from both `priming' words) to 11.11\% and respectively 50\% to 11.11\%. The `priming' words for the `Aspect 3' did not occur at all, thus, do not appear in the table. The table for the `Natural Sciences' cohort shows again the influence of priming in the case of `Aspect 1' and `Aspect 2': 73.33\% to 33.33\%, and 33.33\% to 25\%. For the `Arts and Culture' cohort we observe the effect of priming in the case of `Aspect 1' and `Aspect 2': 60\% to 42.31\%, and 21.43\% to 7.69\%. For the `Aspect 3', only one of the two words occurred, hence only one was considered. 

However, for the `Aspect 2' we do not see a priming effect, when compared with the control group. Quite the opposite, we can observe that the word `beauty' appears many times, with a total number of 53 occurrences and its distribution between the groups relatively even: 50\%, 40.63\%, 35.71\%, and 57.69\%. The word `beauty' seems to be present by default in the vocabulary of this cohort, irrespective of priming. This observation confirms that we did well by choosing a word that is largely used by this population. 

This indicates two other factors that might have had influence on the priming effect. One of these factors is how familiar the respondents are with the priming words. If the words are very little known or not understood, they will not be primed by them, as it is the case of the `Aspect 3' words. In addition, if the respondents have a large vocabulary to their disposal, which can be seen from comparing the `Nr. of unique words' with the `Nr. of words with duplicates' (see Table~\ref{table_firstFiveSuggestedWords}), the System 1 will be less inclined to use the priming words.  Such observations can be made in the case of the `Social sciences' students in comparison with the `Natural sciences' students: 147 unique words compared to 95. We see how the priming was stronger in the latter cohort compared to the former (they strive to find similar words, and the availability heuristic retrieves the primes from the short term memory).

The words the participants choose the most can also be affected by other immediate contextual elements. In the case of the `Natural sciences' cohort the survey was taken by the students as part of a course on logic. This made the word `logic' occur the most for `Aspect 3'.

\subsection{Life-aspects Ranking}\label{subsec_LifeAspects}

The \emph{`\pageNameRanking'} was meant as a control question for the way we identify the background in our cohorts. That is to say, we want to check whether there is a correlation between the self-ranking of the `life-aspects' and what we have considered as the background of the respondents. Moreover, we want to also look at the coded answers from the \emph{`\pageNameExplanation'} compared to the \emph{`\pageNameRanking'} because if the correlation is similar to the one we have observed previously from the background, then this would reinforce our perception of background. 

For creating the three types of cohorts we have considered the educational and professional backgrounds. However, these are only one part of a person's  background, arguably a large part, but yet a larger part is made of the society and culture that the respondents belong to. This is especially so for younger people, such as students. For the \emph{`\pageNameRanking'} we observe influences that come from the socio-cultural as well as educational and professional backgrounds. How strong  these are, and how much they relate to the bias transfer that we have observed before, is what we investigate in this section.

\begin{figure}[t]
\centering
\includegraphics[width=\columnwidth]{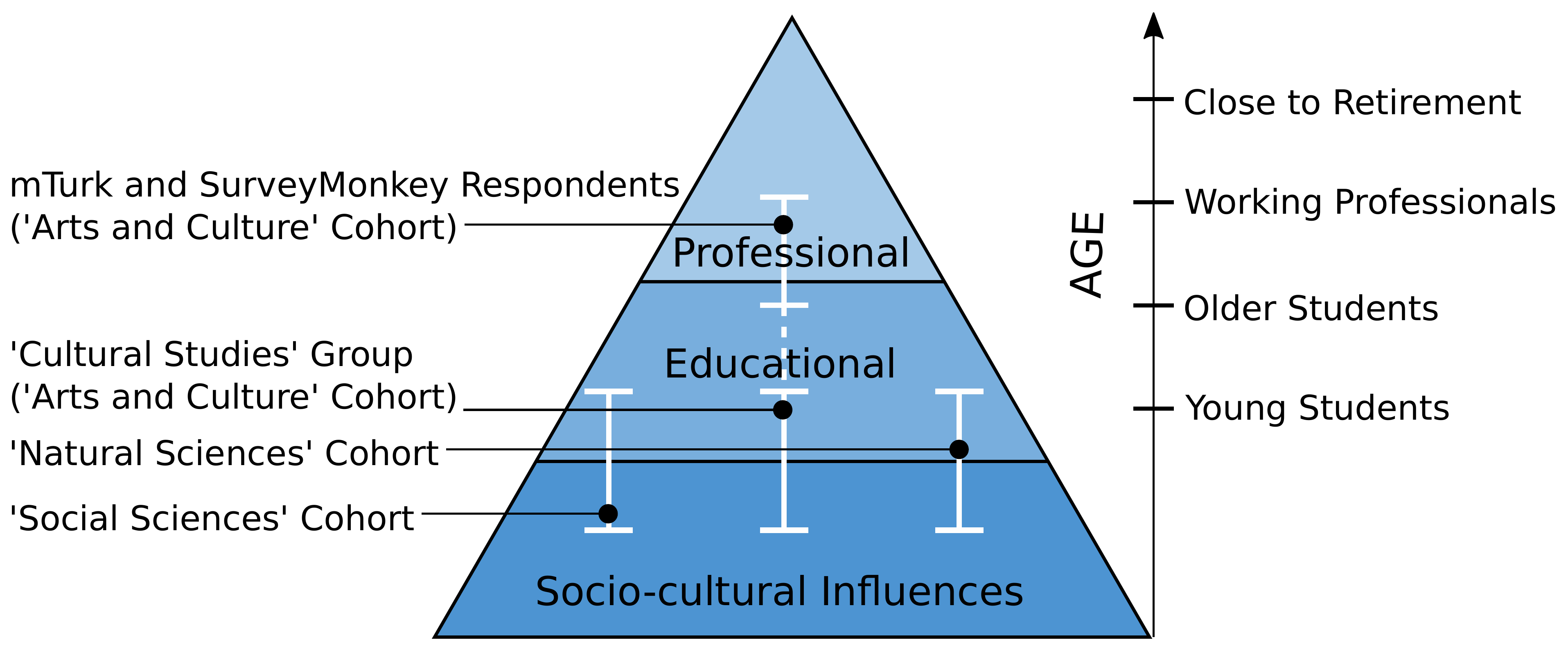}
\caption{Three sources of influence, correlated with the age when they are most strong, for the backgrounds observed for our cohorts, also indicating the age groups observed in the Section~\ref{subsec_Demographics}.} 
\label{fig_backgroundInfluencePyramid}
\end{figure}

We summarize the three types of influences in the Figure~\ref{fig_backgroundInfluencePyramid}, organized as a pyramid as we explain further.

Remember that the names that we gave to the `life-aspects' to be ranked by the participants were each using two of the four words used in Section~\ref{sec_MetaphorsAsPrimingMethod} as priming metaphors, i.e.: `harmony and equality', `aesthetics and arts', and `order and continuity'.  One word from the start of the story and another from the end of the story.

\begin{figure*}[t]
\centering
\includegraphics[width=\textwidth]{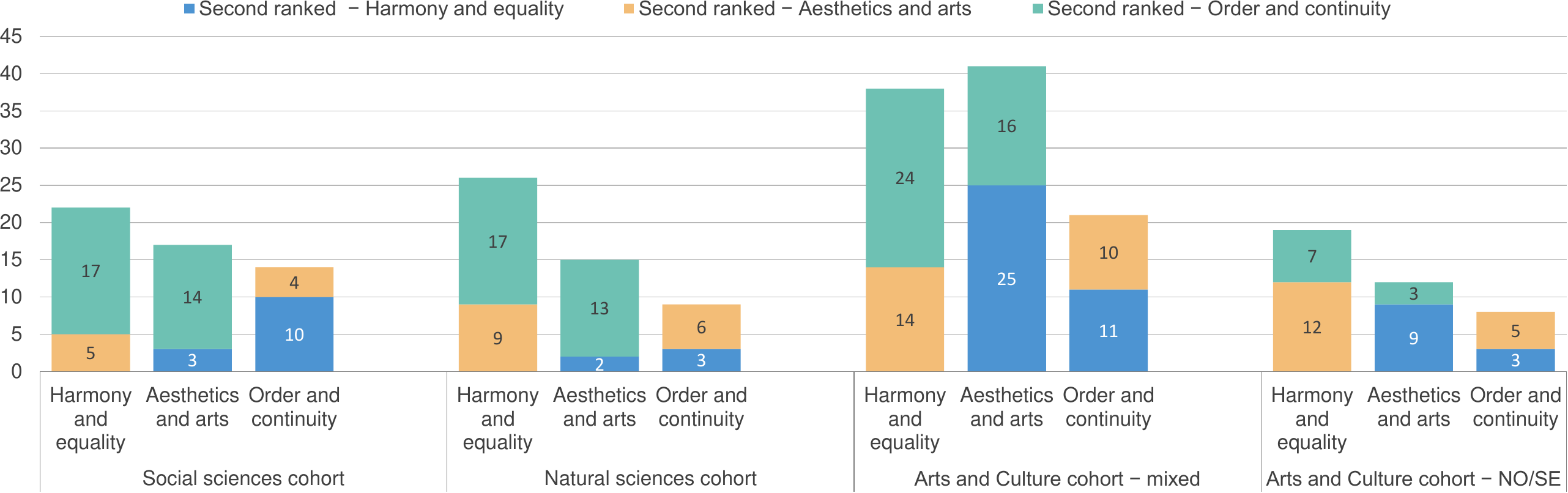}
\caption{Second ranked life-aspect distributed by first ranked and grouped by cohort.} 
\label{chart_secondRankedofthefirstRanked}
\end{figure*}

One's view on life is, among others, highly influenced by society and culture \cite{cialdini2004social,schultz2007constructive,cialdini2009influence}. For children this may be the main influence (e.g., through their parents), whereas for young adults (like many of our respondents who are young students) other factors of their own life-experience start to influence their views, including their education when they are studying and their professional environment when they start working. As presented in the Section~\ref{subsec_Demographics}, our respondents can be grouped in two main  socio-cultural categories based on their location, i.e., one `mixed location' (because the participants have a `mixed' provenance) and one `Scandinavian' (because of the respondents being located in Norway and Sweden). Since in this section we are particularly concerned with the socio-cultural background, and since the Arts and Culture cohort is composed of both respondents with `Scandinavian' and `mixed' provenance, we have decided to regard this cohort as two groups.
{\color{white} \ref{chart_secondRankedofthefirstRanked}}

\begin{figure}[b]
\centering
\includegraphics[width=\columnwidth]{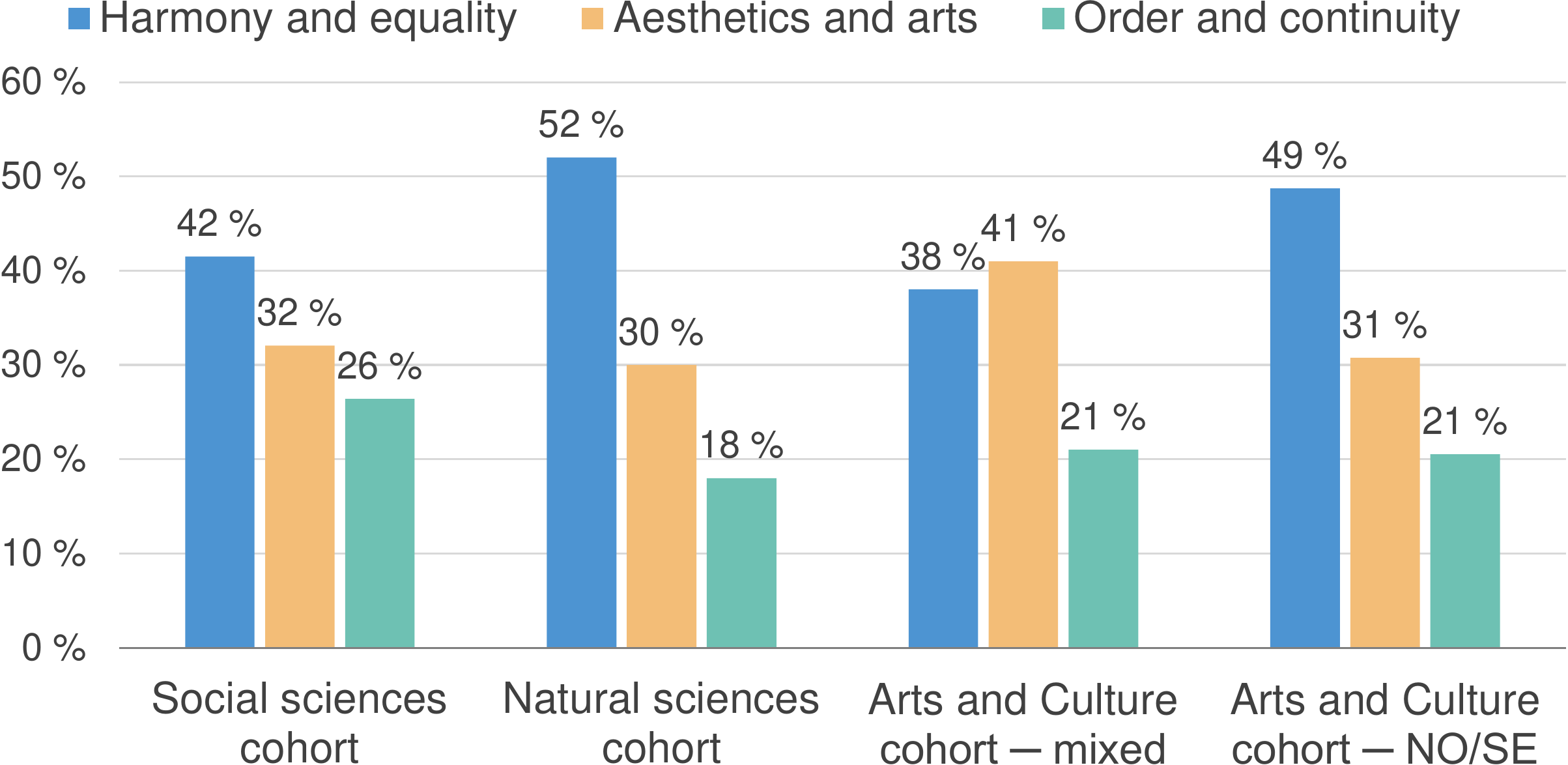}
\caption{First ranked life-aspect in each cohort.} 
\label{chart_firstRanked}
\end{figure}

In the chart from Figure~\ref{chart_firstRanked} 
we have counted the number of respondents that chose to rank first each of the three life-aspects, and we have plotted the respective percentages of the total inside the respective group. It is noteworthy to observe that in the case of the `Social sciences' and `Natural sciences' cohort, as well as the `Scandinavian' group from the Arts and Culture cohort, most \emph{first ranked} is `harmony and equality', followed by `aesthetics and arts', and `order and continuity' is only last. However, in the case of the `mixed location' group from the Arts and Culture cohort most first ranked is `aesthetics and art' whereas `harmony and equality' is slightly less chosen as first ranked.

Our interpretations 
are 
based on the fact that we are knowledgeable when it comes to what characterizes the `Scandinavian' type of culture and society, while having no knowledge about the `mixed location' group. However, our observations for the `mixed location' group are based on their professional background (as opposed to the educational background, in the case of the students' cohorts).

The fact that `Scandinavian' respondents across all cohorts value highest `harmony and equality' can be motivated by socio-cultural influences. These respondents come from a culture that promotes social equality, with high taxes for social-welfare and strong disregard for social unrest. In the case of the `mixed location' respondents, the view on life seems to be influenced strongest by their professional background, maybe also because these are ``working age'' respondents (as shown in the Section~\ref{subsec_Demographics}). 

We also counted the choices of \emph{second ranked} life-aspect (see Figure~\ref{chart_secondRankedofthefirstRanked}), looking inside each of the columns from the graph above. For the `Social sciences' cohort the second ranking was `order and continuity', being on the first rank in both columns with `harmony and equality' and `aesthetics and arts'. Note that `harmony and equality' can be seen both influenced by education and the socio-cultural background. However, in the column with `order and continuity' the most second ranked is the `harmony and equality' which corresponds with the educational background of the respondents. Thus we tend to conclude that the background of this cohort corresponds to the life-aspect `harmony and equality' where both the socio-cultural and educational backgrounds contribute.

In the case of the `Natural sciences' cohort, the second ranked was `order and continuity', which is in accordance with the educational background.

When it comes to the `Arts and Culture' cohort, the second ranked for the `Scandinavian' group is `aesthetics and arts', in both columns that do not correspond to their education i.e., `harmony and equality' and `order and continuity'. This again tells that the educational background influences these respondents, albeit less than their socio-cultural background. 

For the `mixed location' group `order and continuity' was ranked as second most in the `harmony and equality' column, while `harmony and equality' was ranked second most in the `aesthetics and arts' column. In this case we tend to conclude that the `professional' background influences the  view on life of these respondents the most.

In conclusion, we think that the control question about \emph{`\pageNameRanking'} confirms our assumptions about the backgrounds for our three cohorts and the fact that we have associated each of these cohorts with the life-aspect that is most predominant for those respondents. Therefore, we consider adequate the claims that we make throughout the paper where we correlate the background of a cohort with one specific life-aspect, and thus with one specific corresponding bias/rationale.

\section{Results}\label{sec_Results}

\begin{figure*}[tp]
    \centering
    \begin{subfigure}[t]{0.45\textwidth}\vskip 0pt
        \centering
\includegraphics[width=0.9\textwidth]{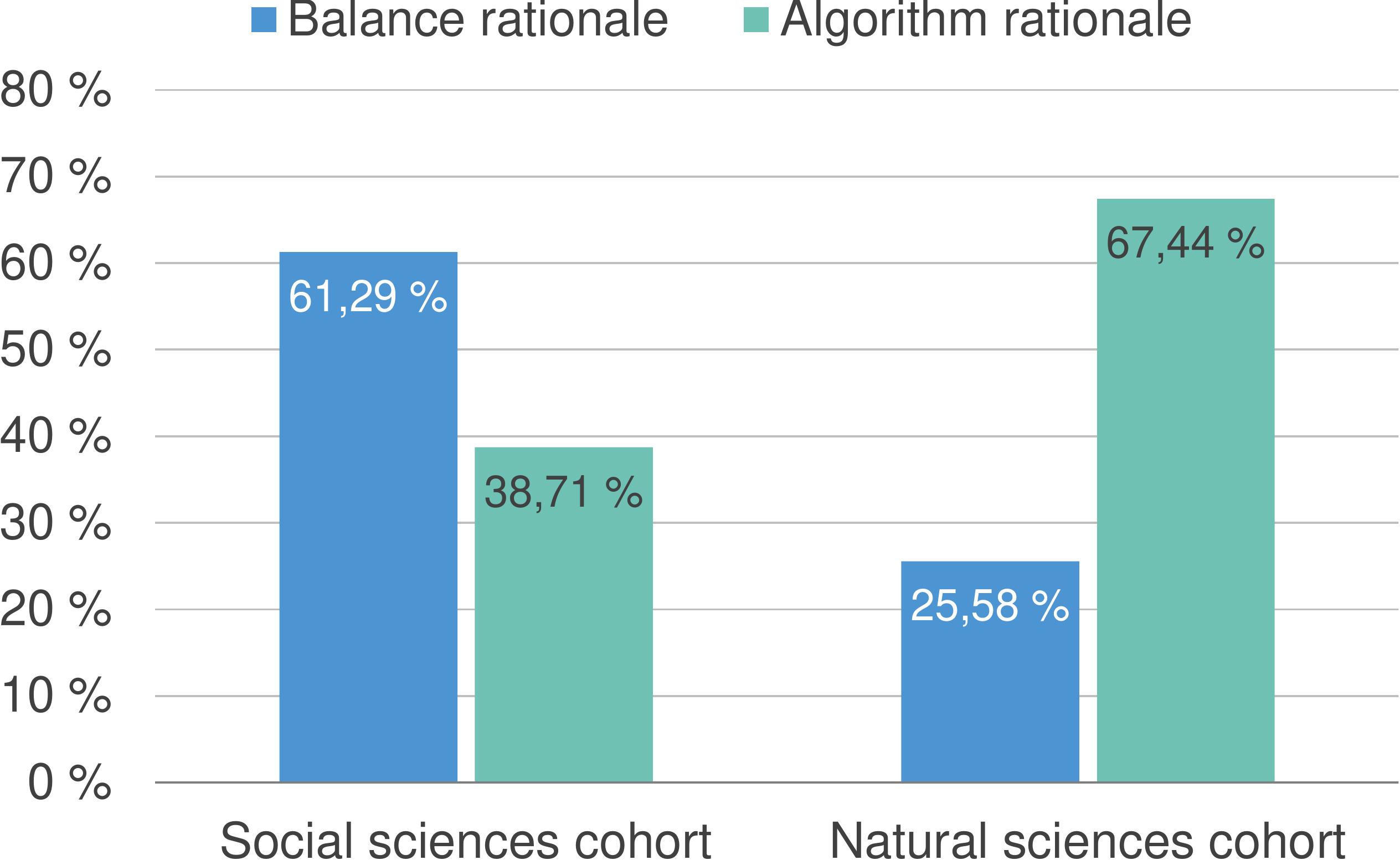}
\caption{\begin{footnotesize}All treatments included.                                 \end{footnotesize}} 
\label{chart_backgroundInfluences_allTreatments}
    \end{subfigure}\hspace{0.2cm}
    \begin{subfigure}[t]{0.45\textwidth}\vskip 0pt
        \centering
\includegraphics[width=0.9\textwidth]{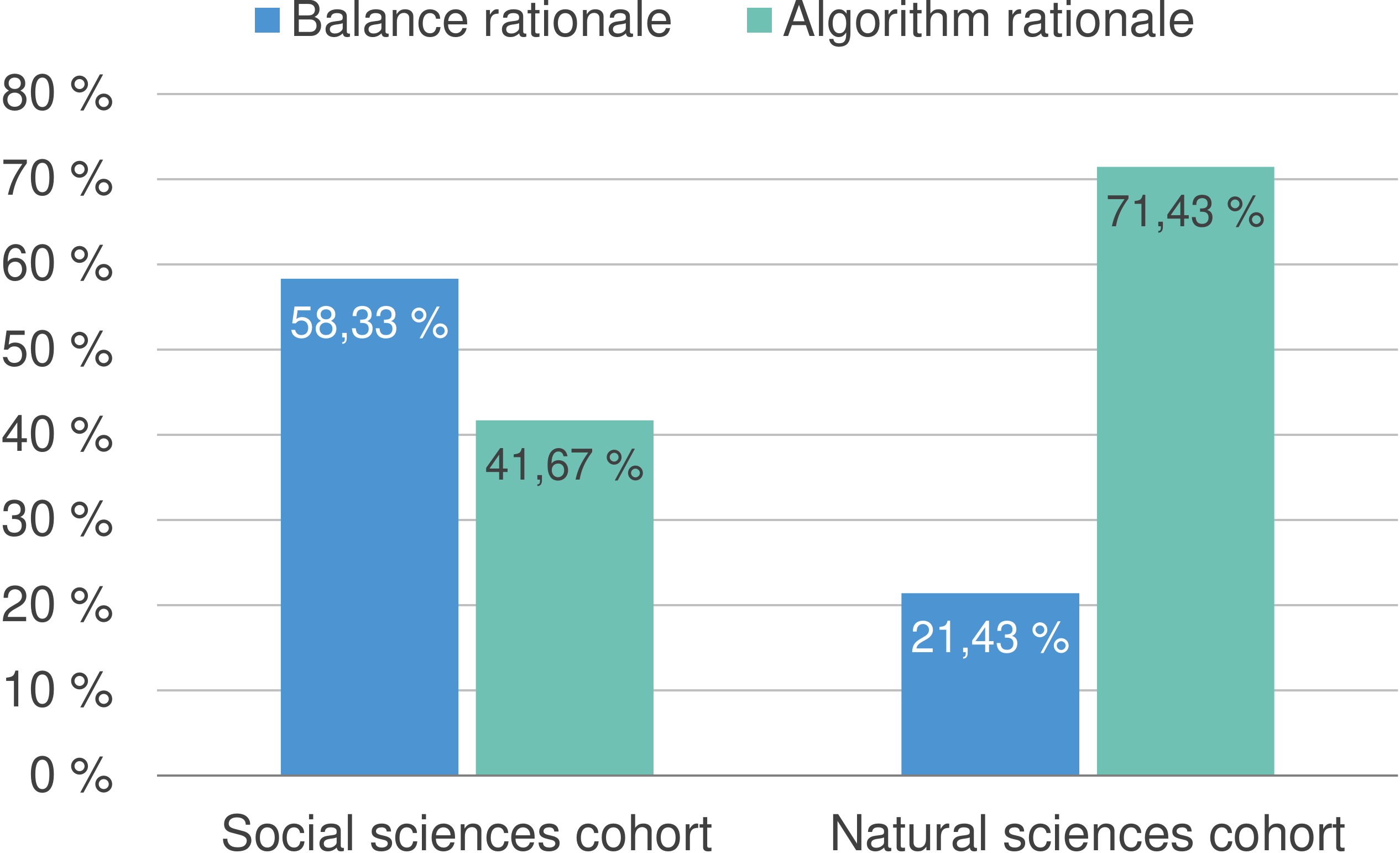}
\caption{\begin{footnotesize}Control group only.                            \end{footnotesize}} 
\label{chart_backgroundInfluences_control}
   \end{subfigure}%
    \caption{Comparison of background influences.}
\label{chart_backgroundInfluences}
\end{figure*}

The data is analysed both quantitatively and qualitatively. The qualitative analysis is done usually to detail the quantitative data, by analysing the participants' responses to the open-ended questions.
{\color{white} \ref{chart_backgroundInfluences}}

Our study is exploratory and tentative, this is why we employ a combination of statistical and descriptive analysis. Statistical analyses were not possible in all situations because of the small number of respondents in those categories. Statistical analyses were possible when the results from all cohorts were put together (analysis across the three cohorts) or in the case of the `Arts and Culture' cohort where we had 139 responses. Results from the descriptive analysis capture several systematic tendencies of the responses, which we detail below.

\subsection{Influences from the Cultural Background}\label{subsec_Results_InfluencesBackground}

Students with a cultural background from social sciences differed significantly from students with a cultural background from computer science. `Social sciences' students were significantly more prone than computer science students to describe their choices matching the rationale `balance' from Section~\ref{sec_BiasRevealing}.\ref{rationaleBalance}, whereas computer science students were significantly more prone than `Social sciences' students to describe their choices matching the rationale `algorithm' from Section~\ref{sec_BiasRevealing}.\ref{rationaleAlgorithm}: $X^{2}(1,N=71)=8.1686$, $p<.05$ (with calculated p-value of .004262). The results support the hypothesis that the cultural background influences people when they carry out programming tasks under conditions of uncertainty.

The statistical significance test, as well as the graph in the Figure~\ref{chart_backgroundInfluences_allTreatments} consider the total number of responses, from all four treatments. The same observations about the cultural background influence are confirmed also when looking only at the control group (see the graph in the Figure~\ref{chart_backgroundInfluences_control}), though a statistical test is not relevant in this case, given the small number of responses. 
For both graphs the percentages are calculated from the `sensical' answers only.

\begin{figure}[hb]
\centering
\includegraphics[width=0.8\columnwidth]{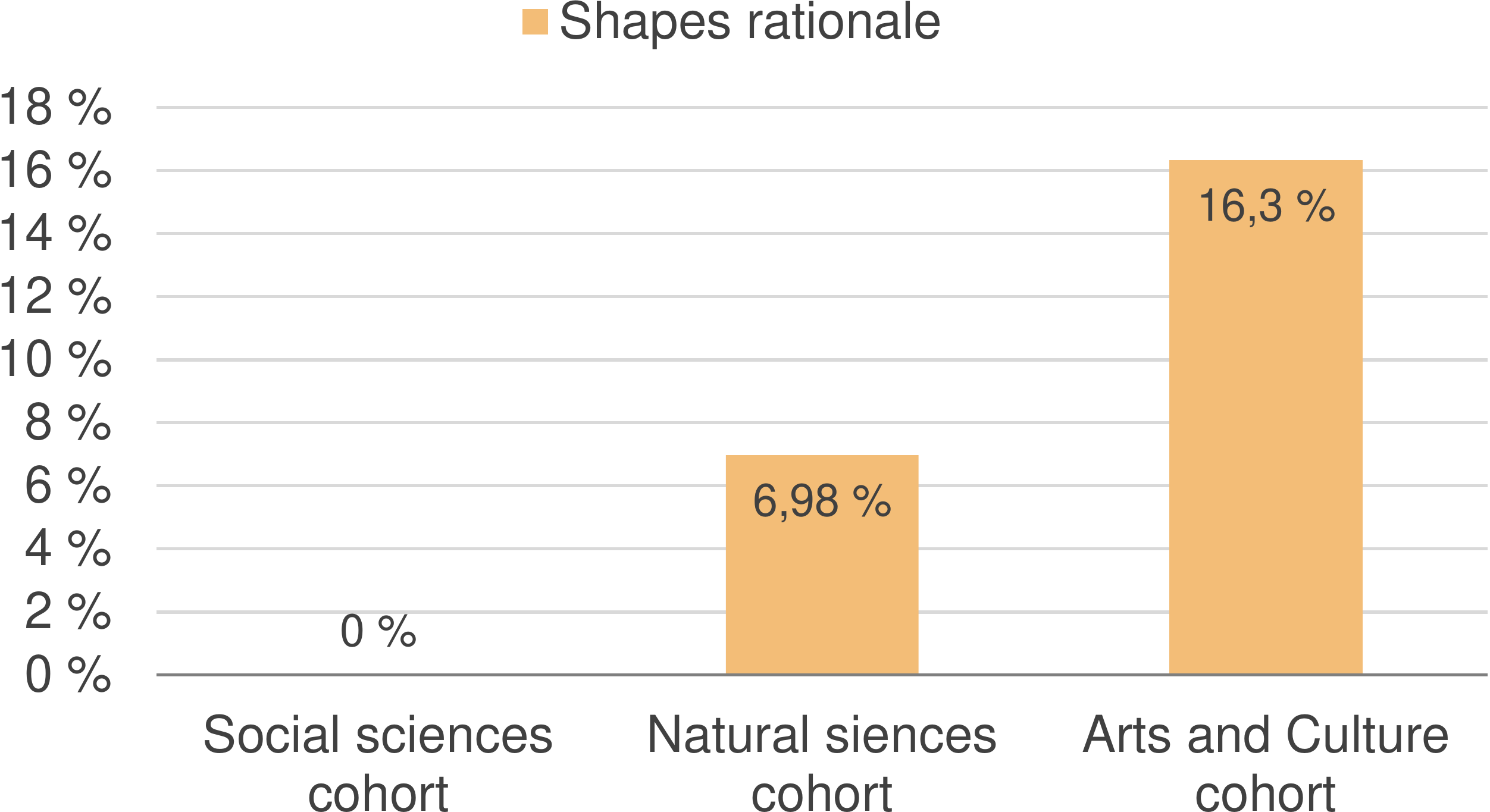}
\caption{Comparison of number of answers categorized in the `shapes' rationale.} 
\label{chart_backgroundInfluences_shapes}
\end{figure}

When analysing the results from the `Arts and Culture' cohort in comparison with the other two cohorts (see Figure~\ref{chart_backgroundInfluences_shapes}), we see that the influence of their artistic background makes them choose much more the `shapes' rationale from Section~\ref{sec_BiasRevealing}.\ref{rationaleShapes}; i.e., 16\% compared to 7\% in the `Natural sciences' cohort and 0\% in the `Social sciences' cohort.

\begin{figure}[t]
\centering
\includegraphics[width=0.55\columnwidth]{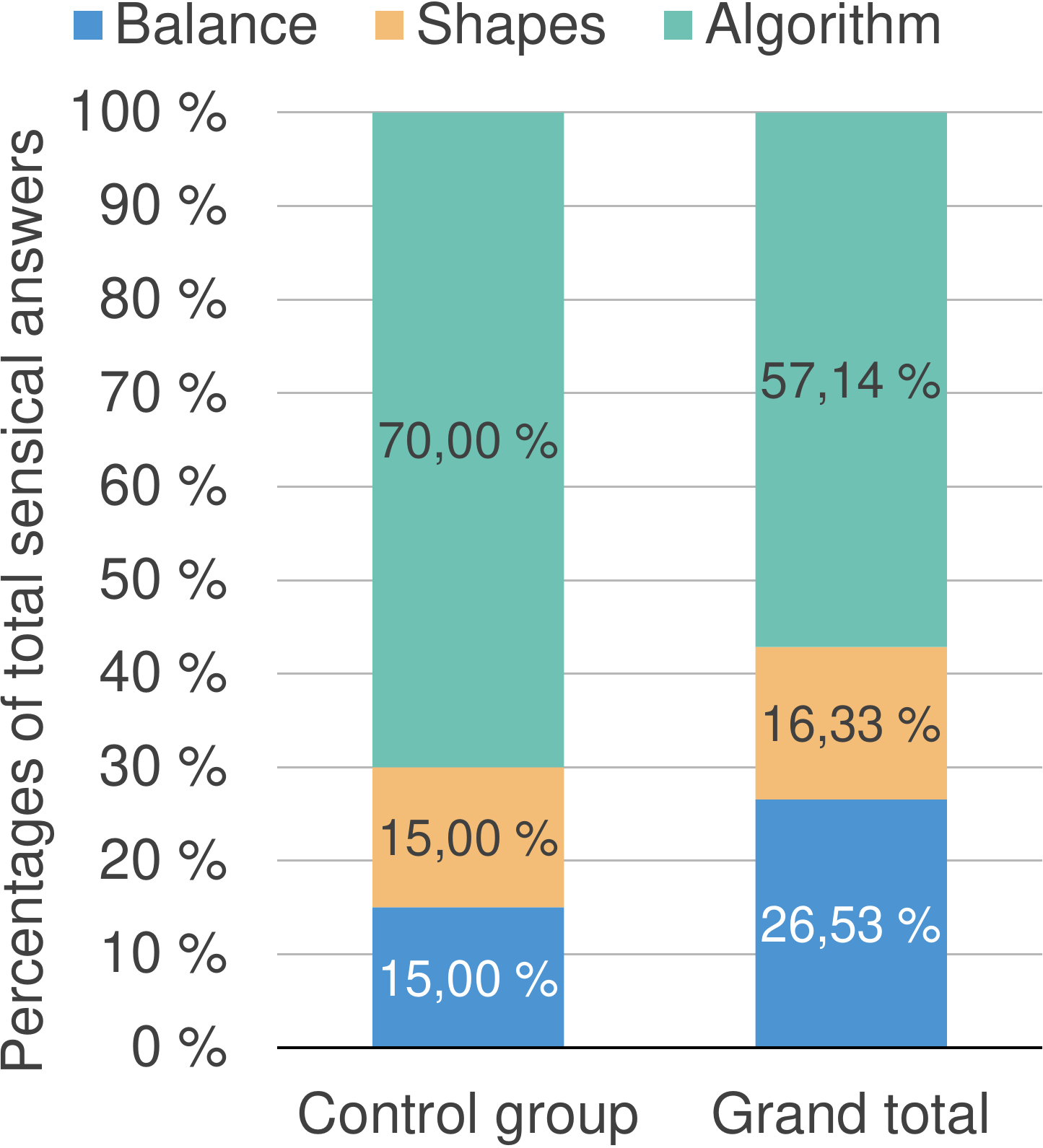}
\caption{Arts and Culture cohort -- percentage of answers distributed by rationale.} 
\label{chart_backgroundInfluences_artsandCultureCohort}
\end{figure}

Analysing further this cohort by itself, independently of the results obtained for the other cohorts,  we see in the Figure~\ref{chart_backgroundInfluences_artsandCultureCohort} that the answers conforming to the `algorithm' rationale are dominant; both when looking at all responses as well as only at the control group. This dominance could be explained by the fact that the respondents tried to comply with the nature and requirements of the exercise, i.e., a programming task where they were asked to assume the role of a programmer. One example of an answer from this cohort confirms this affirmation:\textit{ ``It was always drilled into my head in school, that when it came to math (which I assume is what most programming deals with) that the right side is always the right way\dots `right side right way' that's my reasoning here.''} The respondent tries in this case to bring in to his/her help the math knowledge s/he has from the school, as s/he assumes that informatics \textit{``deals with''}  mathematics. Another example is \textit{``I can't think of a better explanation but to involve mathematics in this game\dots ''}.

Moreover, when analysing qualitatively the answers to the \emph{`\pageNameExplanation'} question we found a considerable number of respondents that brought the game aspect of the task into their reasoning (more than 30 out of 110 explanations of the `Arts and Culture' cohort), i.e., they think in terms of programming a game. This is also an indication that these respondents focused on the task at hand, seeing the puzzle as part of this game programming task -- as they have been asked to -- and did not try to solve the puzzle per se. This increases our confidence in the fact that there was no debiasing happening, and that the respondents did not recognize that the task was in fact meant to reveal a background bias, let alone one of our three rationales or cohort backgrounds that we have assumed. Another aspect that could trigger debiasing is the fact that our puzzle does not have a `correct' answer wrt. the letter placement. However, we have found only two responses that have identified this fact (\textit{``[...]because of both dotted boxes are the correct answer. However, I feel[...]''} from pID M:11270469031, third cohort, and \textit{``[...]there isn't enough information for me to decide, so it is kind of a guess.''} from pID M:11270119183, third cohort); therefore, we rule out this debiasing possibility as well.

\subsection{Influences from the Priming Metaphors}\label{subsec_Results_InfluencesMetaphors}

The qualitative analysis of the respondents' answers to the \emph{`\pageNameExplanation'} shows three instances where the priming metaphors influence the answers of the participants. Here are three examples from the `Arts and Culture' cohort that show how the respondents quote directly the primes from the \emph{`\pageNameStory'} to help in arguing their reasoning behind the choice of letter placement. All three answers fall into the rational category related to the prime; and all invoke only some of the four words that we used for priming. Moreover, these words are taken both from the start and end of the story, which confirms our decision of using several words placed at different points inside the \emph{`\pageNameStory'}. 

\begin{itemize}
\item pID M:11271203853, third cohort: ``If this game is based on the philosopher's tenet of \emph{balance and harmony}, then[...]''\footnote{The words are found at the start of the story.}
\item pID M:11270378880, third cohort: ``The player should be rewarded when he/she places the letter on the right side because that is in keeping with the \emph{continuity and linear structure} of the game.''\footnote{There are actually three of the priming words mentioned here: ``linearity and continuity'' (from the end of the story) and ``structure'' from the start of the story. However, the respondent puts together ``linear structure''.}
\item pID 103, second cohort: ``The philosopher thought \emph{balance and equality} were important, and the player should therefore be rewarded for restoring the balance between the number of letters on the right and left sides.''\footnote{One word from the start of the story and one from the end.}
\end{itemize}

Since these examples are very few, they do not warrant a conclusion of \emph{conscious} transfer of priming bias, which is exactly the point with priming techniques, i.e., that people who are influenced through priming generally do \emph{not} realize it, and thus one does not normally see the priming expressed per se in the respondents arguments. Instead, the respondents being primed would make use of one or more of the heuristics that we mentioned in Sections~\ref{sec_CognitiveBiases} and \ref{sec_MetaphorsAsPrimingMethod}, e.g., the availability heuristic is most often used when people make quick judgements; we have encountered one respondent in which this heuristic has obviously manifested, pID M:11272410463, third cohort: \textit{``I choosed right previously but actually left makes more sense.  Balancing the sides; 4 letters on the left, 4 letters on the right.''}. 

Heuristics are also used substantially in situations of uncertainty, which is the case for our puzzle since we ask participants to find one `solution' to this new puzzle, which at the same time does not have one single correct answer, as any argument would be acceptable. In cases of uncertainty two additional heuristics are usually employed, namely the representativeness heuristic and the anchoring heuristic. If the problem at hand is new, then the mind tries to find another previously encountered problem that, to some extent, has some similarities. This is the case with the puzzle that we devised, aiming to trigger associations with aspects from the cultural/educational background of the person, e.g., `Natural sciences' respondents were expected to cling on to algorithms and the alphabet as an ordered source of indexing in mathematics, thus continuing along the line in our puzzle. The anchoring heuristic is even more important for priming since it is often employed when no useful information is readily available for the problem at hand, so the mind looks into the immediate context (e.g., physical, s.a., surroundings, or temporal, s.a., information received in the recent past, from the short-term memory) for clues. In our case the mind would anchor into the \emph{`\pageNameStory'} metaphor, and maybe draw on the meaning of one of the four priming words. 

In analyzing the explanations/responses we observed to some extent influences of our experimental manipulations, albeit not reaching statistical significance. Thus, since we can neither rule out a Type I error (i.e., failing to reject the research hypothesis, i.e., no priming effect) nor a Type II error (failing to reject the null hypothesis, i.e., there exists an effect, but we were not able to elicit it), the influence of contextual metaphors need to be further researched. In the rest of this section we present results from quantitative analyses of the priming effect and whether or not this transferred to the programs, i.e., found in the answers to the \emph{`\pageNameExplanation'} and \emph{`\pageNameAlternative'} questions.

\begin{figure}[t]
\centering
\includegraphics[width=\columnwidth]{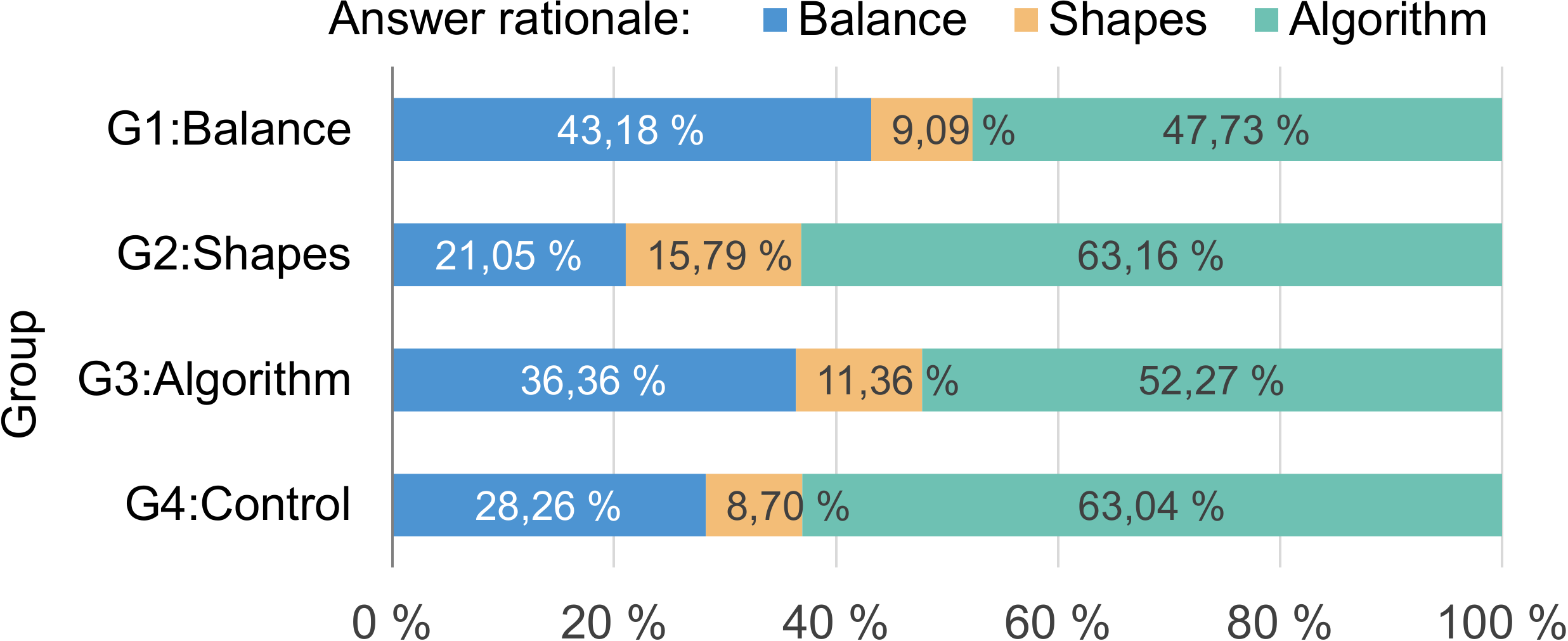}
\caption{Effect of priming, irrespective of background -- respondents from all cohorts put together.} 
\label{chart_primingEffect_allCohorts}
\end{figure}

The graph in the Figure~\ref{chart_primingEffect_allCohorts} shows the influence of the three groups of priming metaphors when the responses from all the cohorts are put together. This shows the priming effect irrespective of the participants' background. We compare each group with the control group. 

First of all, we observe that the `algorithm' group gives answers that cannot be readily seen as being influenced by priming. The same inconclusive observation is found also when looking inside each cohort, comparing  the `algorithm' group there with the respective control group. However, this is conforming with the observations made in Section~\ref{subsec_WordsSuggestions}, where the words used for priming are little known or maybe not understood by the participants, and thus cannot have an impact on their choice. However, one needs to take this conclusion with a grain of salt because the priming metaphor, depending on the anchoring heuristic, has a temporal flavour as it is stronger closer to the time of the priming; i.e., in our case the \emph{`\pageNameExplanation'} question is very close to the priming metaphor, whereas the \emph{`\pageNameWords'} question is farther away, maybe with a delay of a few minutes. This can mean that even if we do not see an effect of the metaphor in the \emph{`\pageNameWords'} question one can still have some effect in the \emph{`\pageNameExplanation'} question. Moreover, this can be compounded by other factors as well, such as for the \emph{`\pageNameWords'} question we are looking only for two of the words whereas in the \emph{`\pageNameExplanation'} question all our four priming words are in effect; or by the semantics of the words, which can have different meanings in different context, thus possibly causing one influence on the programming task and another influence on the synonyms generation task. Therefore, we focus in the rest of the section on the other two groups of priming. 

Secondly, when we analyse the other two groups we clearly observe priming influences, albeit of different kinds as explained further. For the `balance' group we see that the `balance' rationale increases from 28.26\% (for the control group) to 43.18\% (for the `balance' group), whereas in the case of the `shapes' group the `shapes' rationale increase from 8.70\% to 15.70\%; irrespective of the background. 

Besides these observations about the general strength of the priming metaphors, we are to a greater extent interested in their interactions with the educational/profession\-al background of the participants, as discussed in the previous subsection. Since we already established that the results related to the `algorithm' metaphors are not reliable, we exclude them from our further investigation. 

\begin{figure}[!b]
    \centering
    \begin{subfigure}[t]{\columnwidth}\vskip 0pt
        \centering
\includegraphics[width=\columnwidth]{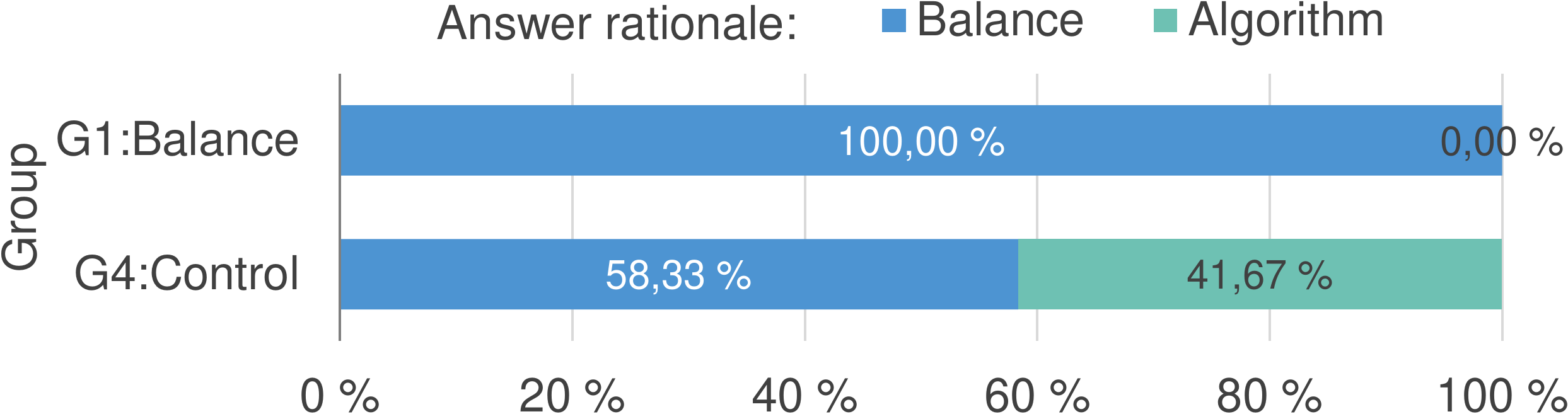}
\caption{\begin{footnotesize}Social sciences cohort.                                 \end{footnotesize}} 
\label{chart_primingEffect_socialSciences}
    \end{subfigure}
    \vspace{0.2cm}
    
    \begin{subfigure}[t]{\columnwidth}\vskip 0pt
        \centering
\includegraphics[width=\columnwidth]{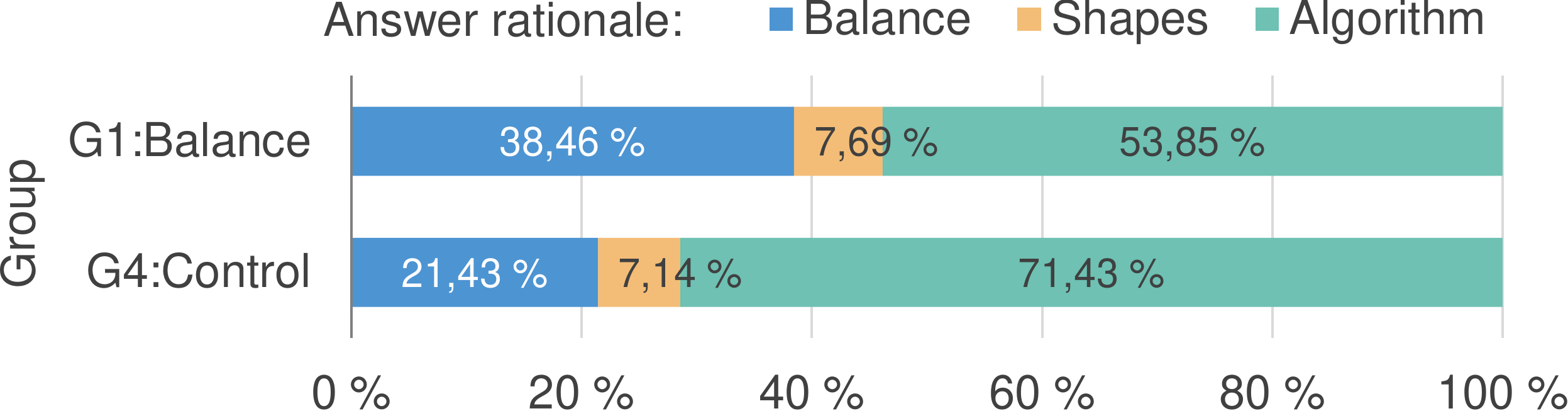}
\caption{\begin{footnotesize}Natural sciences cohort.                            \end{footnotesize}} 
\label{chart_primingEffect_naturalSciences}
   \end{subfigure}
   \vspace{0.2cm}
   
   \begin{subfigure}[t]{\columnwidth}\vskip 0pt
        \centering
\includegraphics[width=\columnwidth]{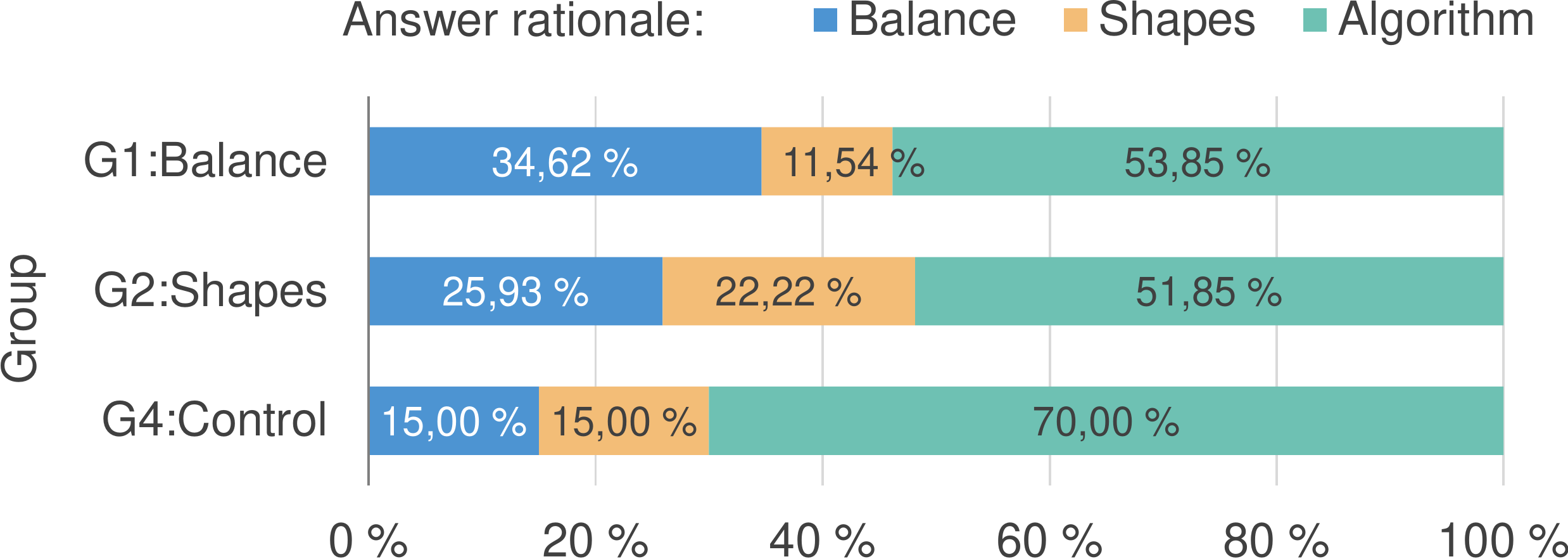}
\caption{\begin{footnotesize}Arts and Culture cohort.                            \end{footnotesize}} 
\label{chart_primingEffect_artsandCulture}
   \end{subfigure}%
    \caption{Effect of priming, inside each cohort.}
\label{chart_primingEffect}
\end{figure}

In the chart from the Figure~\ref{chart_primingEffect_socialSciences}, related to the `Social sciences' cohort, we observe that the background of the respondents is further strengthened by the priming metaphors, with an increase from  58.33\% to 100\%. Quite the opposite, in the case of the `Natural sciences' cohort (see Figure~\ref{chart_primingEffect_naturalSciences}), we see a weakening effect of their background since the `algorithms' rationale decreases from 71.43\% to 58.85\% in the group that was influenced with the `balance' metaphor, in favor of the `balance' rationale. For the `Arts and Culture' cohort we again see (Figure~\ref{chart_primingEffect_artsandCulture}) that the `shapes' metaphor strengthens their background since the `shapes' rationale increases from 15\% to 22.22\% inside the group that was primed with the `shapes' metaphor. For this cohort also the `balance' metaphor has an influence (from 15\% to 34.62\%), due to the fact that this metaphor's words were well chosen, as we have observed in Section~\ref{subsec_WordsSuggestions}.

We can conclude that the contextual metaphors that have been deemed as strong enough in the control question \emph{`\pageNameWords'} are also found to have an effect in strengthening or weakening the influence from metaphors in the cultural background of the respondents. Contextual metaphors have a strengthening effect when the words are representative of the respective cultural background, e.g., the `balance' rationale is strengthened by the priming metaphor words \emph{`harmony and balance; equality and fairness'}, whereas the `shapes' rationale by the \emph{`aesthetics and beauty; forms of arts'} metaphor words. On the other hand, when the contextual metaphor is well chosen, such as for the `balance' group, it weakens the effect from the cultural background of the respondents, as is the case with the `balance' metaphor, which when applied in the `Natural sciences' cohort, it increases the respective rationale.

\section{Conclusions and Discussions}

The aim of this study, as well as its implications, are manifold. The study can be categorized as both 
(i) 
a comparative/experimental study of how biases from cultural and contextual metaphors can be transferred from programmers to programs, and 
(ii) 
an exploratory study on how to develop ergonomically valid and reliable instruments, procedures and testing conditions to empirically study such biases transfer.
As such, this paper is a foundation for future research endeavours to \emph{improve} and \emph{diversify} these instruments, procedures and testing conditions. 

The strengths of this work reside in its exploratory nature in studying a hitherto not researched phenomenon, namely the transfer of human biases from the (not necessarily expert) programmer to the artefact that is developed (or configured).
Concretely, we have exposed (in Section~\ref{sec_Results}) interesting aspects of our main hypothesis, namely that machine bias may originate not only from biased data, but also from the programmer's biases in terms of influences from the cultural background as well as contextual influences from the programming environment.

Interestingly, under conditions of uncertainty (e.g., in the absence of instructions or specifications, something which is often the case for ubiquitous systems programming carried out increasingly by non-experts), we observe that the programmers' cultural background influences the choices they make and are subsequently transferred from the programmer to the program artifact. Thus, cultural metaphors in terms of irrelevant and inappropriate influences on the programming task at hand, represent instances of biases that are being transferred from humans to machines. This implies that human culture `transfers' to machines through the humans that program these, thus representing a strong source of bias. 

Equally interesting, attempts to moderate the strong influence from the cultural metaphors by means of experimentally introducing `hidden' (i.e., not consciously detected) contextual metaphors, were only successful to a certain extent. When the priming metaphor was chosen well (as in the case of `philosopher story' related to the `balance' rationale; with words that were easy to understand and rather common in a standard vocabulary) we saw influences in both directions of strengthening the cultural background as well as moderating it, each time tipping the balance of answers in the direction of the metaphor. These findings are orthogonal to what traditional and current machine bias research suggests, i.e., that machine bias originates from data, and thus our findings provide new insights into the origins of bias in the wide spreading AI and decision-support systems.

We believe that the present study shows how various aspects regarding design, instruments, and procedures can be successfully explored and controlled, and consequently incorporated in future studies that could (i) extend the present study by exploring related causes and mechanisms that lead to the transfer of bias from programmers to programs, as well as (ii) improve the designs, instruments and procedures in order to undertake this expanded endeavour.

\subsection{Discussions and Limitations}

One potential limitation of our study is the relatively small numbers of explanations/responses that we obtained in the experimental groups, despite the fact that we have had quite a large number of participants, i.e., 242 completely sensical explanations/answers that we were able to interpret and categorize, plus 55 more that were not completely nonsensical, but  nevertheless uninterpretable and uncategorizable. This limitation is explained, however, by our intention to perform a manifold exploratory study, investigating several related aspects and research questions, as well as providing a transparent foundation for further future improvements of design, instruments and procedures. As a consequence we arrived at a limited number of sensical and interpretable explanations `inside' each of the respective cohorts. However, a major intention in our study is to allow future researchers to take our study as a point of departure and carry out more controlled studies of some of the single aspects that we identified.  In order to achieve increased confidence in the results that we have presented here, it would be valuable to perform our survey with even more participants, maybe in the order of $1300+$ (estimating, according to the observations from Section~\ref{subsec_SensicalVsNonsensical}, with a 4-to-1 ratio of sensical vs. nonsensical results). Such a number would allow for all our groups to be populated with $30+$ responses, something which would allow for an even more valuable use of statistical significance tests.

It proved difficult to achieve our number of 300 participants for this type of study; however, if one would want to restrict a similar study to only professionals, one of the many online forums or communities for programmers could be a good place to recruit participants (e.g., on \url{freelancer.com} or \url{stackoverflow.com}).

One interesting speculative observation that we would like to make out of our results regards a potential effect resulting from the difference between (i) interpreting information based on its \emph{structure} and thus as something systemic that is `detached' from having individual characteristics, versus (ii) interpreting information based on its \emph{content} and thus as having individual characteristics. For example, subject-programmers that chose the rationales of `balance' or `algorithm' may view information (as the one coming from the `game board' puzzle picture that we showed them) merely as representing structure and may thus have disregarded the notion that data could also have individual characteristics in addition to being part of an overall structure. Contrary to this, respondents that chose `shapes' may in fact have acknowledged the notion that data do have individual characteristics and are thus not `only' part of an overall structure `outside' the data's individual characteristics. Interestingly, subjects in the arts \& culture cohort provided explanations in terms of `shapes' substantially more often than subjects in the `Social sciences' cohort and the `Computer science' cohort. This could indicate that people with a cultural background (judging from  their education and/or profession) from arts and culture are more prone than others to view data as representation of individuals that have unique characteristics, rather than viewing data only as being part of an overall structure. In other words, people with a background in arts and culture may possibly exhibit a more `human' interpretation of data, or at least they may be more prone than people from other cultural backgrounds to acknowledge data as `individual' rather than `systemic'. 

\subsection{Possible Future Research Directions}

Future studies are invited to investigate more deeply any of the aspects that we have explored and tentatively concluded from. One venue for future research would be to refine our study design's ability to elicit cultural or contextual influences in an even more fine-grained manner, specifically by improving our instruments and procedures. 

One possibility is to perform similar studies focused on specific categories of subjects that can be seen as programmers, e.g., one playfull possibility could be to study children as programmers -- programming languages/environ\-ments specific for children abund, s.a. Google's Blockly \cite{trower2015creating,weintrop2017comparing} or MIT's Scratch \cite{resnick2009scratch,maloney2010scratch,armoni2015scratch}. Studies on biases in adults are more available \cite{klaczynski1997goal,klaczynski2000personal,bruine2007individual} whereas studies on biases in children are less \cite{baron1993decision,klaczynski1997bias}. One could argue that this is because children are not biased; others could claim that ethical considerations make such studies of children too difficult to carry out; yet others could argue that biases in children are distinctly different from biases in adults, given the differences in mental representations from children and adults. However, we think that it is important to test the age aspect in biases transferred to programs, given the ubiquity and pervasiveness of IoT-programming in everyday life for all age groups.

One useful refinement of our work could be to study professional programmers in a professional environment,  both (i) classical programming environments, such as in a programming company guided by software development life cycle methods and tools, maybe focusing on current emerging programming cultures like Scrum or DevOps; and (ii) non-expert programming environments, such as complex configurations, DSLs, graphical programming, or curating of big data. It is possible to study different questions in this setting, for example:

\begin{itemize}
\item \emph{On what avenues biases arrive into the programming environment?}
We have assumed that it is a result of underspecified requirements; i.e., when some functionality is left open and the programmer does not have the resources to find more specific details. This is a most common form of uncertainty in programming; but there are others as well, and which of these give way to biases is important to know so that one can build debiasing techniques, maybe even incorporated in the tools of the programmers, like in IDEs (integrated development environments).
\item \emph{Are expert programmers, when under the scrutiny of their tools and methods, like testing suites, still transferring their biases into their programs?}
Is this happening only when they are given choices, or also in other situations (e.g., even when fully specified requirements are given; or when working with big data)?
\item \emph{Are programmers immune to biases because of their education or because of their work (e.g., because at work their `mind-set' is a ``mathematical'' one)?} 
\end{itemize}

One good source of alternative investigations can be the study of specific biases in specific situations or social activities where software is paramount. One example can be biases related to privacy in the big data economy (sometimes called the `surveillance capitalism' \cite{zuboff2019age}), e.g.: Are privacy related concepts or views from the cultural background -- which is specific to the programmer -- transferred to the software -- which is used on an international scale? One can imagine a programmer coming from a cultural background that always promotes the slogan \textit{``You have zero privacy; get over it!''}, or another programmer from a background that \textit{``is entrenched by rules and regulations about who/how any form of private electronic data can be used''}. Are such different cultural views transferred to the software built by these two different programmers? What is the global influence of such bias transfers? In this setting, one could alternatively study biases coming from the user of the software (not the programmer) to see whether the user biases (call them `wishes' or `needs') are transferred to the software through specifications elicitation, user stories, and other interaction design methods \cite{interdesign11book,lazar2017research} that are now a popular way of developing software systems.

We have studied two sources of biases, namely cultural biases and priming metaphors, that we consider situated at the two extremes on the vertical axis from Figure~\ref{fig_paperOverview}, which indicates the strength of the bias, and also a temporal aspect regarding the persistence of these biases (e.g., priming may not be as strong as the culture, and acts on a short time scale, usually minutes after the priming is applied). One could study other sources of influence that can generate biases through the metaphors that they induce in the human, which would lie on our vertical axis in between the ones studied in this paper. Examples of influencing methods relevant for our study can be propaganda (i.e., misinformation \cite{Mintz2012misinformation,kumar2014misinfo} and disinformation \cite{graham03disinfo}), which may be done on limited but considerable stretches of time; or working cultures which can influence a programmer in different ways when changing jobs. 

\section*{Acknowledgements}

The authors would like to thank the student volunteers, their tutors and student administrative personell from Bj\o{}rknes College, from Institute for Informatics at the University of Oslo, and from  
Oslo Art School (\url{http://www.oslokulturskole.no}) 
and 
Norwegian Academy of Music (\url{https://nmh.no/}) 
for their help with our studies. We would also like to thank John S\"{o}ren Pettersson from the Karlstad University (Sweden) for allowing us to utilize their Usability laboratory and the eye tracking equipment for our pilot testing.

\section*{References}


\newpage
\appendix

\onecolumn

\section{The Questionnaire}\label{app_TheQuestionnaire}

\begin{figure*}[!h]
\centering
\includegraphics[width=0.9\textwidth]{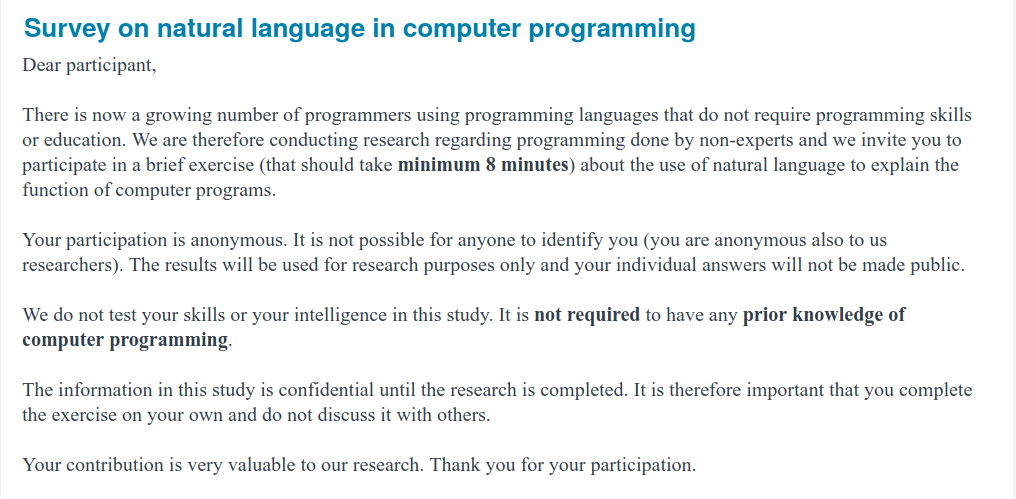}
\caption{Page 1 \emph{`\pageNameIntro'}.} 
\label{fig_IntroPage}
\end{figure*}

\begin{figure*}[!h]
\centering
\includegraphics[width=0.9\textwidth]{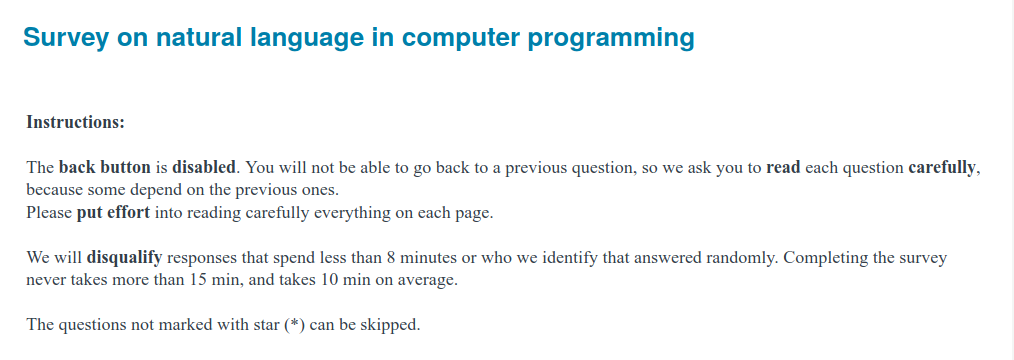}
\caption{Page 2 \emph{`\pageNameInstructions'}.} 
\label{fig_InstructionsPage}
\end{figure*}

\begin{figure*}[hp]
\centering
\includegraphics[width=0.9\textwidth]{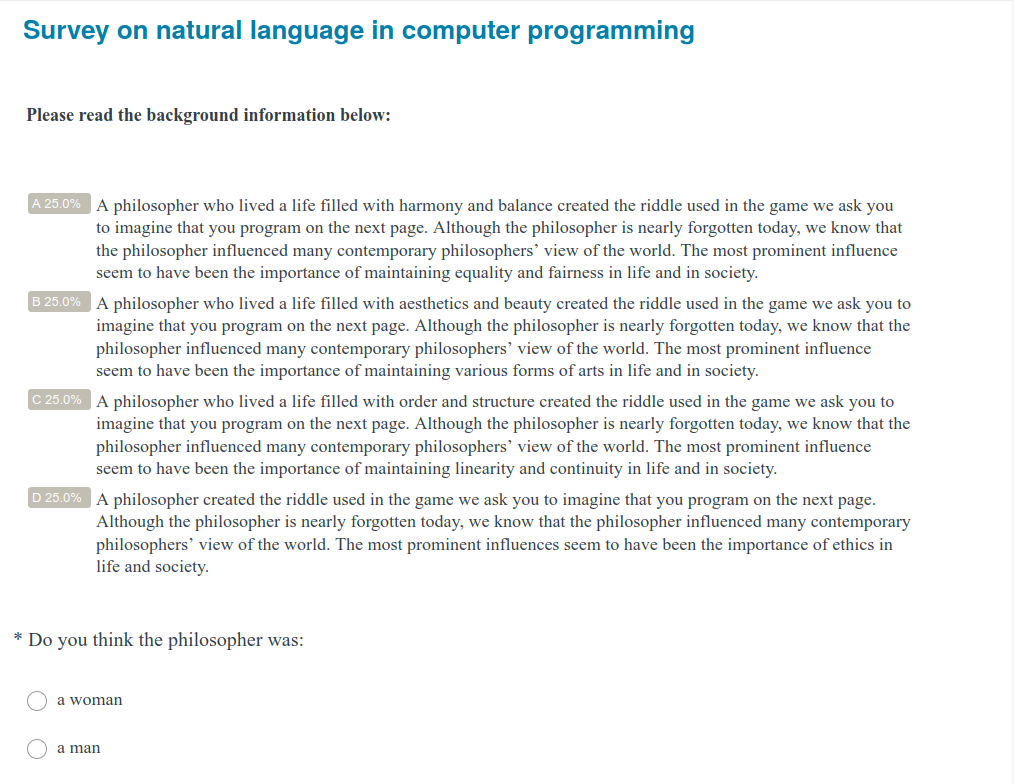}
\caption{Page 3 \emph{`\pageNameStory'}.} 
\label{fig_StoryPage}
\end{figure*}

\begin{figure*}[hp]
\centering
    \includegraphics[width=0.9\textwidth]{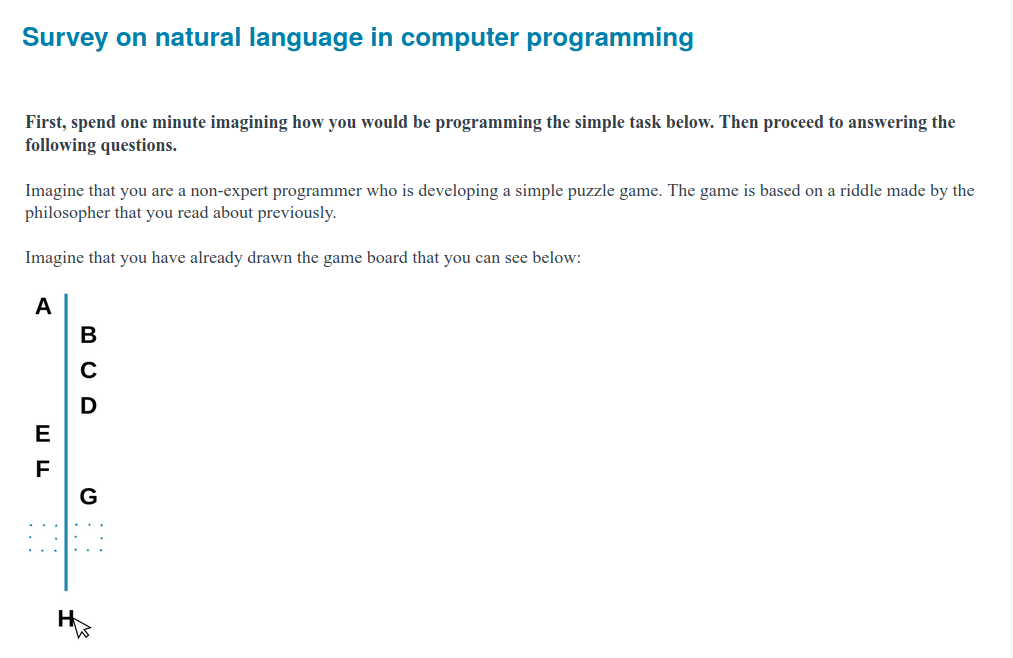}
    \includegraphics[width=0.9\textwidth]{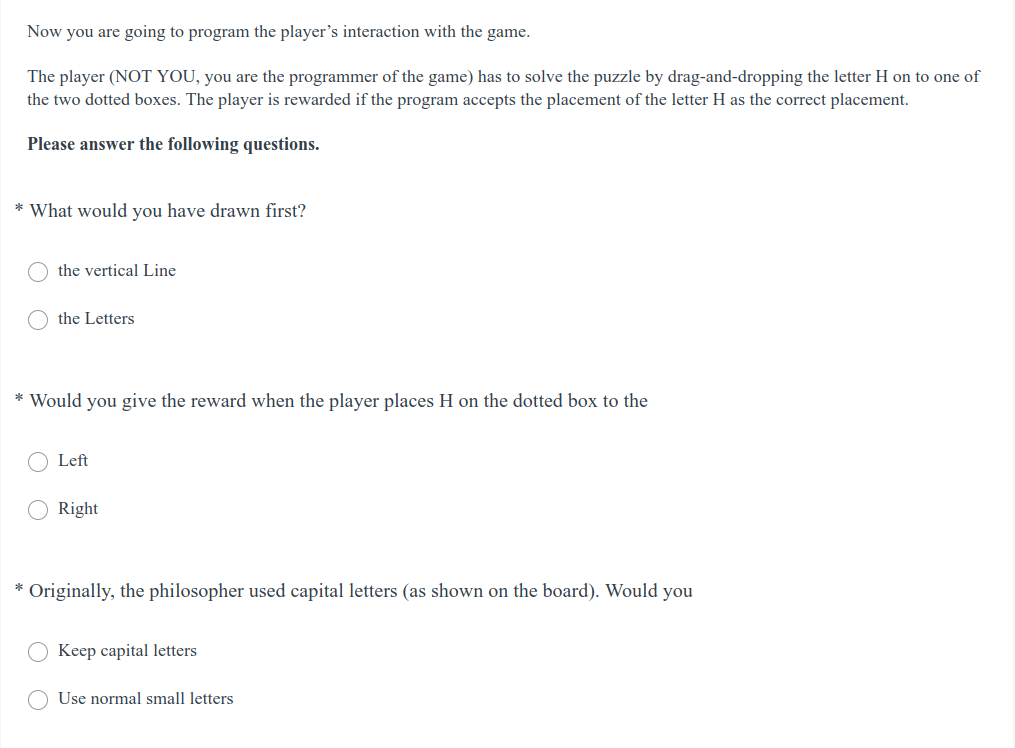}
    \caption{Page 4 \emph{`\pageNameTask'}.}
    \label{fig_TaskPage}
\end{figure*}

\begin{figure*}[hp]
\centering
\includegraphics[width=0.9\textwidth]{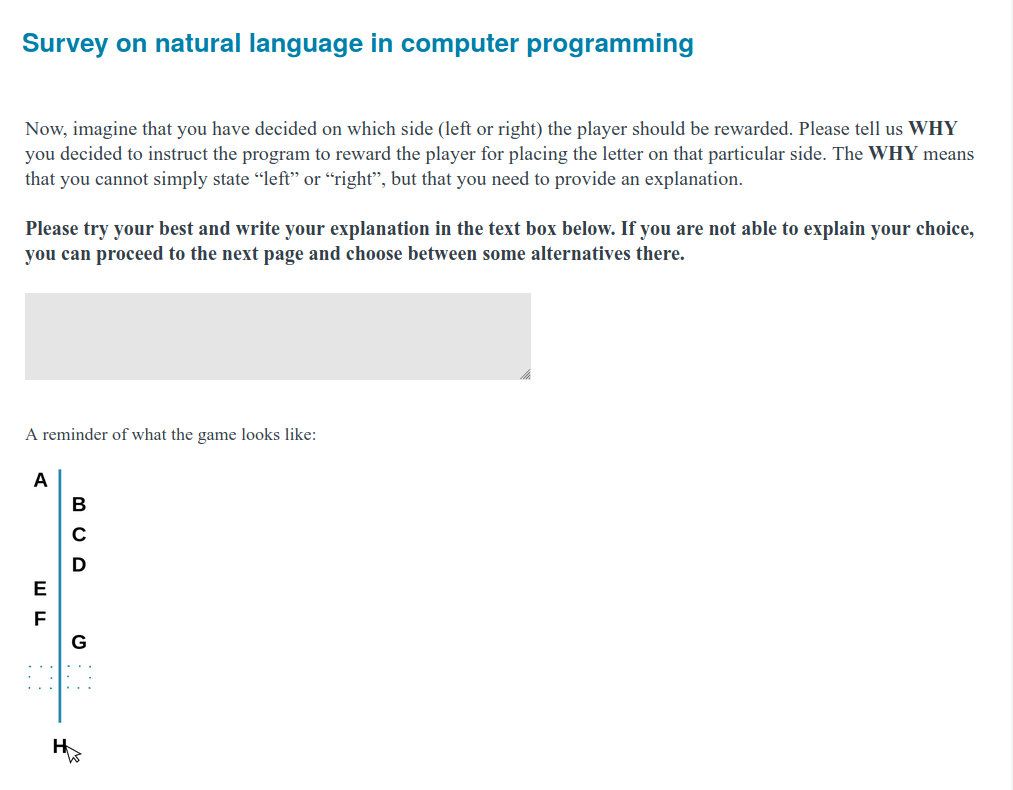}
\caption{Page 5 \emph{`\pageNameExplanation'}.} 
\label{fig_ExplanationPage}
\end{figure*}

\begin{figure*}[hp]
\centering
\includegraphics[width=0.9\textwidth]{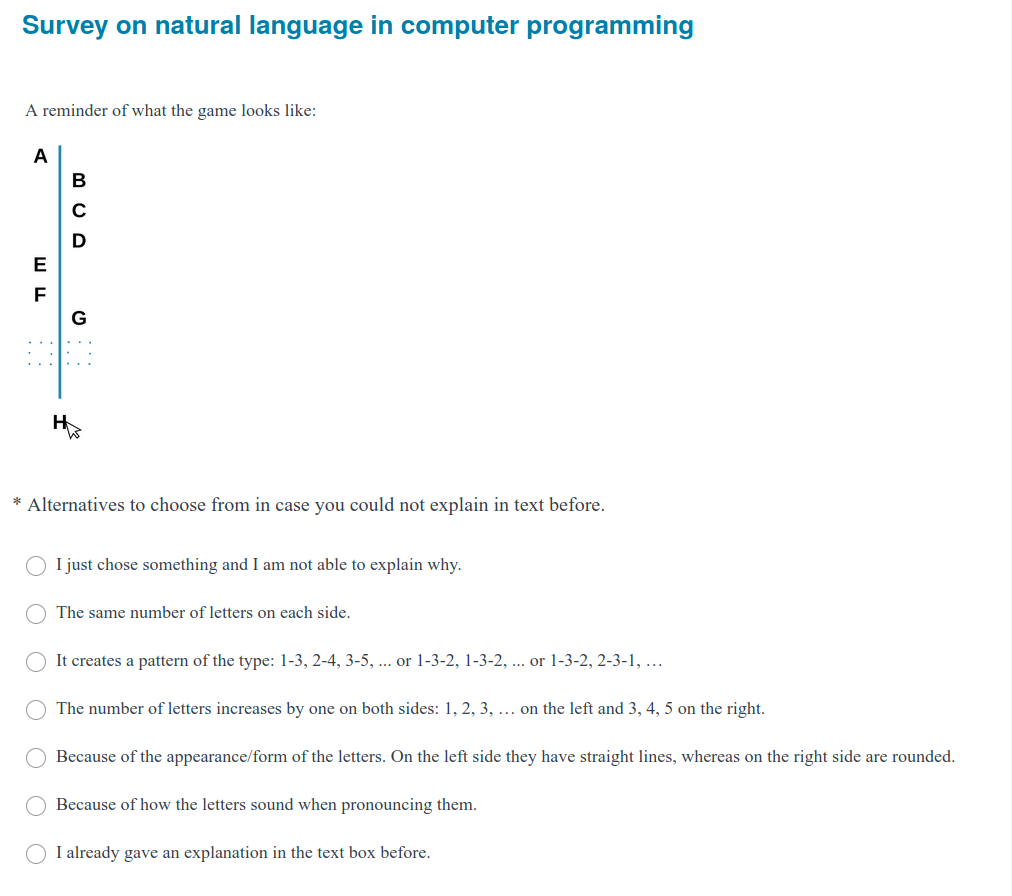}
\caption{Page 6 \emph{`\pageNameAlternative'}.} 
\label{fig_AlternativePage}
\end{figure*}

\begin{figure*}[hp]
\centering
\includegraphics[width=0.9\textwidth]{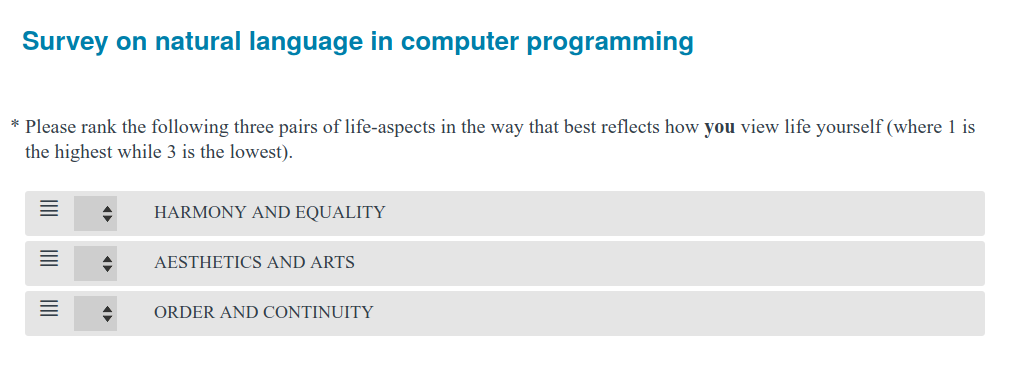}
\caption{Page 7 \emph{`\pageNameRanking'}.} 
\label{fig_RankingPage}
\end{figure*}

\begin{figure*}[hp]
\centering
\includegraphics[width=0.9\textwidth]{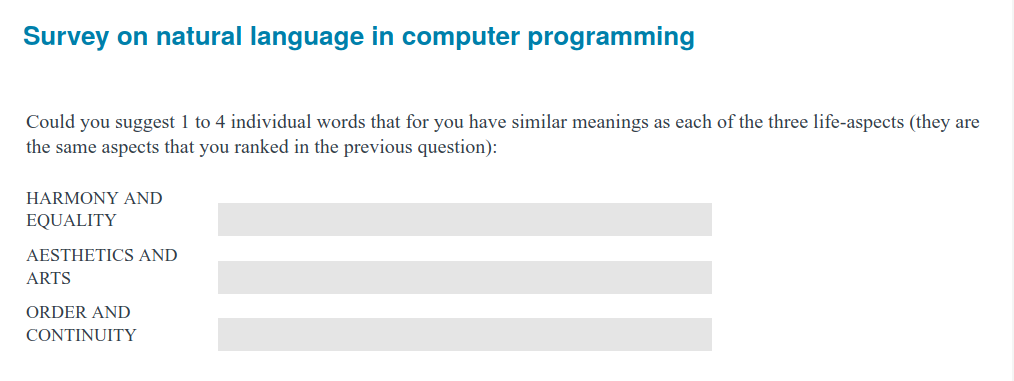}
\caption{Page 8 \emph{`\pageNameWords'}.} 
\label{fig_WordsPage}
\end{figure*}

\begin{figure*}[hp]
\centering
    \includegraphics[width=0.9\textwidth]{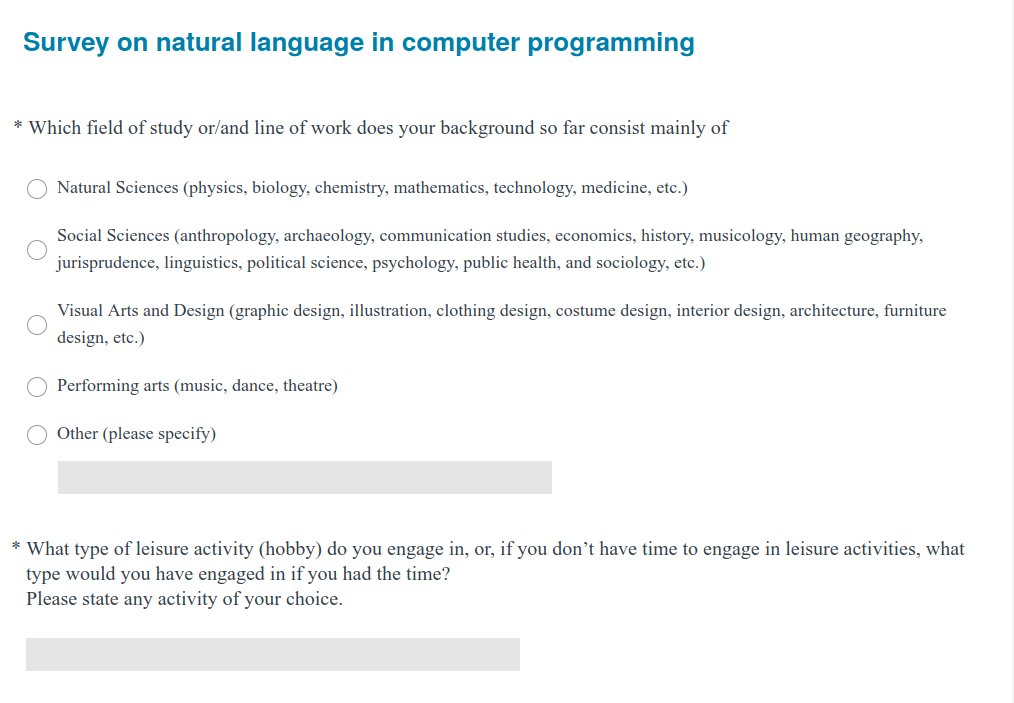}
    \includegraphics[width=0.9\textwidth]{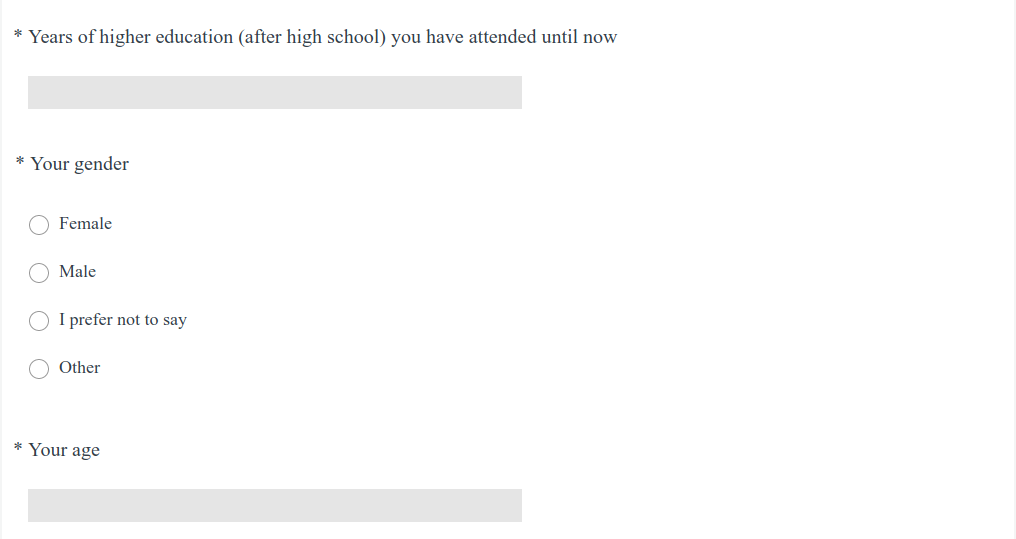}
\caption{Page 9 \emph{`\pageNameDemographics'}.} 
\label{fig_DemographicsPage}
\end{figure*}

\end{document}